%
%
\input harvmac
\newcount\yearltd\yearltd=\year\advance\yearltd by 0

\noblackbox

\input epsf

\def\tilde{\widetilde}
\def\hat{\widehat}
\newcount\figno
\figno=0
\def\fig#1#2#3{
\par\begingroup\parindent=0pt\leftskip=1cm\rightskip=1cm\parindent=0pt
\baselineskip=11pt
\global\advance\figno by 1
\midinsert
\epsfxsize=#3
\centerline{\epsfbox{#2}}
\vskip 12pt
{\bf Figure\ \the\figno: } #1\par
\endinsert\endgroup\par
}
\def\figlabel#1{\xdef#1{\the\figno}}
\def\encadremath#1{\vbox{\hrule\hbox{\vrule\kern8pt\vbox{\kern8pt
\hbox{$\displaystyle #1$}\kern8pt}
\kern8pt\vrule}\hrule}}

\def\threesidefig#1#2#3#4#5{
\par\begingroup\parindent=0pt\leftskip=1cm\rightskip=1cm\parindent=0pt
\baselineskip=11pt \global\advance\figno by 1 \midinsert
\centerline{
\epsfxsize=#2
\epsfbox{#3}
\epsfxsize=#2
\epsfbox{#4}
\epsfxsize=#2
\epsfbox{#5}
}
\vskip 12pt
{\bf Figure\ \the\figno: } #1\par
\endinsert\endgroup\par
}

\def\apm{{\alpha^{\prime}}}

 \def\a{{\alpha}}
 
 \def\frac#1#2{{#1\over #2}}
 \def\l{{\lambda}}

 \def\CO{{\cal O}}
 
 \def\Ph{{\Phi }}

 \def\CD{{\cal D}}

 \def\apm{{\a^{\prime}}}
 \def\r{\rightarrow}

\def\lp{{\lambda^{\prime}}}

\def\tit{{\tilde{t}}}
\def\IZ{\relax\ifmmode\hbox{Z\kern-.4em Z}\else{Z\kern-.4em Z}\fi}
\def\IR{\relax{\rm I\kern-.18em R}}
\def\tl{{\hat{\l}}}

\lref\glpaper{ O.~Aharony, J.~Marsano, S.~Minwalla and T.~Wiseman,
``Black hole - black string phase transitions in thermal $1+1$ dimensional
supersymmetric Yang-Mills theory on a circle,''
Class.\ Quant.\ Grav.\  {\bf 21}, 5169 (2004)
[arXiv:hep-th/0406210].
}

\lref\sundborg{
B.~Sundborg,
``The Hagedorn transition, deconfinement and $N = 4$ SYM theory,''
Nucl.\ Phys.\ B {\bf 573}, 349 (2000)
[arXiv:hep-th/9908001].
}

\lref\hawkingpage{
S.~W.~Hawking and D.~N.~Page,
``Thermodynamics of black holes in anti-de Sitter space,''
Commun.\ Math.\ Phys.\  {\bf 87}, 577 (1983).
}

\lref\witteno{
E.~Witten,
``Anti-de Sitter space and holography,''
Adv.\ Theor.\ Math.\ Phys.\  {\bf 2}, 253 (1998)
[arXiv:hep-th/9802150].
}

\lref\wittent{
E.~Witten,
``Anti-de Sitter space, thermal phase transition, and confinement in  gauge
theories,''
Adv.\ Theor.\ Math.\ Phys.\  {\bf 2}, 505 (1998)
[arXiv:hep-th/9803131].
}

\lref\adsrev{
O.~Aharony, S.~S.~Gubser, J.~M.~Maldacena, H.~Ooguri and Y.~Oz,
``Large N field theories, string theory and gravity,''
Phys.\ Rept.\  {\bf 323}, 183 (2000)
[arXiv:hep-th/9905111].
}

\lref\susskind{
L.~Susskind,
``Matrix theory black holes and the Gross Witten transition,''
arXiv:hep-th/9805115.
}

\lref\Staudacher{
W.~Krauth and M.~Staudacher,
``Eigenvalue distributions in Yang-Mills integrals,''
Phys.\ Lett.\ B {\bf 453}, 253 (1999)
[arXiv:hep-th/9902113].
}

\lref\Hotta{
T.~Hotta, J.~Nishimura and A.~Tsuchiya,
``Dynamical aspects of large $N$ reduced models,''
Nucl.\ Phys.\ B {\bf 545}, 543 (1999)
[arXiv:hep-th/9811220].
}

\lref\Aoki{
H.~Aoki, S.~Iso, H.~Kawai, Y.~Kitazawa and T.~Tada,
``Space-time structures from IIB matrix model,''
Prog.\ Theor.\ Phys.\  {\bf 99}, 713 (1998)
[arXiv:hep-th/9802085].
}

\lref\Austingthesis{
P.~Austing,
``Yang-Mills matrix theory,''
arXiv:hep-th/0108128.
}

\lref\Austing{
P.~Austing and J.~F.~Wheater,
``Convergent Yang-Mills matrix theories,''
JHEP {\bf 0104}, 019 (2001)
[arXiv:hep-th/0103159].
}

\lref\pol{Polyakov,
\hbox{} }

\lref\creutz{
M.~Creutz,
``On Invariant Integration Over SU(N),''
J.\ Math.\ Phys.\  {\bf 19}, 2043 (1978).
}
\lref\pol{Polyakov,
\hbox{hep-th/0003160}}

\lref\gw{
D.~J.~Gross and E.~Witten,
``Possible third order phase transition in the large $N$ lattice gauge theory,''
Phys.\ Rev.\ D {\bf 21}, 446 (1980).
}
\lref\wigner{Weigner}
\lref\itziksonzuber{Itzikson and Zuber}

\lref\AmbjornXW{
J.~Ambjorn and S.~Wolfram,
``Properties Of The Vacuum. 1. Mechanical And Thermodynamic,''
Annals Phys.\  {\bf 147}, 1 (1983).
}

\lref\sz{
  G.~W.~Semenoff, O.~Tirkkonen and K.~Zarembo,
  ``Exact solution of the one-dimensional non-Abelian Coulomb gas at  large
  N,''
  Phys.\ Rev.\ Lett.\  {\bf 77}, 2174 (1996)
  [arXiv:hep-th/9605172].
}

\lref\jh{
J. Leyland-Harris, ``Analysis of Phase Transitions in Weakly Coupled
Large N Gauge Theories in 0+1 Dimensions'', UBC Undergraduate Thesis,
April 2004.
}

\lref\martineco{
M.~Li, E.~J.~Martinec and V.~Sahakian,
``Black holes and the SYM phase diagram,''
Phys.\ Rev.\ D {\bf 59}, 044035 (1999)
[arXiv:hep-th/9809061].
}

\lref\martinect{
E.~J.~Martinec and V.~Sahakian,
``Black holes and the SYM phase diagram. II,''
Phys.\ Rev.\ D {\bf 59}, 124005 (1999)
[arXiv:hep-th/9810224].
}

\lref\Nekrasov{
G.~W.~Moore, N.~Nekrasov and S.~Shatashvili,
``D-particle bound states and generalized instantons,''
Commun.\ Math.\ Phys.\  {\bf 209}, 77 (2000)
[arXiv:hep-th/9803265].
}

\lref\GrossTU{
D.~J.~Gross,
``Two-dimensional QCD as a string theory,''
Nucl.\ Phys.\ B {\bf 400}, 161 (1993)
[arXiv:hep-th/9212149].
}

\lref\GrossHU{
D.~J.~Gross and W.~I.~Taylor,
``Two-dimensional QCD is a string theory,''
Nucl.\ Phys.\ B {\bf 400}, 181 (1993)
[arXiv:hep-th/9301068].
}

\lref\GrossYT{
D.~J.~Gross and W.~I.~Taylor,
``Twists and Wilson loops in the string theory of two-dimensional QCD,''
Nucl.\ Phys.\ B {\bf 403}, 395 (1993)
[arXiv:hep-th/9303046].
}

\lref\MigdalAB{
A.~Migdal,
``Recursion equations in gauge theories,''
Sov.\ Phys.\ JETP.\ {\bf 42}, 413 (1975)
}

\lref\ItzyksonFI{
C.~Itzykson and J.~B.~Zuber,
``The Planar Approximation. 2,''
J.\ Math.\ Phys.\  {\bf 21}, 411 (1980).
}

\lref\SamuelVK{
S.~Samuel,
``U(N) Integrals, 1/N, And The Dewit-'T Hooft Anomalies,''
J.\ Math.\ Phys.\  {\bf 21}, 2695 (1980).
}

\lref\thooft{ G.~'t Hooft, ``A planar diagram theory for strong
interactions,'' Nucl.\ Phys.\ B {\bf 72}, 461 (1974).}

\lref\tHooftUJ{ G.~'t Hooft, ``A property of electric and magnetic
flux in nonabelian gauge theories,'' Nucl.\ Phys.\ B {\bf 153}, 141
(1979).
}

\lref\tHooftHY{ G.~'t Hooft, ``On the phase transition towards
permanent quark confinement,'' Nucl.\ Phys.\ B {\bf 138}, 1 (1978).
}

\lref\sphere{
O.~Aharony, J.~Marsano, S.~Minwalla,
K.~Papadodimas and M.~Van Raamsdonk, ``The Hagedorn / deconfinement
phase transition in weakly coupled large N gauge theories,''
Adv.\ Theor.\ Math.\ Phys.\  {\bf 8}, 603 (2004)
[arXiv:hep-th/0310285].
}

\lref\ItzhakiDD{ N.~Itzhaki, J.~M.~Maldacena, J.~Sonnenschein and
S.~Yankielowicz, ``Supergravity and the large N limit of theories
with sixteen  supercharges,'' Phys.\ Rev.\ D {\bf 58}, 046004 (1998)
[arXiv:hep-th/9802042].
}

\lref\adscft{ J.~M.~Maldacena, ``The large $N$ limit of
superconformal field theories and supergravity,'' Adv.\ Theor.\
Math.\ Phys.\  {\bf 2}, 231 (1998) [Int.\ J.\ Theor.\ Phys.\  {\bf
38}, 1113 (1999)] [arXiv:hep-th/9711200].}

\lref\MaldacenaIM{
  J.~M.~Maldacena,
  ``Wilson loops in large N field theories,''
  Phys.\ Rev.\ Lett.\  {\bf 80}, 4859 (1998)
  [arXiv:hep-th/9803002].
}

\lref\ReyIK{
  S.~J.~Rey and J.~T.~Yee,
  ``Macroscopic strings as heavy quarks in large N gauge theory and  anti-de
  Sitter supergravity,''
  Eur.\ Phys.\ J.\ C {\bf 22}, 379 (2001)
  [arXiv:hep-th/9803001].
}

\lref\BarbonCR{ J.~L.~F.~Barbon, I.~I.~Kogan and E.~Rabinovici, ``On
stringy thresholds in SYM/AdS thermodynamics,'' Nucl.\ Phys.\ B {\bf
544}, 104 (1999) [arXiv:hep-th/9809033].
}

\lref\MaldacenaBW{
  J.~M.~Maldacena and A.~Strominger,
  ``AdS(3) black holes and a stringy exclusion principle,''
  JHEP {\bf 9812}, 005 (1998)
  [arXiv:hep-th/9804085].
}

\lref\DijkgraafFQ{
  R.~Dijkgraaf, J.~M.~Maldacena, G.~W.~Moore and E.~P.~Verlinde,
  ``A black hole farey tail,''
  arXiv:hep-th/0005003.
}

\lref\SusskindDR{ L.~Susskind, ``Matrix theory black holes and the
Gross Witten transition,'' arXiv:hep-th/9805115.
}

\lref\BanksVH{ T.~Banks, W.~Fischler, S.~H.~Shenker and L.~Susskind,
``M theory as a matrix model: A conjecture,'' Phys.\ Rev.\ D {\bf
55}, 5112 (1997) [arXiv:hep-th/9610043].
}

\lref\kleb{ K.~Demeterfi, I.~R.~Klebanov and G.~Bhanot,
``Glueball spectrum in a (1+1)-dimensional model for QCD,''
Nucl.\ Phys.\ B {\bf 418}, 15 (1994) [arXiv:hep-th/9311015].
}

\lref\threelooppaper{
O.~Aharony, J.~Marsano, S.~Minwalla,
K.~Papadodimas and M.~Van Raamsdonk,
  ``A first order deconfinement transition in large N Yang-Mills theory on a
  small $S^3$,''
  Phys.\ Rev.\ D {\bf 71}, 125018 (2005)
  [arXiv:hep-th/0502149].
}

\lref\RusakovRS{
  B.~E.~Rusakov,
  ``Loop Averages And Partition Functions In U(N) Gauge Theory On
  Two-Dimensional Manifolds,''
  Mod.\ Phys.\ Lett.\ A {\bf 5}, 693 (1990).
}

\lref\WittenWE{
  E.~Witten,
  ``On quantum gauge theories in two-dimensions,''
  Commun.\ Math.\ Phys.\  {\bf 141}, 153 (1991).
}

\lref\FineZZ{
  D.~S.~Fine,
  ``Quantum Yang-Mills On The Two-Sphere,''
  Commun.\ Math.\ Phys.\  {\bf 134}, 273 (1990).
}

\lref\BlauMP{
  M.~Blau and G.~Thompson,
  ``Quantum Yang-Mills theory on arbitrary surfaces,''
  Int.\ J.\ Mod.\ Phys.\ A {\bf 7}, 3781 (1992).
}

\lref\WittenDF{
  E.~Witten,
  ``Constraints On Supersymmetry Breaking,''
  Nucl.\ Phys.\ B {\bf 202}, 253 (1982).
}

\lref\SchnitzerQT{ H.~J.~Schnitzer, ``Confinement / deconfinement
transition of large N gauge theories with  N(f) fundamentals: N(f)/N
finite,'' Nucl.\ Phys.\ B {\bf 695}, 267 (2004)
[arXiv:hep-th/0402219].}

\lref\SemenoffBS{ G.~W.~Semenoff, ``Matrix model thermodynamics,''
arXiv:hep-th/0405107.
}

\lref\LiuVY{ H.~Liu, ``Fine structure of Hagedorn transitions,''
arXiv:hep-th/0408001.
}

\lref\HadizadehBF{ S.~Hadizadeh, B.~Ramadanovic, G.~W.~Semenoff and
D.~Young, ``Free energy and phase transition of the matrix model on
a plane-wave,'' Phys.\ Rev.\ D {\bf 71}, 065016 (2005)
[arXiv:hep-th/0409318].
}

\lref\SpradlinPP{ M.~Spradlin and A.~Volovich,
Nucl.\ Phys.\ B {\bf 711}, 199 (2005) [arXiv:hep-th/0408178].
}

\lref\SpradlinSX{ M.~Spradlin, M.~Van Raamsdonk and A.~Volovich,
``Two-loop partition function in the planar plane-wave matrix
model,'' Phys.\ Lett.\ B {\bf 603}, 239 (2004)
[arXiv:hep-th/0409178].
}

\lref\AlvarezGaumeFV{ ~Alvarez-Gaume, C.~Gomez, H.~Liu and S.~Wadia,
``Finite temperature effective action, AdS(5) black holes, and 1/N
expansion,'' Phys.\ Rev.\ D {\bf 71}, 124023 (2005)
[arXiv:hep-th/0502227].
}

\lref\GomezReinoBQ{ M.~Gomez-Reino, S.~G.~Naculich and
H.~J.~Schnitzer, ``More pendants for Polya: Two loops in the SU(2)
sector,'' arXiv:hep-th/0504222.
}

\lref\BasuPJ{ P.~Basu and S.~R.~Wadia, ``R-charged AdS(5) black
holes and large N unitary matrix models,'' arXiv:hep-th/0506203.
}

\lref\AharonyCX{ O.~Aharony, S.~Minwalla and T.~Wiseman,
``Plasma-balls in large N gauge theories and localized black
holes,'' arXiv:hep-th/0507219.
}

\lref\FuruuchiSY{ K.~Furuuchi, E.~Schreiber and G.~W.~Semenoff,
``Five-brane thermodynamics from the matrix model,''
arXiv:hep-th/0310286.
}

\lref\DumitruNG{ A.~Dumitru, R.~D.~Pisarski and D.~Zschiesche,
``Dense quarks, and the fermion sign problem, in a SU(N) matrix
model,'' arXiv:hep-ph/0505256.
}

\lref\BarabanschikovRI{ A.~Barabanschikov, L.~Grant, L.~L.~Huang and
S.~Raju, ``The spectrum of Yang Mills on a sphere,''
arXiv:hep-th/0501063.
}

\lref\DumitruGD{ A.~Dumitru, J.~Lenaghan and R.~D.~Pisarski,
``Deconfinement in matrix models about the Gross-Witten point,''
Phys.\ Rev.\ D {\bf 71}, 074004 (2005) [arXiv:hep-ph/0410294].
}

\lref\MeisingerRM{ P.~N.~Meisinger and M.~C.~Ogilvie, ``Polyakov
loop models, Z(N) symmetry, and sine-law scaling,''
arXiv:hep-ph/0409136.
}

\lref\DumitruHP{ A.~Dumitru, Y.~Hatta, J.~Lenaghan, K.~Orginos and
R.~D.~Pisarski, ``Deconfining phase transition as a matrix model of
renormalized Polyakov loops,'' Phys.\ Rev.\ D {\bf 70}, 034511
(2004) [arXiv:hep-th/0311223].
}

\lref\FuruuchiQM{ K.~Furuuchi, ``From free fields to AdS: Thermal
case,'' arXiv:hep-th/0505148.
}



\def\my_Title#1#2{\nopagenumbers\abstractfont\hsize=\hstitle\rightline{#1}%
\vskip .5in\centerline{\titlefont
#2}\abstractfont\vskip.5in\pageno=0}

\my_Title {\vbox{\baselineskip12pt \hbox{WIS/21/05-AUG-DPP}
\hbox{\tt hep-th/0508077}}} {\vbox{\centerline{The Phase Structure of
Low Dimensional}
\vskip 5pt \centerline{Large $N$ Gauge Theories on Tori}}}

\centerline{Ofer Aharony$^{a}$, Joseph Marsano$^{b,c}$,
Shiraz Minwalla$^{c,b}$,
}
\centerline{Kyriakos Papadodimas$^{b,c}$, Mark Van Raamsdonk$^{d}$ and
Toby Wiseman$^{b}$}

\medskip

\centerline{\sl $^{a}$Department of Particle Physics, Weizmann Institute of
Science, Rehovot 76100, Israel}
\centerline{\sl $^{b}$Jefferson Physical Laboratory, Harvard University,
Cambridge, MA 02138, USA}
\centerline{\sl $^{c}$Department of Theoretical Physics, Tata Institute of
Fundamental Research,}
\centerline{\sl Homi Bhabha Rd, Mumbai 400005, India}
\centerline{\sl $^{d}$Department of Physics and Astronomy,
University of British Columbia,}
\centerline{\sl Vancouver, BC, V6T 1Z1, Canada}
\medskip


\medskip
\medskip

\noindent

In this paper we continue our study of the thermodynamics of large $N$
gauge theories on compact spaces. We consider toroidal
compactifications of pure $SU(N)$ Yang-Mills theories and of maximally
supersymmetric Yang-Mills theories dimensionally reduced to $0+1$ or
$1+1$ dimensions, and generalizations of such theories where the
adjoint fields are massive. We describe the phase structure of these
theories as a function of the gauge coupling, the geometry of the
compact space and the mass parameters. In particular, we study the
behavior of order parameters associated with the holonomy of the gauge
field around the cycles of the torus. Our methods combine analytic
analysis, numerical Monte Carlo simulations, and (in the maximally
supersymmetric case) information from the dual gravitational theories.

\vskip 0.5cm \Date{August 2005}



\newsec{Introduction}

In this paper we study one and two dimensional Euclidean $SU(N)$
Yang-Mills theories compactified on a circle and on a torus,
respectively. This is relevant in particular for studying these
theories at finite temperature. In the 't Hooft large $N$ limit, we
determine the phase diagrams of these theories as a function of masses
(of adjoint matter fields), coupling constants and compactification
parameters. The systems we study exhibit rich dynamics; in particular,
they undergo sharp large $N$ phase transitions upon varying
parameters. These phase transitions are associated with the
spontaneous breakdown of the $\IZ_N$ symmetry $W \r e^{{2 \pi i
\over N}} W$, where $W$ is the holonomy along a non-contractible cycle
of the compactification manifold.

Our analysis employs different techniques in different regimes of
parameter space. When all matter fields are very massive compared to
the scale set by the gauge coupling (more precisely, when $m^{4-d} \gg
\lambda$, where $m$ is the mass scale of the matter fields, $d$ is the
dimension of space-time and $\lambda \equiv g_{YM}^2 N$ is the 't
Hooft coupling), the gauge theories we study are weakly
coupled\foot{Recall that the massless gauge field has no
dynamical degrees of freedom.}, and
may reliably be analyzed in perturbation theory. Already in this
analytically tractable regime our systems undergo sharp phase
transitions and display a rich phase structure. Outside this
perturbative regime, the phase diagrams we study can be constrained by
the requirement that they reproduce well known results in special
limits. The strong coupling behavior of maximally supersymmetric
Yang-Mills theories may be analyzed using generalizations of the AdS/CFT
correspondence. Finally, it is sometimes practical to employ Monte
Carlo simulations to supplement the information from our other
techniques. Employing all these methods, we are able to present a
reasonably complete picture of the phase diagram of the systems we
study.

The investigations reported in this paper are similar in spirit to
the recent study of Yang-Mills theory on $S^p \times S^1$
\refs{\sundborg, \sphere}. It may be recalled that the phase
transitions discovered in \refs{\sundborg, \sphere} may be thought
of as a weak coupling continuation of the gravitational Hawking-Page
transition in $AdS_5 \times S^5$. At least some of the phase
transitions we study in this paper also have gravitational
analogues. For example, as we have already reported in the letter
\glpaper\ (building on the work of \refs{\susskind, \martineco,
\martinect}), the phase transition in maximally supersymmetric
Yang-Mills theory on $T^2$ is holographically dual to a Gregory-Laflamme
black hole/black string phase transition in type II string theory.
See section 7 for other examples.

The structure of this paper is as follows. In section 2, we discuss
various general properties of low dimensional gauge theories on tori,
and in particular the order parameters that distinguish the
various phases that appear. In section 3, we review the behavior of
Yang-Mills theory dimensionally reduced to 0 dimensions; this is
relevant to the small volume limit of the higher dimensional cases. In
section 4, we use analytic and numerical techniques to map out the
phase diagram of pure $p+1$-dimensional gauge theory dimensionally
reduced to 0+1 dimensions, and its generalizations with masses for
the adjoint scalars. In section 5, we study the thermodynamics of 1+1
dimensional Yang-Mills theory with massive or massless adjoint scalars
on a spatial circle by studying the partition function of Euclidean
two dimensional Yang-Mills theory
on $T^2$. In the case where the scalars are very
massive, we obtain a fairly complete picture of the rather rich phase
structure through analytic analysis, leading to the phase diagrams in
figures 14 and 15. In sections 6 and 7, we repeat the analysis of
sections 4 and 5 for the maximally
supersymmetric Yang-Mills theories in $0+1$ and $1+1$ dimensions,
respectively. For these theories, knowledge of the dual gravitational
theories provides additional information about the strong coupling
behavior of the gauge theory. We find that the supersymmetric theories
show rather different qualitative behavior from the non-supersymmetric
theories; for the two dimensional case this behavior is summarized in
the phase diagrams of figures 18 and 20.

\newsec{Low Dimensional Yang-Mills Theory on Tori: Generalities}

Massless vector fields have no propagating degrees of freedom in one
and two dimensions. As a consequence, pure Yang-Mills theory in these
dimensions is exactly solvable, and exhibits relatively tame
dynamics. However, the same theory displays rich dynamical
behavior when coupled to matter fields. In this paper we will study
large $N$ $SU(N)$ Yang-Mills theories coupled to matter fields in the adjoint
representation. We begin in this section with an overview of the
properties of these theories.

\subsec{Quantum mechanical gauge theories ($d=1$)}

Different $SU(N)$ quantum mechanical gauge theories with adjoint
scalar fields can behave in at least three qualitatively distinct
ways in the IR. In some theories an attractive effective potential
(at long range) between the eigenvalues of the scalar fields ensures
that the vacuum state is normalizable, and that the spectrum is
gapped. In other theories, the scalar potential has exactly flat
directions and the spectrum is ungapped. In yet other theories the
long-range scalar effective potential is repulsive; such theories
lack a vacuum state.

Theories with a mass gap (see section 4 below for a detailed study
of one class of examples) may be thought of as one dimensional
analogues of confining theories in $d=4$. In the 't Hooft large $N$
limit with fixed $\lambda \equiv g_{YM}^2 N$ \thooft, the spectrum
of such theories is expected to display a Hagedorn-like growth in
the density of states, with a high energy density of states $\rho(E)
\sim \exp(E/T_H)$ for energies which remain finite in the large $N$
limit (see section 2 of \sphere\ for a simple example). It follows
that, upon heating, these theories undergo a `deconfinement' phase
transition at or below their effective Hagedorn temperature $T_H$.
In other words, the Euclidean partition function on $S^1$ undergoes
a phase transition when the circumference of the compactification circle
is larger than or equal to
the inverse Hagedorn temperature. The free energy jumps from
$\CO(1)$ at low temperatures to $\CO(N^2)$ at high temperatures
(where the theory is weakly coupled), so $\lim_{N\to \infty}
F(T)/N^2$ may be viewed as an order parameter for the deconfinement
transition. As in four dimensions, the expectation value of the
Polyakov loop $\vev{\tr(P\exp(i\oint A_0))}$ in the Euclidean theory
is another order parameter for this phase transition \foot{The
Polyakov loop vanishes at large circle radius but is non-zero at
small radius (there is a subtlety in this statement; see footnote 5
below). Intuitively, the Polyakov loop vanishes at low temperatures
in a theory with a mass gap because only an infinite number of
adjoint photons can effectively screen a fundamental charge; the
mass gap ensures that such a configuration has infinite energy. See
section 5.7 of \sphere\ for a more careful and detailed
discussion.}.

Theories of the second type (those with a well-defined vacuum but an
ungapped spectrum -- see section 6 below for a detailed study of an
example) may be thought of as the analogues of conformal theories in
4 dimensions. We expect such theories to be `deconfined' at all
temperatures. If such theories undergo phase transitions as a
function of temperature (this may or may not happen), the two phases are
distinguished by an order parameter more sensitive than the
expectation value of the Polyakov loop or $F(T)/N^2$, each of which
is nonzero in both phases.

We will not consider theories that lack a vacuum state in this
paper.

\subsec{$d=2$ gauge theories}

We begin by considering $1+1$ dimensional large $N$ $SU(N)$
Yang-Mills theories, with
purely adjoint matter content, on $\IR \times $time. Our general
expectations for the thermodynamics of such systems are analogous to
those for quantum mechanical systems reviewed in the previous
subsection. Gapped theories are expected to have a Hagedorn growth
in their density of states for energies which remain finite in the
large $N$ limit (see \refs{\sphere, \kleb} for examples). These
theories are expected to undergo a deconfining phase transition at
or below the Hagedorn temperature. The free energy is expected to be
of order $\CO(N^2)$ (and the Polyakov loop nonzero) at high
temperatures, while the free energy is expected to be $\CO(1)$ (and
the Polyakov loop vanishes) at low temperatures. On the other hand
the free energy in ungapped theories is expected to be $\CO(N^2)$
(and the Polyakov loop expected to be nonzero) at all temperatures
\foot{This is true generically, but it is not true in some
supersymmetric examples in which the low temperature free energy is
of order $N$.}.

However, field theories in $d=2$ have additional structure absent in
$d=1$. It is possible, and rather natural, to study the
thermodynamics of such theories compactified on a spatial circle. In
Euclidean space this thermodynamics is described by the partition
function of the corresponding Yang-Mills theory on a $T^2$. This
opens the possibility for a much richer phase structure in these
models, as we explain in the rest of this subsection.

Recall that we are interested in $SU(N)$ theories which only have
fields in the adjoint representation. For such theories, the gauge
group is really $SU(N)/\IZ_N$, since the gauge transformations in
the $\IZ_N$ center of the gauge group act trivially. Whenever such a
theory is compactified on a torus, it possesses a $(\IZ_N)_{0}
\times (\IZ_N)_{1}$ global symmetry. This global symmetry is
generated by gauge transformations $G$ that are not periodic on the
torus but instead obey $G(x^\mu+p e_0 + q e_1)= \alpha^p \beta^q
G(x^\mu)$, where $e_0$ and $e_1$ are the fundamental cycles of the
torus (along the $0$ and $1$ directions) and $\alpha$ and $\beta$ are
both gauge transformations in the center $\IZ_N$. All local
gauge-invariant operators are uncharged under this global symmetry, but
fundamental Wilson lines that wrap around a $(p,q)$ cycle before
closing carry charge $(p,q)$. Thus, non-zero
expectation values for such Wilson loops break this global symmetry. In
particular, an expectation value for the  Wilson loop $W_{\mu}$
\foot{Where $W_{\mu}(\vec{x}) = {1\over N}\vev{\tr (P e^{i \oint
A_{\mu}})}$ around the circle in the $\mu$ direction, for
$\mu=0,1$.} spontaneously breaks the $(\IZ_N)_{\mu}$
symmetry\foot{Actually we should be more careful here -- in fact
there is no symmetry breaking at finite volume. The closest one can
come to a $\IZ_N$ symmetry breaking is to have $N$ different saddle
points (related by the $\IZ_N$ symmetry) dominating the path
integral, but the path integral sums over all of them. Thus, there
will not really be an expectation value for $W_{\mu}$, but just for
$|W_{\mu}|^2$ (for example). For the most part we will leave this
issue implicit in our discussions.}, where $\mu=0,1$.

As we will see later in this paper, in the decompactification limit
$R_1 \to \infty$ the symmetry $(\IZ_N)_1$ is never broken so $W_{1}$
always vanishes. The Polyakov loop $W_0$ is
the sole order parameter for the system in
this limit, making contact with the
discussion at the beginning of this subsection. At finite $R_1$,
$W_0$ and $W_1$ are both nontrivial order parameters, allowing for
intricate two dimensional phase diagrams with the four possible
phases separated by phase transition lines. Of course, such a phase
diagram in a system that is ungapped in the decompactification ($R_1
\to \infty$) limit (see section 7 for an example) has
qualitative differences from its counterpart in a theory that is
gapped in the same limit (see section 5 for examples).

In the large $N$ limit, instead of just considering the Wilson loop
$W_{\mu}$ in the fundamental representation, it is often useful to
study  the full holonomy matrix $U_{\mu}(\vec{x}) = P e^{i \oint
A_{\mu}}$, which is a unitary matrix whose trace (in some
representation) gives the Wilson loop (in that representation). The
set of eigenvalues of this matrix (which live on the unit circle) is
gauge-invariant, and in the large $N$ limit their distribution along
the unit circle is some continuous function. The $(\IZ_N)_{\mu}$
global symmetry described above shifts the phases of
all the eigenvalues of $U_{\mu}$ by an
angle of $2\pi/N$. In the $\IZ_N$-symmetric phase the large $N$
eigenvalue distribution is constant, while otherwise it is generally
maximized at some particular value, spontaneously breaking the
$\IZ_N$ symmetry. Detailed analysis of the eigenvalue distributions
sometimes permits
a sharp distinction between two phases with the same symmetry
breaking pattern, allowing for the possibility of even more
intricate phase diagrams (interpolating between a larger number of
phases) than those described above.

\subsec{On dimensional reduction}

One of the methods we will use in our analysis is dimensional reduction.
Whenever one of the circles in the problem is very small
compared to the other scales, the theory may be approximated by
a theory in one dimension less. So,
the functional integral of a $(d+1)$-dimensional gauge theory
will approximate, in the limit of a very small circumference $R_{d+1}$ for
one of the compact circles,
that of a $d$-dimensional gauge theory, with a
coupling constant $\lambda_d = \lambda_{d+1} / R_{d+1}$.  When the
small circle has periodic boundary conditions for fermions,
the lower-dimensional gauge theory has
precisely the same field content as the original theory, except that
the zero mode of one of the components of the vector field becomes a
scalar field.  When the fermions have anti-periodic boundary conditions
on the small circle (as for thermal boundary conditions on a temporal circle)
they are projected out, leaving only the bosonic fields.
Naively, the lower dimensional gauge theory is valid whenever the
Kaluza-Klein scale $1/R_{d+1}$ is much larger than the other
compactification scales $1/R_i$ and $T$, and than the dynamical
scale $\Lambda_d = \lambda_d^{1/(4-d)}$; the latter condition is the
same as $R_{d+1} \ll \lambda_{d+1}^{-1/(3-d)} = 1 / \Lambda_{d+1}$.

This argument is naive since an $SU(N)$ gauge theory on a torus
actually has excitations whose classical energies scale as
$1/(NR_{d+1})$, coming from configurations for which gauge-invariant sets
of eigenvalues are periodic only up to a permutation (these configurations
play a major role in M(atrix)
theory). This scale goes to zero in the 't Hooft large $N$ limit, so
naively we can never ignore all the KK modes in this limit. The
importance or otherwise of these modes (and so the validity of
dimensional reduction) is a dynamical issue determined by the saddle
point eigenvalue distribution of the Wilson line operator around the
corresponding circle. The light modes described earlier in this
paragraph are present only when the Wilson line eigenvalues are
spread over the circle (the `shift' distribution in M(atrix) theory
language). When the eigenvalues of the Wilson line operator are
sufficiently clumped, these light modes are absent and dimensional
reduction is justified. In this paper we will only use dimensional
reduction when this criterion is met.

\newsec{Bosonic Matrix Integrals ($d=0$)}

In this section we briefly review the behavior of bosonic $SU(N)$
matrix integrals; we will find the results of this section useful in
our analysis below.

Consider the matrix integral\foot{Note that in zero dimensions the
only difference between a gauged matrix integral and a non-gauged
integral is the volume of $SU(N)$, so we will not need to distinguish
the two.}
\eqn\mtint{Z=\int \CD \Ph_i \exp \left[
-{N \over {2 \lambda_0}} \Tr \left( \sum_{i} {m_i^2 } \Ph_i^2 -
\sum_{i <j} [\Ph_i, \Ph_j]^2 \right) \right]. }
where $i,j =
1,\ldots,p$, ($p>2$) and the $\Ph_i$ are $N\times N$ Hermitean bosonic
matrices, in the 't Hooft large $N$ limit in which $N$ is taken to
infinity with constant $\lambda_0$. Redefining variables $\Ph_i=
{\sqrt{\lambda_0} \over m_i} \ph_i$, we find
\eqn\mtintn{Z=\int \CD \ph_i \exp
\left[ -N \Tr \left( \sum_{i} {\ph_i^2 \over 2} - {\lambda_0 \over {2
m_i^2 m_j^2}} \sum_{i <j} [\ph_i, \ph_j]^2 \right) \right]. }
Let us first
study the limit ${ \lambda_0 / m_i^4} \to 0$ (for all $i$).  In
this limit the integral factorizes into a product of $p$ identical
integrals, each of which is easily solved by saddle points. The
saddle point eigenvalue distribution for (say) $\ph_1$ is given by
the Wigner semi-circle law,
\eqn\eigenvaluedist{ \rho(x) = {2 \over
\pi}\sqrt{1- x^2},}
and we have
\eqn\trphsq{{1\over N}{ \vev{\tr \sum_i
\Ph_i^2}}={\lambda_0 \over 4}\sum_i{1 \over m_i^2}.}

In the simple limit considered above, suitably normalized gauge
invariant expectation values (like  ${1\over N}{\Tr \sum_i \Ph_i^2}$)
are completely determined by a saddle point with sharp edges,
and are independent of $N$. This follows from rather general
considerations (factorization and 't Hooft scaling), and so generalizes
to most of the systems we study in this paper.

\fig{Graph showing $\vev{{1\over N}{\Tr \sum_i \Ph_i^2}}$ for the massless
matrix integral \mtint\ with $\lambda_0 = 1$ for various values of
$p$ from 3 to 10. The red points correspond to $N = 10$,
the blue to $N = 20$.}{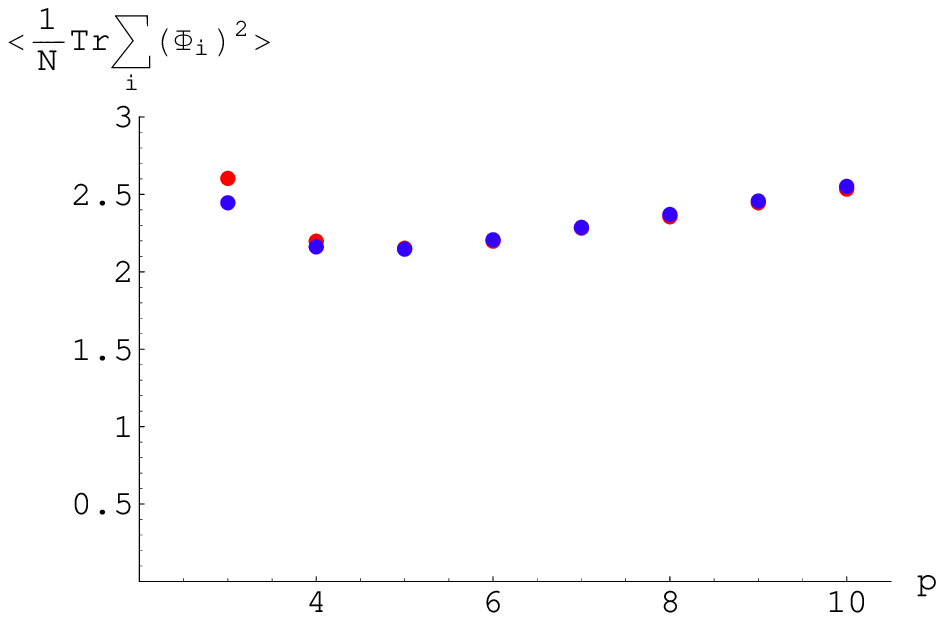}{4truein}
\figlabel{\integralwidth}

\threesidefig{Plots showing the distribution of eigenvalues for the
massless matrix integral with $\lambda_0 = 1$ for $p=3,5,10$ (left
to right), generated by Monte Carlo simulations for $N = 20$.}{2.5truein}
{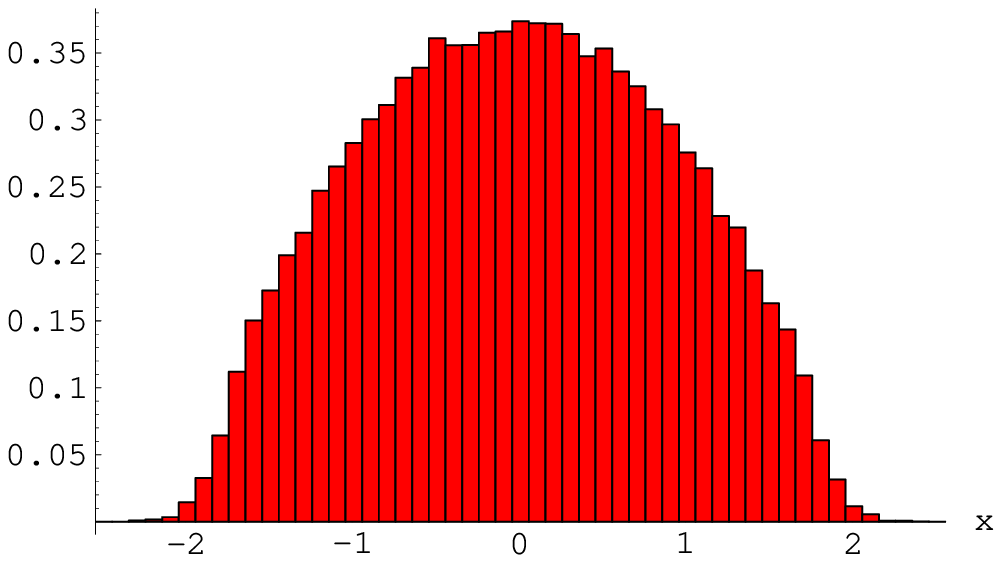}{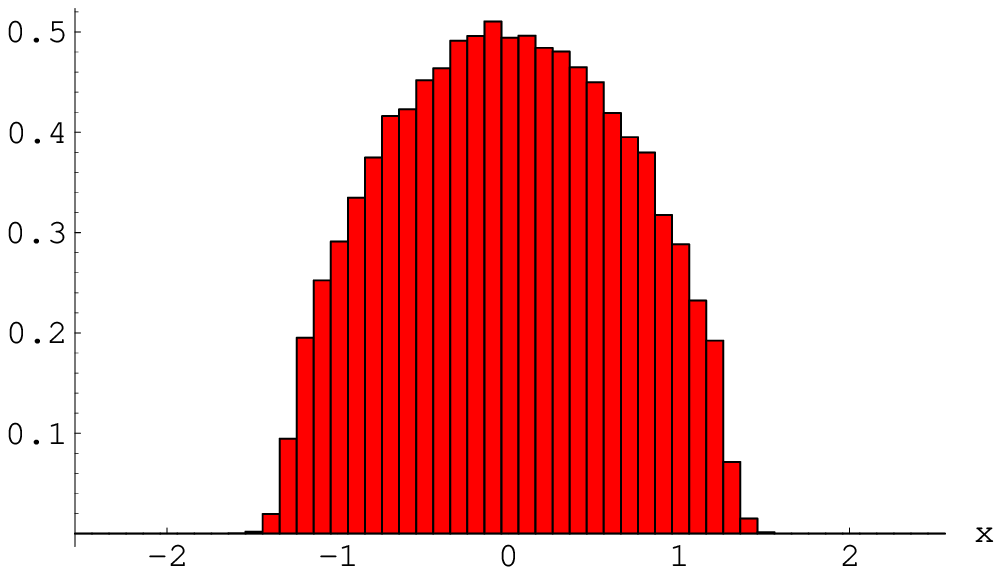}{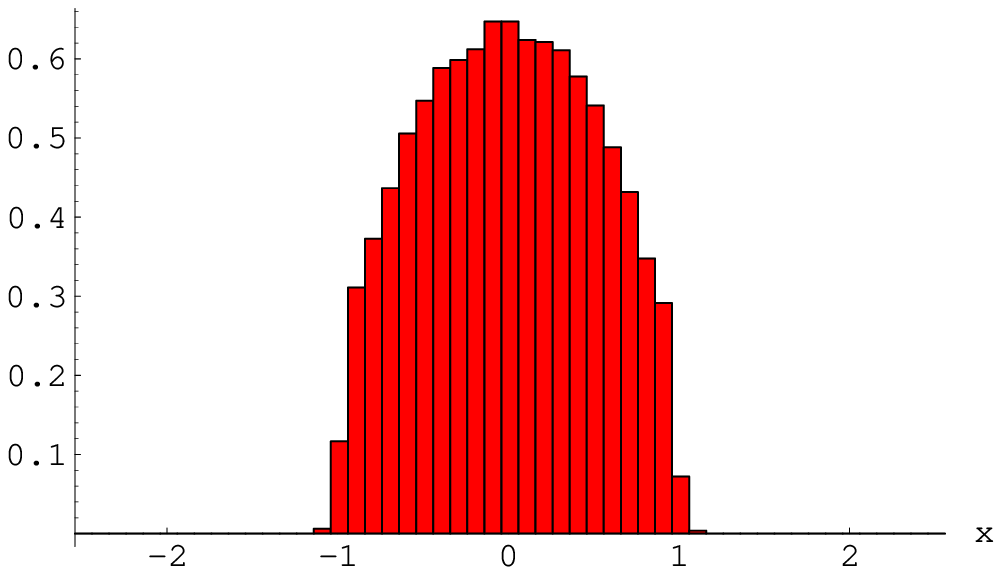}
\figlabel{\integraldist}

We now turn to the opposite, massless limit of \mtint. It follows from
a simple rescaling (similar to the one we used above) that the
eigenvalues are localized on a length scale $a = K \lambda_0^{{1\over
4}}$ at the corresponding saddle point; Monte Carlo simulations
demonstrate that $K$ is of order unity (see Appendix A.1 for more
details and references), and the values of $\vev{{1\over N}{\Tr \sum_i
\Ph_i^2}}$ are shown in figure \integralwidth\ for various values of $p$.
Sample eigenvalue distributions which demonstrate a qualitatively
similar form to \eigenvaluedist, including its sharp edge, are shown
in figure \integraldist.

It natural to guess that \mtint\ is dominated by a saddle point in
which the eigenvalues of the scalars are sharply localized for all
values of $m_i$ and $\lambda_0$. The localization length scale varies
smoothly from approximately $\sqrt{\lambda_0}/m$ at large $m$ to
approximately $\lambda_0^{{1 \over 4}}$ at small $m$. This expectation
is easily verified (at sample values of parameters) by a Monte Carlo
simulation.

\newsec{Bosonic Gauge Theories in One Dimension}

In this section we study the quantum mechanical $SU(N)$ Yang-Mills
theory coupled to $p$ adjoint scalar fields ($p \geq 2$), whose Euclidean
action is given by
\eqn\eym{S = {N \over {2 \lambda_1}} \int dt~ \tr \left(\sum_i D_0
\Phi_i D_0 \Phi_i + \sum_i M^2 \Phi_i^2 - \sum_{i,j}
{1 \over 2} [\Phi_i, \Phi_j]^2\right),}
where $D_0$ is a covariant derivative involving the non-dynamical
gauge field $A_0$, we assumed equal masses for simplicity,
and $i,j=1,\cdots,p$. This theory is believed to
possess a unique normalizable vacuum state, and to have a mass gap in
its spectrum, at all values of $M$ including $M=0$. We will study the
Euclidean partition function of this theory on a circle of circumference $R$,
which is the same as the thermal partition function of the quantum
mechanical system at temperature $T=1/R$. This theory
is characterized by two dimensionless parameters,
which we take to be the inverse radius in units of the coupling, $\tit =
1 / (R \lambda_1^{1 \over 3})$, and the mass in units of the coupling, $m = M /
\lambda_1^{1 \over 3}$. In this section we will determine the phase
diagram of \eym\ as a function of these parameters.

Two asymptotic regions of this phase diagram are amenable to analytic
analysis. First, in the limit of large $\tit$, \eym\ effectively reduces
to the matrix integral \mtint\ (as discussed in section 2); from the
analysis of the previous section it then follows that the eigenvalues
of the holonomy matrix\foot{By an abuse of notation, we will sometimes
refer to these as the eigenvalues of the Polyakov loop operator
$W_0 = {1\over N} \tr(U_0)$.} $U_0 = P e^{i\oint A_0}$
are clumped at all values of $m$ at large $\tit$.

At large $m$, \eym\ may be analyzed in perturbation theory. In the next
subsection we demonstrate that, in this limit, our system undergoes a
first order deconfinement transition at $\tit_c\sim m /
\ln(p)$, and we determine the first correction (in $1/m$) to this
phase transition curve. For $\tit>\tit_c$, the eigenvalues of the Polyakov
loop operator are clumped and the $\IZ_N$ invariance is broken. For
$\tit<\tit_c$ the eigenvalues of the Polyakov loop operator are uniformly
distributed and $\IZ_N$ invariance is restored.

The analysis of the two limits described above (and especially the
analytic form of the large $m$ phase transition curve given in the
next subsection) suggests that the phase diagram of \eym\ takes the
form shown in figures 3 and 4, with the phase boundary extending to some
fixed temperature for nonzero mass. In sections 4.2 and 4.3 below, we will use
Monte Carlo simulations to plot out this phase diagram in detail,
confirming this prediction and giving a quantitative picture of the
phase diagram for all values of the mass.

In section 4.4, we discuss the phase diagram for a closely related
theory in which one scalar remains massless while the rest have mass
$M$; this theory is
relevant to the high temperature limit of 1+1 dimensional gauge
theories with massive adjoint scalars (to be discussed in section
5). We find similar qualitative behavior, except that the transition
temperature for large mass becomes $\tit \sim m / \ln(m)$.

\subsec{The large mass limit}

When the masses in \eym\ are sufficiently large ($m \gg 1$), \eym\
is weakly coupled at all temperatures, and the thermodynamic
behavior can be studied in perturbation theory. As discussed in
\refs{ \sundborg, \sphere}, it is straightforward to integrate out
all massive degrees of freedom in the weakly coupled limit to obtain
an effective action in terms of the sole light degree of freedom,
namely the Wilson line of the gauge field about the circle
\eqn\defofu{
U = P e^{i \oint_0^R dt A_0} \; .
}
The partition function then takes the form of a unitary matrix model,
\eqn\seffu{
Z = \int DU e^{-S_{eff}(U)},
}
where
\eqn\weakeffo{ S_{eff}(U) = p \sum_{n=1}^{\infty} \; {x^n\over n} \; \tr(U^n)
\tr(U^\dagger
{}^n) + {\cal O}(1/m^3), }
with $x \equiv e^{-MR}=e^{-m/\tit}$. In the weak coupling ($m \to
\infty$) limit, this matrix model (for $p>1$) undergoes a large $N$
phase transition as a function of $\tit$, from a low $\tit$ phase dominated
by a saddle point in which the eigenvalues of $U$ are distributed
uniformly around the circle, to a high $\tit$ phase in which the
eigenvalues of $U$ are clumped in the saddle-point configuration.

In the strict $m \to \infty$ limit, the phase transition occurs at the
point $\tit = m / \ln(p)$, and it is (weakly) of first order. However, as
discussed in section 6 of \sphere, the nature of the phase transition
at large but finite $m$ depends on higher order terms in the effective
potential, arising from two and three-loop diagrams. In appendix C,
we compute the effective action \weakeffo\ to three-loop order, and
find that after integrating out all scalars as well as all $n>1$
Fourier modes of the eigenvalue distribution function (these are all
massive near the transition), we are left with an effective action for
the lowest Fourier mode $u_1 = {1 \over N} \tr(U)$, of the form
\eqn\weakeff{ S_{eff} (u_1) = N^2(m_1^2(x, 1/m^3) |u_1|^2 + {1 \over
m^6} b(x, 1 / m^3)
 |u_1|^4 + {\cal O}(1 / m^{12})), }
where the leading order expressions for $m_1^2$ and $b$ are given in
appendix C. Since we find $b<0$ at the transition temperature where
$m_1^2$ switches sign, this effective action describes a phase
transition which continues to be first order also for large finite $m$
(as at $m \to \infty$) (see
\sphere), and occurs at the critical temperature
\eqn\crittemp{\eqalign{
\tit_c = & m/\ln(p)
+ {1 \over {4 m^2}} {{(p-1)(2p+1)} \over {p \,\ln(p)}} \cr &- {1
\over {128 m^5}} {(p-1) \over {p^3 \,\ln(p)}} (\ln(p)(21p^2+6p)+ 48 p^4 +20
p^3 -17 p^2-20p-4) + {\cal O}(1/m^8), }}
which is slightly below the Hagedorn temperature of this theory.

In summary, when $m \gg p^{1 \over 3}$ (so that perturbation theory
is reliable), \eym\ undergoes a first order phase transition at
$\tit_c$ given by \crittemp. For $\tit<\tit_c$ the Polyakov line eigenvalue
distribution is uniform. For $\tit>\tit_c \approx m/\ln(p)$, the eigenvalues
of the Wilson line are clumped.

\fig{A plot showing the prediction of a first order phase transition at
large $m >> p^{1/3}$ from
 perturbation theory. The small circle
limit suggests localization of the Wilson
loop eigenvalues at large $\tit$, for fixed $m$.
Then, extrapolating the blue
phase boundary would naively indicate that it meets the
$\tit$ axis. }{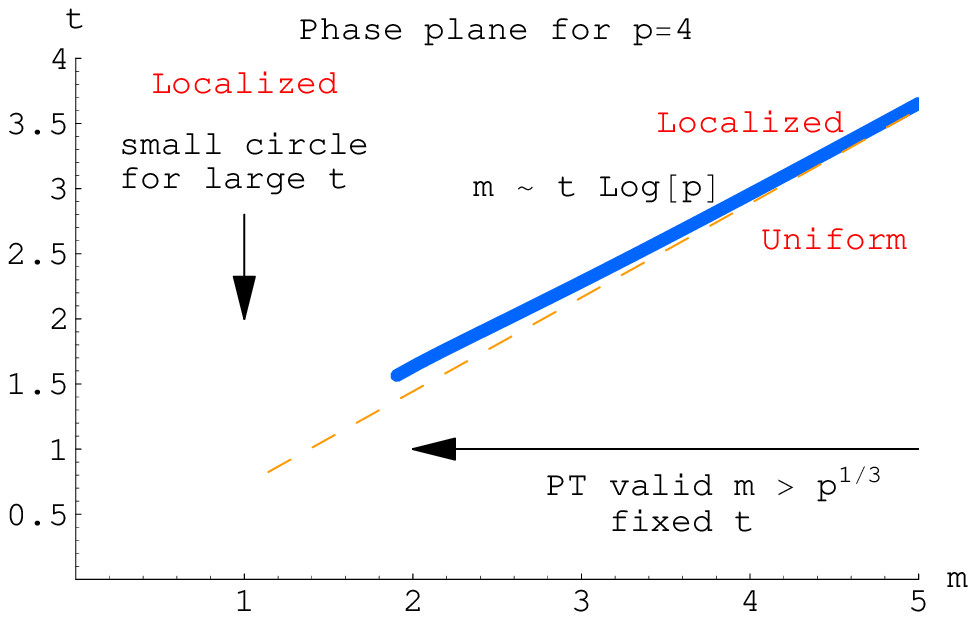}{4truein}
\figlabel{\phaseA}

It is tempting to extrapolate these results beyond the validity of
perturbation theory. As displayed in figure \phaseA, plotted for
$p=4$, the phase transition line asymptotes to the
line $\tit = m/\ln(p)$ for large $m$, and gradually rises above it as $m$
decreases, until perturbation theory is no longer valid (at roughly $m
\approx p^{1\over 3}$).  A naive extrapolation of our weak coupling results
suggests that the phase transition curve will hit the vertical axis
($m=0$) at a finite temperature.

\fig{A plot showing the conjectured phase boundary translated
into the $\lambda_1, M^2$ plane, choosing units with $R = 1$.  The Monte
Carlo data presented below confirms this
picture.}{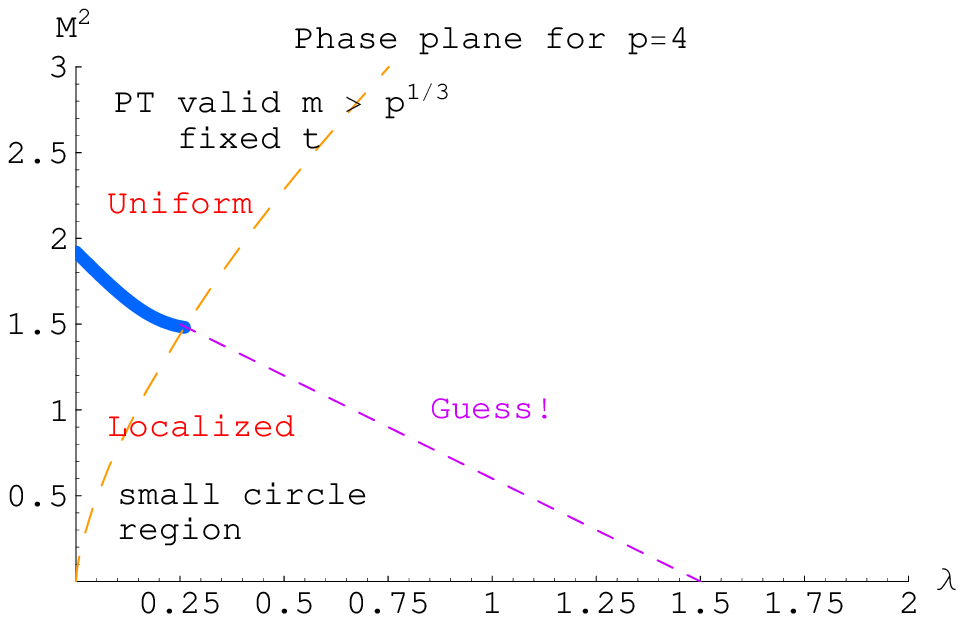}{4truein}
\figlabel{\phaseB}

In the rest of this section we will use Monte Carlo techniques to
demonstrate that this guess is indeed correct.

\subsec{Results for $m=0$ from Monte Carlo simulations}

The bosonic one-dimensional model \eym\ is rather simple to implement
numerically using elementary Monte Carlo methods. In this section we
will discuss the numerical results for the behavior of the Wilson
line eigenvalues in the cases $p=2,4,9$, characterized by $u_n={1\over
N}|\tr(U^n)|$ (which vanish for all non-zero values of $n$ in the uniform
distribution). For these diverse values of $p$ we will see that they
all have the same qualitative behavior, and we will link the results
to the analytic limits discussed above. Details of the method may be
found in Appendix E.

In the massless limit, \eym\ is simply the dimensional reduction of
the pure $SU(N)$ gauge theory in $(p+1)$ dimensions to one dimension. In
this limit
\eym\ has a single dimensionless parameter $\lambda_1 R^3 = 1/\tit^3$. As
we noted above, at large $\tit$ this system reduces to the matrix
integral of the previous section (with masses set to zero), and so
it lies in a phase with a clumped distribution of eigenvalues for the
Polyakov line. We now describe results of a Monte
Carlo analysis we have performed on this theory.

\fig{Figure of $\vev{u_1}$ as a function of $\lambda_1$ for $p=9$ with
various values of $N$. We choose units with $R = 1$, and thus $\lambda_1 =
1/\tit^3$. We see that as $N$ increases the points to the right of $\lambda_1
\simeq 1.4$ appear to decrease, consistent with $1/N$ scaling. To the
left of this value the points appear to tend to a limiting curve. This
indicates a sharp discontinuity in $u_1$ at infinite $N$, with $u_1$
being an order parameter. Statistical error bars are comparable to the
point sizes.}{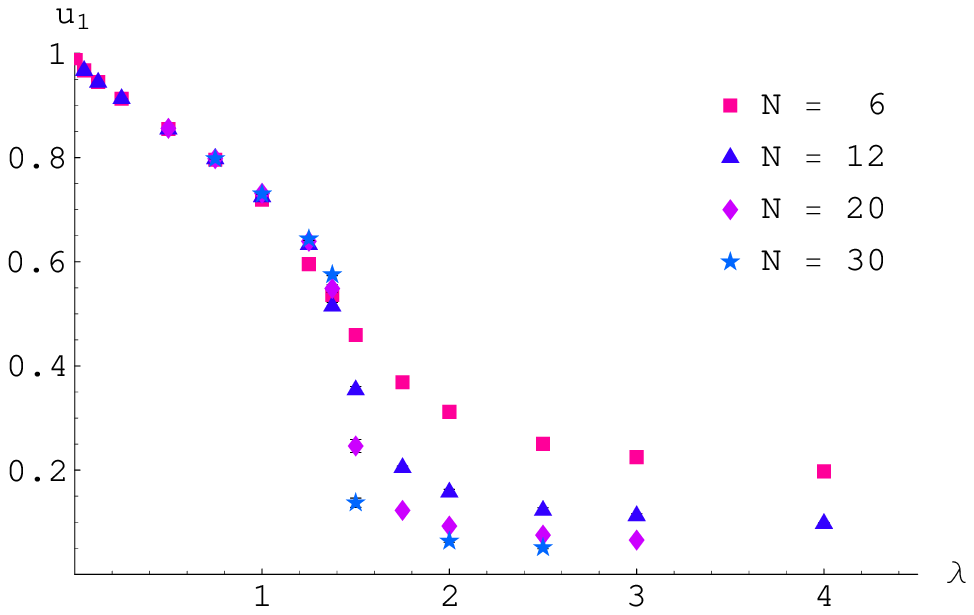}{4truein}
\figlabel{\masslessA}

\fig{Figure of $\vev{u_2}$ as a function of $\lambda_1$ for $p=9$ with
various values of $N$.
Again a discontinuous behavior is indicated in the large $N$ limit.
As above, we choose $R = 1$ units. }{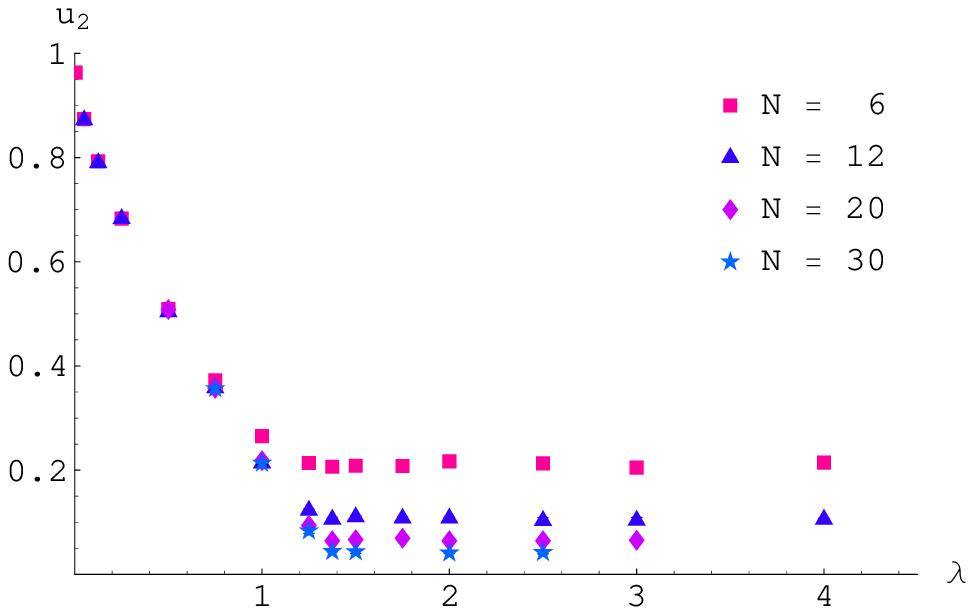}{4truein}
\figlabel{\masslessB}

In figures \masslessA\ and \masslessB\ we plot the Monte Carlo
results for $\vev{u_1}$ and $\vev{u_2}$
in the theory with $p=9$, for various values
of $N$. These results
were previously reported in \glpaper; as we have reported there, our
data strongly suggests a large $N$ transition at $1 / \tit^3
\simeq 1.4$. As is apparent from figures \masslessA\ and \masslessB,
the low $\tit$ (or large $\lambda_1$) phase has $u_1=0$ and $u_2=0$ (more
precisely, $u_1$ and $u_2$ are of order $1/N$ for finite $N$).
Clearly, $u_1$ and $u_2$ are nonzero (and hence the eigenvalue distribution
is non-uniform) in the high temperature phase. All this is perfectly in
line with the conjecture made in the previous subsection.

As discussed in \glpaper, we have not been able to clearly resolve
the order of the phase transition studied in this subsection.
Unfortunately, the phase transition of our system appears to
lie very near the boundary between first and second order behavior (see
\glpaper). In this situation it is difficult to numerically
distinguish between the two reasonable possibilities (see Appendix E),
which are either a first order phase transition or a second order
phase transition followed by another, third order phase
transition \sphere. However, it is clear from the data that if the second
scenario is correct then the two phase transitions must be very close
together.

We have repeated the analysis of this section at $p=2,4$. We find
qualitatively identical behavior to that reported in the previous
paragraphs, but this time with the phase transition at
$\lambda_{crit} R^3=1 / \tit_c^3
\simeq 0.4, 0.9$, respectively.

\subsec{Monte Carlo simulations with non-zero mass}

We have also performed Monte Carlo simulations of \eym, as a
function of $\tit$, at various different values of $m$. The results of
these simulations (which we present in this subsection) smoothly
interpolate between the analytic prediction of subsection 4.1 and
the $m=0$ results of the previous subsection, allowing us to fill
out the phase diagram of \eym\ and to confirm the guess displayed in
figure \phaseB.
\fig{Figure of $\vev{u_1}$ in the $\lambda_1, M$ plane (in $R = 1$ units).
The red (top) surface is for $N = 6$, blue is $N = 12$ and green
(bottom) $N = 20$. We
see a transition line bounding the origin where outside the line $u_1$
decreases with increasing $N$ (approximately consistent with $1/N$ scaling),
and inside the line a limit is approached. The transition line
appears to connect the massless transition to the large $m$
transition. The black mesh of the surfaces gives the sampling
density. Thus, the (small) statistical error can be seen as the
`roughness' of the surface.}{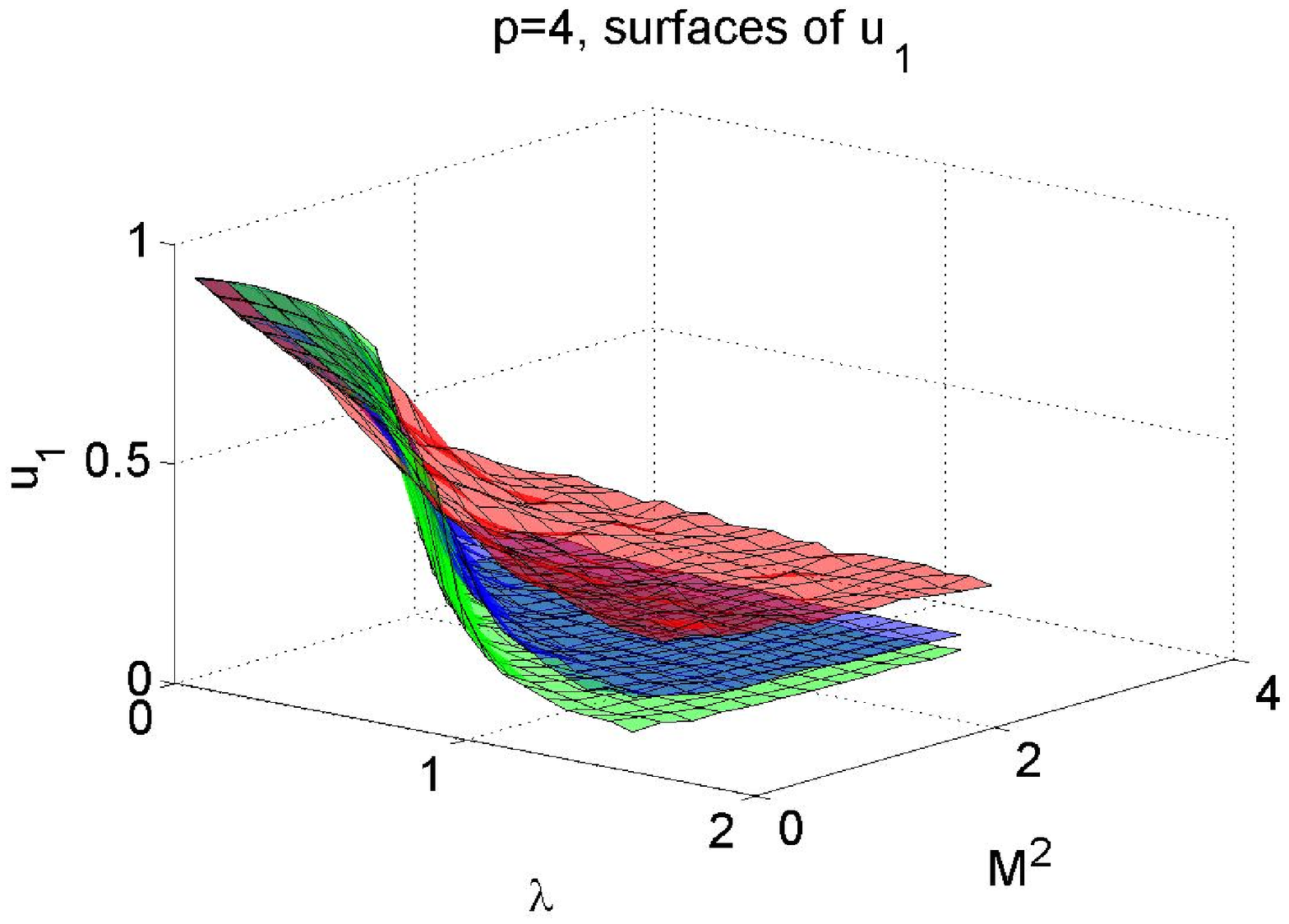}{4truein}
\figlabel{\surfA}
\fig{Figure of $\vev{u_2}$ in the $\lambda_1, M$ plane, employing
the same conventions as in the previous figure.}{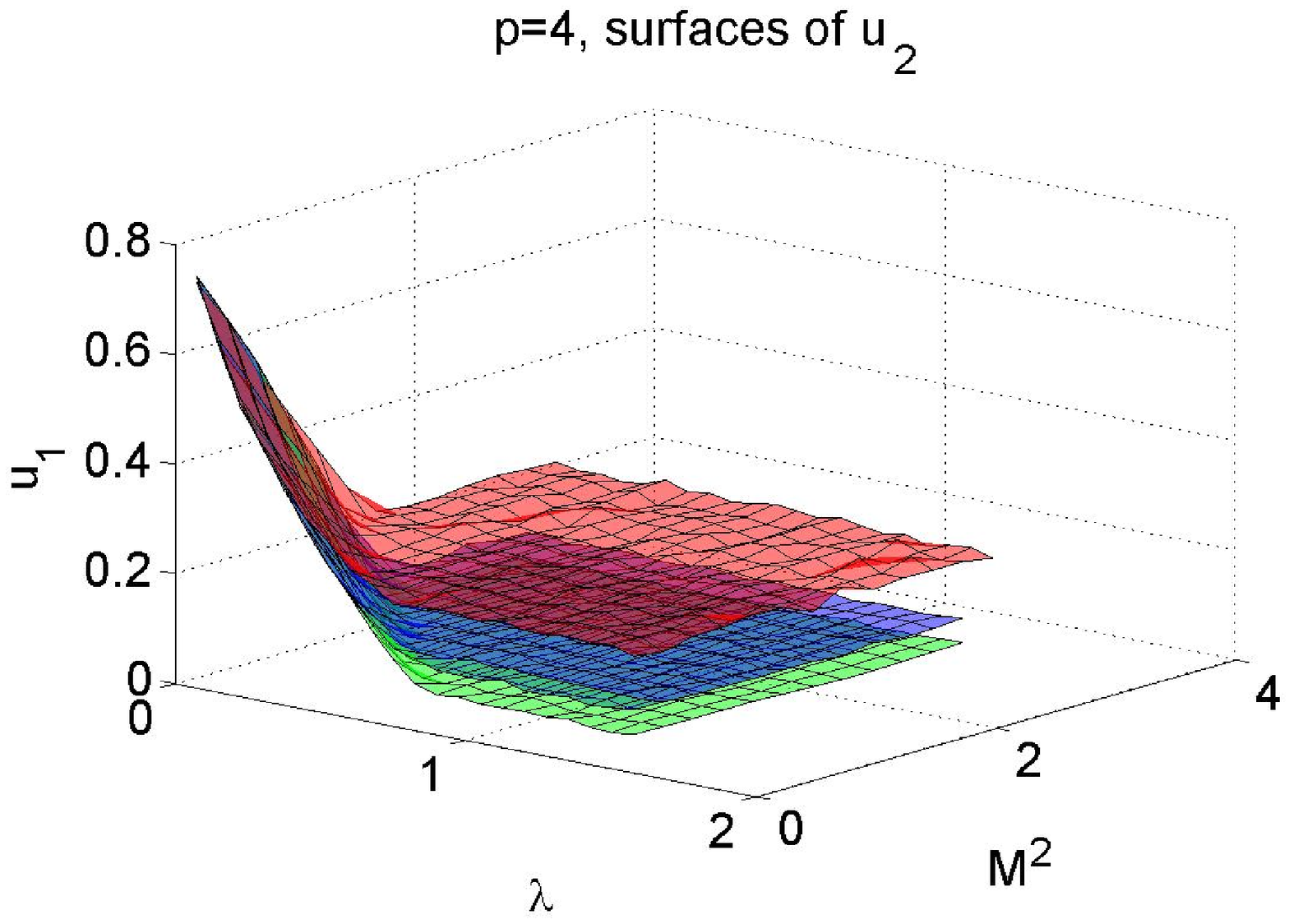}{4truein}
\figlabel{\surfB}

In figures \surfA\ and \surfB\ we present three dimensional plots of
$\vev{u_{1}}$ and $\vev{u_2}$ as functions of $\lambda_1 R^3 =1 / \tit^3$
and $M R=m / \tit$, for the case $p=4$ \foot{These graphs are best viewed
in colour.}. An examination of figures \surfA\ and \surfB\ reveals
that the $m$, $\tit$ plane may be divided into two regions. In the first
(uniform) region $u_1$ and $u_2$ decrease as $N$ increases; this
decrease is approximately fit by a $1/N$ decay to zero. In the
second (clumped) region $u_1$ and $u_2$ asymptote to fixed,
$m$ and $\tit$ dependent, nonzero values. These two regions are divided by
a transition line. As in the case of $m=0$, $u_1$ appears to jump
discontinuously (or at least with a very sharp slope)
across this transition line in the limit $N \to \infty$.

\fig{Contours of $\Delta^{12,20} \vev{u_1}$ for the $p=2$ theory, in the
$\lambda_1, M$ plane (in $R = 1$ units). The green region is predicted
to be in the uniform eigenvalue phase, and the red region in the
localized phase. The blue curve gives the large $m$ perturbative
prediction, where the perturbative expansion is valid. We clearly see
that the phase boundary connects the two
axes. }{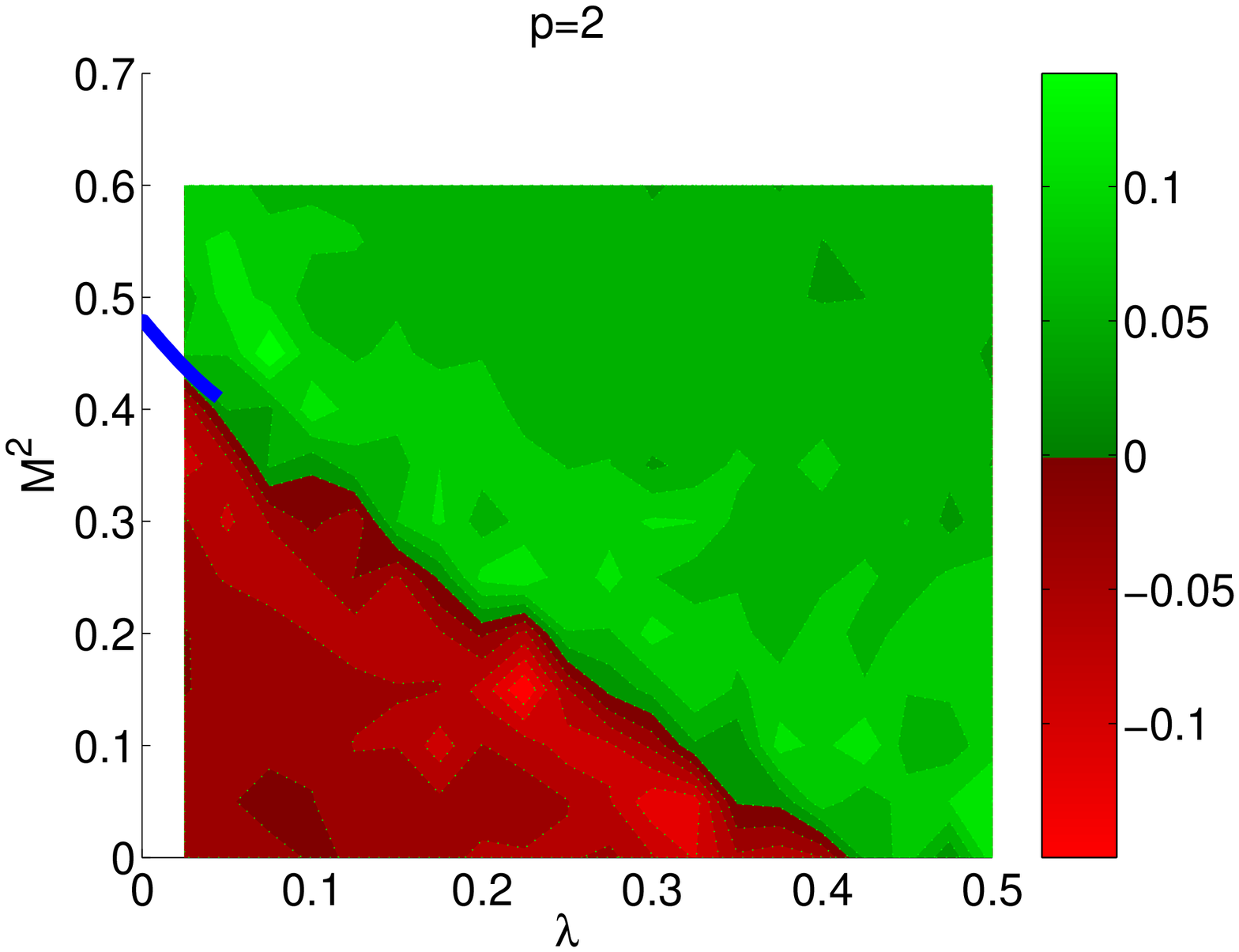}{4truein}
\figlabel{\contourA}

\fig{Contours of $\Delta^{12,20} \vev{u_1}$ for the $p=4$ theory, with the
same conventions as above.}{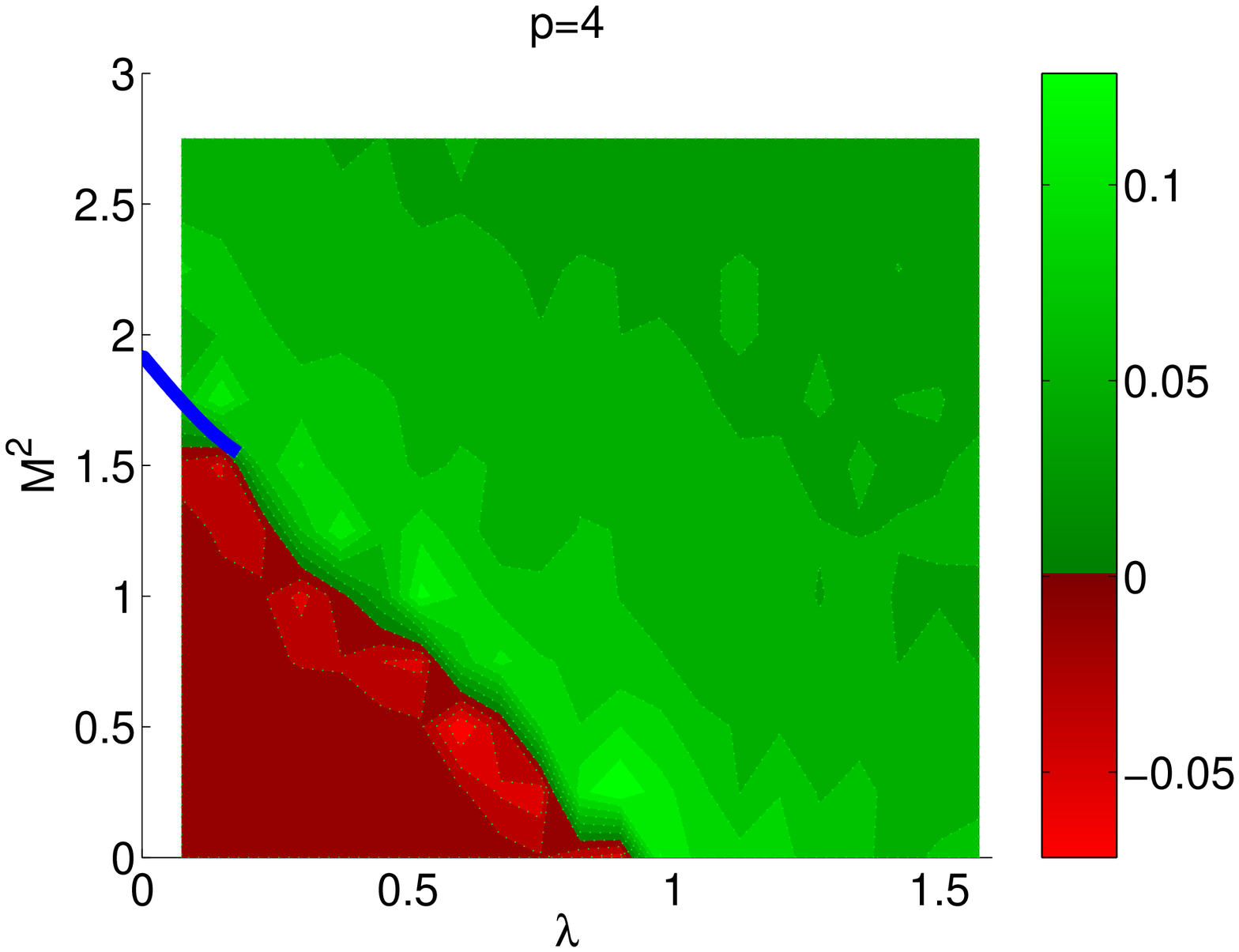}{4truein}
\figlabel{\contourB}

\fig{Contours of $\Delta^{6,12} \vev{u_1}$ for the $p=9$ theory, with the
same conventions as above.}{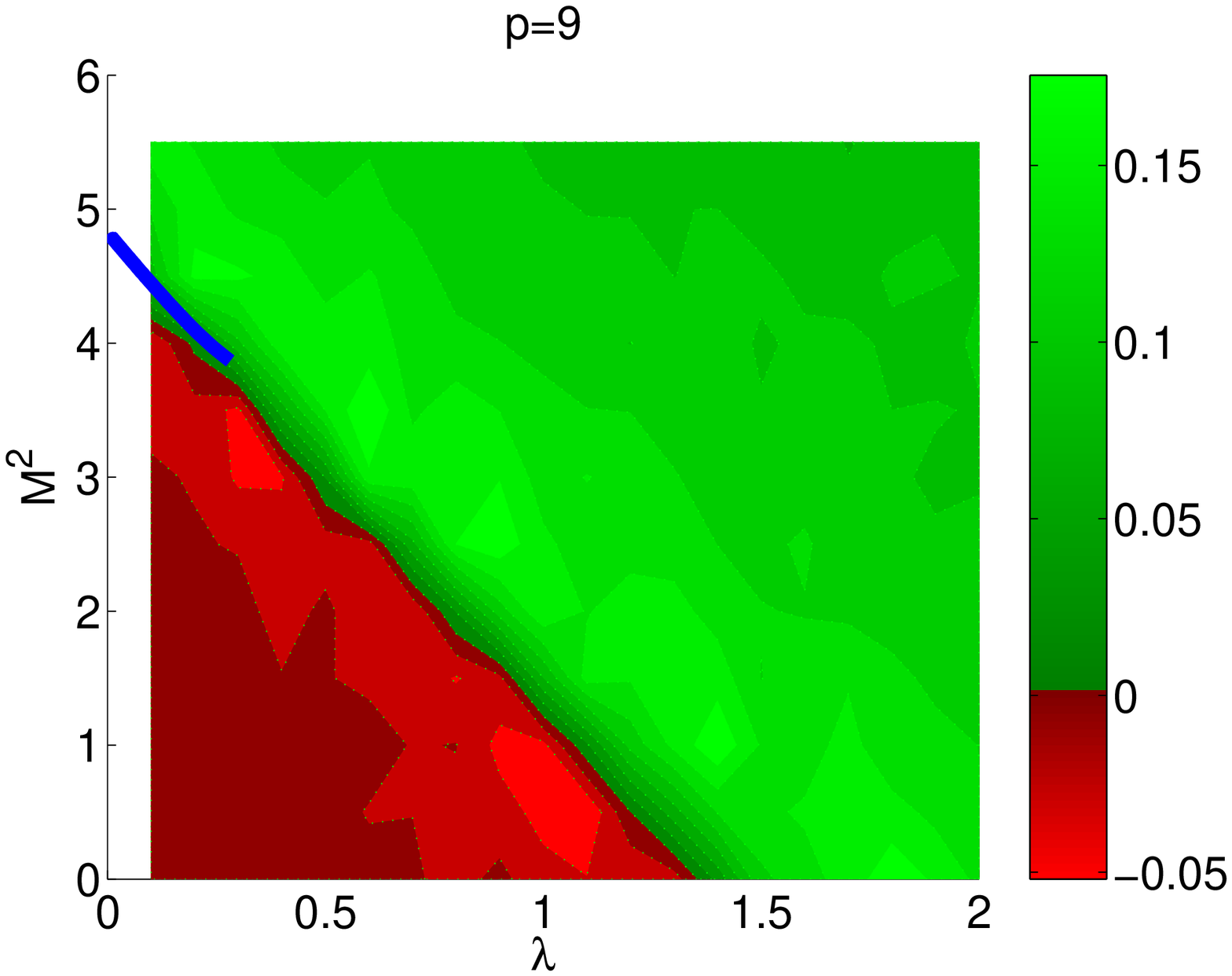}{4truein}
\figlabel{\contourC}

In figures \contourA, \contourB, \contourC\ we plot this
transition line as a function of $m$ and $\tit$ (see below for more
details on how exactly this transition line was obtained from our
data). In each case this line interpolates smoothly between the
analytic prediction of subsection 4.1 and the results of the
previous subsection. As in the previous subsection, our numerics are
unable to definitively establish the order of the transition;
however they are certainly consistent with the simplest conjecture,
which is that the transition (which the computations of subsection 4.1
have established to be of first order at large $m$) remains of first order
for all values of $m$.

To end this subsection we give a precise definition of the
transition lines plotted in figures \contourA, \contourB, \contourC.
We characterize the position of the transition by computing the
difference between the values of $\vev{u_1}$ for different values of
$N$.  `Outside' the phase transition boundary $u_1$ decreases with
increasing $N$, while `inside' it stays roughly constant (with a small
increase). We may estimate the phase boundary as the line where these
differences are zero, namely the function $u_1$ is neither increasing nor
decreasing with increase of $N$. Of course, this estimate to the phase
transition line becomes more accurate as we increase $N$.

\fig{Figure of $\Delta^{6,20} \vev{u_1}$ (blue) and $\Delta^{12,20} \vev{u_1}$
(red) in the $\lambda_1, M$ plane for the $p=4$ theory (taking $R=1$ units).
We take the
zero contour to measure the position of the phase boundary.
}{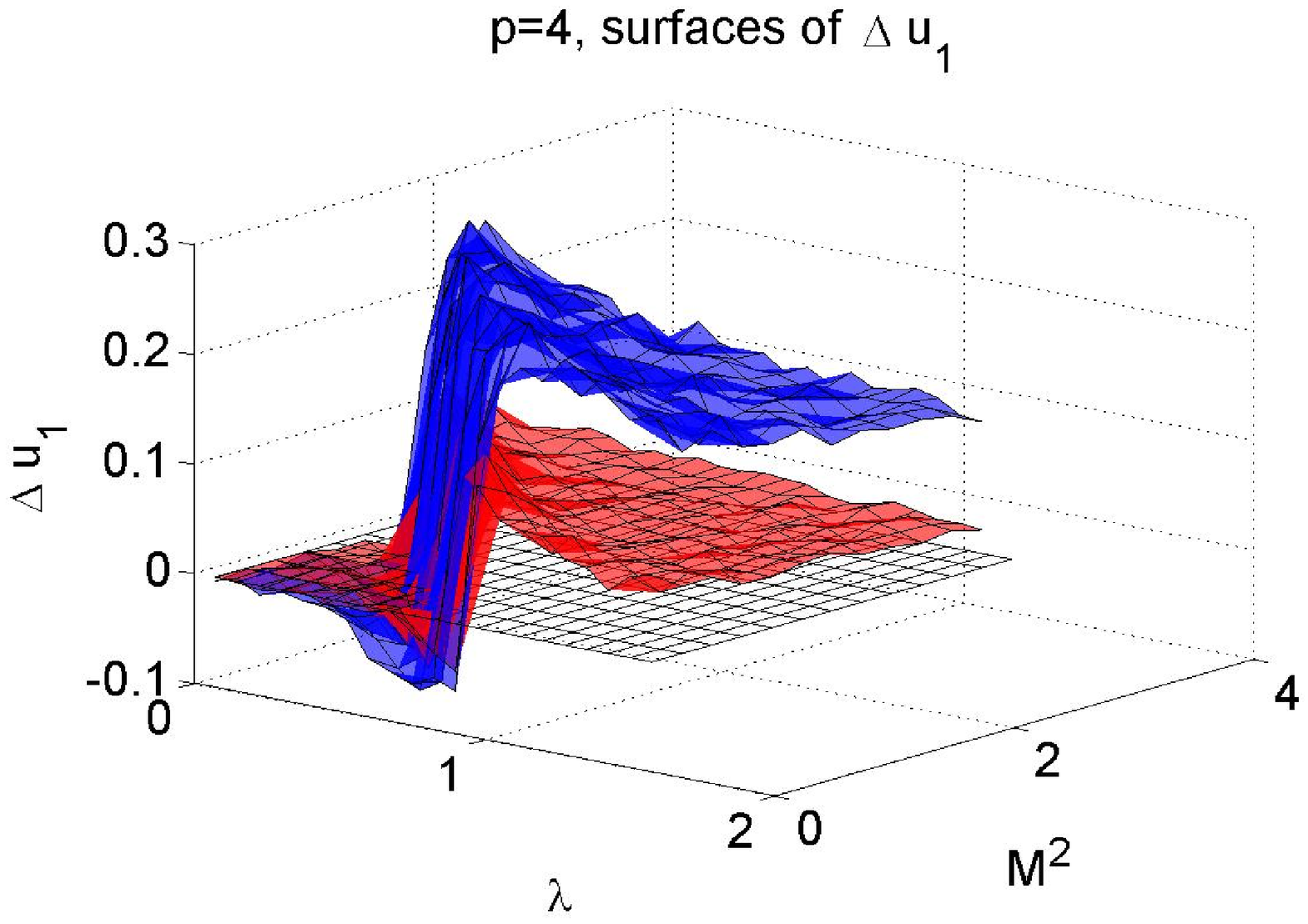}{4truein} \figlabel{\surfC} \

To illustrate our procedure, in figure \surfC\ we plot
$\Delta^{6,20} \vev{u_1} \equiv \vev{u_1}^{(N=6)}-\vev{u_1}^{(N=20)}$
and $\Delta^{12,20}
\vev{u_1} \equiv \vev{u_1}^{(N=12)}-\vev{u_1}^{(N=20)}$.
Both surfaces in the plot give a consistent phase boundary location.

\subsec{One massless and $p$ massive scalars}

Before concluding this section, we briefly consider a generalization
of \eym\ for which one of the scalars (which we call $\phi$) remains
at zero mass, while the other $p$ scalars are massive.
This arises in studying the high temperature limit of a
1+1 dimensional gauge theory with massive scalars, for which the
temporal component of the gauge field leads to a massless scalar in
the dimensionally reduced theory.

The theory behaves in a qualitatively similar manner to the massive
theory considered above. For $m \to 0$, the two theories are the same,
while at large $m$, we will see that both theories have a phase
boundary at a temperature which goes to infinity as $m$ goes to
infinity, though with different dependence on $m$.

In this limit, it is again appropriate to integrate out the massive
scalars at one loop to generate an effective potential for $\phi$ and
the holonomy matrix $U$. The $\phi$ independent terms will be given by
\weakeffo\ as before, with corrections negligible as long as
$m \gg p^{1/3}$.
At very large $m$, the most important $\phi$-dependent term will be a
mass term\foot{Here, $\phi$ is normalized to have a kinetic term $\int
dt \tr({1 \over 2}( D_0 \phi)^2)$.}
\eqn\phidepterm{
S_{eff} = \int dt {\lambda_1 p \over 2 M} \tr( \phi^2).  }
The leading
$\phi^4$ interaction comes with a coefficient $g_\phi^2 \sim \lambda_1^2
/N M^3$, and can be ignored relative to the quadratic terms when
integrating out $\phi$ as long as
\eqn\contwo{
m_\phi/(g_\phi^2 N)^{1 \over 3} \gg 1 \qquad \Rightarrow \qquad m \gg 1.
}
There will be additional terms involving both $\phi$ and $U$, but
these are nonlocal and will be suppressed exponentially by powers of
$e^{-R M}$ for large $R$. We should be able to ignore these if
\eqn\conthree{
RM \gg 1 \qquad \Rightarrow \qquad \tit \ll m.
}
When all these conditions are satisfied, we may integrate out the
scalar $\phi$ at one loop (using only the kinetic term and mass term)
to obtain additional terms in the effective action for $U$. The result
(including terms we have already from integrating out the
scalars of mass $M$) is
\eqn\effmm{ S_{eff}(U) = \sum_{n=1}^{\infty} \; {1 \over n}
(p x^n + x^{n\epsilon}) \; \tr(U^n)
\tr(U^\dagger {}^n), }
where $x = e^{-M R}=e^{-m/\tit}$ as before and $\epsilon$ is the ratio of the
induced $\phi$ mass to the $X$ masses,
\eqn\epsval{
\epsilon = {\sqrt{p} \over m^{3 \over 2}} \ll 1.
}
This theory has a deconfinement phase transition when
\eqn\mmtrans{\eqalign{
&px + x^\epsilon = 1\cr
\Rightarrow &\ px + \epsilon \ln(x) = 0 \cr
\Rightarrow &\ x = {\epsilon \over p} ( \ln({p \over \epsilon}) +
{\cal O}(\ln(\ln({p \over \epsilon})))) \cr
\Rightarrow &\ \tit ={2 \over 3} {m \over \ln(m)}\left(1 +
{\cal O}\left({\ln(\ln(m))
\over \ln(m)}\right)\right).
}}
Note that at this temperature, \contwo\ and \conthree\ are both
satisfied, so our analysis is self consistent. Thus, we conclude that
at large $m$, the phase transition temperature goes as $\tit = {2
\over 3} m/\ln(m)$.

\newsec{Two Dimensional Bosonic Gauge Theories on $T^2$}

In this section we study the two dimensional large $N$
$SU(N)$ Yang-Mills theory
coupled to $p$ Hermitian scalar fields in the adjoint representation,
\eqn\ymlmsm{S=\frac{N}{2\lambda}\int d^2 x ~\tr\left\{F_{12}^2+
\sum_I \left[ \left(\del_{\mu}\Phi^I-i\left[A_{\mu},\Phi^I\right]\right)^2 +
 m_I^2 \Phi_I^2 \right] -\sum_{I<J}\left[\Phi^I,\Phi^J\right]^2\right\},}
with $I,J=1,\cdots,p$.  Viewed as a $1+1$ dimensional quantum field
theory, this theory is believed to possess a unique normalizable
vacuum state and a Hagedorn growth in its density of states (see
\kleb\ for the special case $p=1$). We will study the partition
function of the Euclidean theory on a rectangular torus with radii
$R_1$ and $R_2$, as a function of the geometry and masses (in units where
the coupling is fixed). We will
investigate this system in two separate limits; first when all scalars
are massless, and second when all scalars have large mass.

In this case, the geometry has two non-contractible cycles, so any
saddle point configuration will be characterized by two holonomy
matrices which we call $U$ and $V$. At each point in parameter space,
we will ask whether the eigenvalues of these matrices in the saddle
point configuration are clumped or unclumped (uniformly distributed on
the unit circle), and so obtain a phase diagram containing four possible
phases. The phase diagram will turn out to be similar in the massless
and large mass limits, and so we believe it is likely to have the same
structure for any value of the mass.

\subsec{The massless theory}

We first consider the theory \ymlmsm\ in the limit where the scalar
masses are zero.  We wish to understand the phase diagram of this
theory on a rectangular torus, as a function of its two dimensionless
parameters, which we can choose to be $r_1 = R_1 \sqrt{\lambda}$ and
$r_2 = R_2 \sqrt{\lambda}$, the circumferences of the spatial and temporal
circles in units where the coupling is set to one. Since there is
nothing to distinguish the two circles, the phase diagram must be
symmetric under the exchange $r_1 \leftrightarrow r_2$.

Let us first review what we know about the theory in the limit $r_1 \to
\infty$ keeping $r_2$ fixed; this limit gives the
decompactified system at a temperature $1 / R_2$. At low temperatures
this system is believed to confine, with a Hagedorn tower of glueball
states (see \kleb\ for a proof for the case $p=1$). As the temperature
is raised, the system undergoes a sharp deconfinement transition at a
temperature of order $\sqrt{\lambda}$. In Euclidean space the order
parameter for this phase transition is the expectation value of the
Wilson line around the $x_2$ circle; the eigenvalues of this Wilson
line are clumped at high temperatures but uniformly smeared over the
circle at low temperatures. Thus, for large $r_1$, we expect a
deconfinement transition line which asymptotes to a constant $r_2$ of
order unity. Exchanging the two circles, we must then have also a
phase transition line for large $r_2$ asymptoting to constant $r_1$,
across which the eigenvalues of the Wilson line in the $x_1$ direction
clump as we decrease $r_1$.

We now turn to the opposite limit where the spatial circle $R_1$ is
very small. When the Kaluza-Klein scale $1/R_1$ is much larger than
the scale $\lambda_1^{1/3}$ set by the effective one-dimensional
coupling $\lambda_1 = \lambda / R_1$ and also much larger than the
temperature $1/R_2$, the theory will reduce to the one-dimensional
theory considered in the previous section with $p+1$ massless
scalars. The parameter $\tilde{t}$ of the previous section is given in
terms of the two-dimensional parameters by
\eqn\tildetdef{
\tilde{t} = {1 \over R_2 \lambda_1^{1 \over 3}} = {r_1^{1 \over 3} \over r_2}.
}
Thus, in the regime $r_1 \ll 1$ and $r_1 \ll r_2$ we expect a phase
transition along the curve
\eqn\rtwopt{
r_2 = {1 \over \tilde{t}_c^{(p+1)}} r_1^{1 \over 3}
}
where $\tilde{t}_c^{(p+1)}$ is the critical value of $\tilde{t}$ in the
one-dimensional theory with $p+1$ massless scalars. The eigenvalues of $U$ (the
holonomy around $R_2$) will be clumped or unclumped for values of
$r_2$ below or above this line (again assuming that $r_1 \ll 1$ and
$r_1 \ll r_2$).

Again, we may swap the role of the two circles and conclude that there
is an additional phase transition along the curve
\eqn\rtwoptn{
r_2 = (\tilde{t}_c^{(p+1)})^3 r_1^3
}
such that the eigenvalues of $V$ (the holonomy around $R_1$) are
clumped/unclumped above/below the curve, in the regime of validity
$r_2 \ll r_1$ and $r_2 \ll 1$.

Note that these results rely on the numerics of the previous section, however, in the regime
\eqn\pertrange{
r_1^{1 \over 3} \gg r_2 \gg r_1^3
}
all non-zero modes are weakly coupled and can be integrated out at one
loop (see appendix A) so we are able to verify analytically that both
$U$ and $V$ are clumped in this regime.\foot{Naively, we might expect
that the theory should admit a perturbative analysis as long as the
torus area is small in units of the coupling. In fact, as in finite
temperature computations in 3+1 dimensional theories, this naive
expectation is modified by infrared effects (see \glpaper\ for a
discussion of this point). The correct regime of validity may be
obtained by requiring that the $R_1$ circle should be small in units
of the one-dimensional coupling $\lambda / R_2$ and that the $R_2$
circle should be small in units of the one-dimensional coupling
$\lambda / R_1$, leading to \pertrange. }

\fig{Known aspects of the phase diagram of massless bosonic $SU(N)$
gauge theories with three possible completions. $U$ is the expectation
value of the holonomy in the $x_2$ direction, and $V$ is the
expectation value of the holonomy in the $x_1$
direction.}{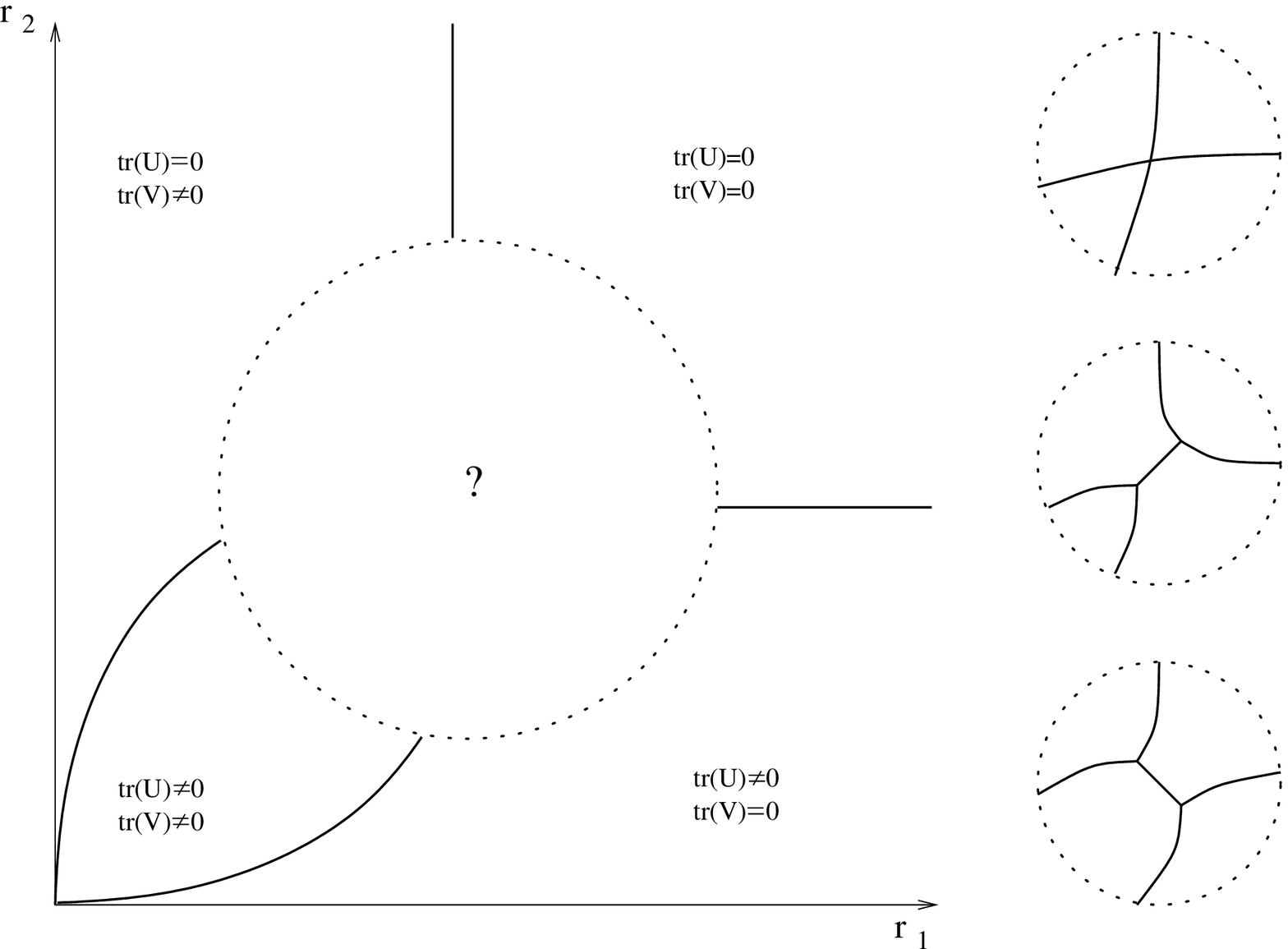}{4truein} \figlabel{\smMgen}

Putting all this together, we conclude that the phase diagram of
\ymlmsm\ for $m \gg 1$  takes the form shown in figure \smMgen.
We have sketched three possible simple completions of this phase
diagram, however, our current lack of understanding of the dynamics at
$r_1,r_2 \sim 1$ prevents us from distinguishing between these three
possibilities.

The phase transitions in figure \smMgen\ are driven by the strongly
coupled dynamics of a field theories' worth of degrees of freedom, and
so are difficult to analytically control. In the following subsections,
we will see that as for the one-dimensional theories considered in
section 4, analytic analysis becomes possible in the opposite limit
when the scalar masses become large, $m \gg 1$, where we take $m_I=M$
and define the
dimensionless mass parameter $m = M / \sqrt{\lambda}$.

\subsec{$M= \infty$}

First, consider the strict limit $M \to \infty$. In this limit all
scalar fields may simply be set to zero and \ymlmsm\ reduces to pure
two dimensional Yang-Mills theory on a torus, an exactly solvable
system.  Over 30 years ago \MigdalAB, Migdal rewrote the partition
function of two dimensional Yang-Mills theory on a torus in terms of
an integral over the two unitary holonomy matrices
\eqn\pureym{Z_{ym}=\int\,DU\,DV\,\sum_Rd_Re^{-\frac{\tilde{\lambda}}{2N}
C_2(R)}\chi_R(UVU^{-1}V^{-1}),}
where
\eqn\tildelamdef{
\tilde{\lambda} = \lambda R_1 R_2 = r_1 r_2.
}
Here, the sum over $R$ runs over all representations of the $SU(N)$ gauge
group, and $d_R$ and  $C_2(R)$ are the dimension and the quadratic Casimir
of the representation $R$
(see Appendix D for a brief derivation). It is not difficult
to integrate out one of the two matrices -- say, $V$ -- in \pureym, to
obtain an effective action for $U$,
\eqn\ymuact{
Z_{ym}=\int\,DU\,\sum_R e^{-\frac{\tilde{\lambda}}{2N}
C_2(R)}\chi_R(U) \chi_R(U^\dagger) \; ,
}
which may be evaluated to obtain \refs{\RusakovRS,\WittenWE,\FineZZ,\BlauMP}
\eqn\ymres{
Z_{ym} = \sum_R e^{-\frac{\tilde{\lambda}}{2N}
C_2(R)} \; .}

For our analysis below, it will be useful to note (see Appendix D)
that at large $N$, \ymuact\ may be written as
\eqn\ymmarkp{Z_{YM}=\int\,DU\,\exp\left(\sum_n
{1 \over n}(-e^{-\tilde{\lambda}n}+
2e^{-\frac{\tilde{\lambda}n}{2}}) \tr(U^n)\tr(U^{-n})\right).}
Equation \ymmarkp\ may be rewritten as an integral over the moments
of the eigenvalue density function as
\eqn\ymurho{Z_{YM}=\int\,du_n d \bar{u}_n \,\exp \left( -N^2 \sum_n {1 \over n}
(e^{-\frac{\tilde{\lambda}n}{2}}-
1)^2|u_n|^2\right),}
where the moments, $u_n$, are related to $U$ by
\eqn\rhondefd{u_n \equiv \frac{1}{N}\tr(U^n).}
Note that the coefficients of $|u_n|^2$ in \ymurho\ are positive
for all $n$; it follows that $Z_{YM}$ is dominated by the saddle
point $u_n=0$ for all $n$, in which the eigenvalue distribution is
uniform. Of course, the same result is true for the eigenvalues
of the holonomy matrix $V$. Thus we conclude that, independent of
the torus radii, the eigenvalues of both holonomy matrices are
unclumped for pure Yang-Mills theory on $T^2$.

Notice that at small ${\tilde \lambda}$ the effective mass for
$u_n$ in \ymurho\ is very small (the mass squared is approximately $n {\tilde
\lambda}^2 / 4$). As a consequence, in this regime a small
attractive perturbation could cause the eigenvalues to clump. We
will now argue that such a perturbation is supplied by the effective
potential generated by scalar fields $\Phi_I$ at large but
finite mass.

\subsec{Large $M$, noncompact limit}

We first note that when $R_1$ is larger than any other length scale in
the problem, the theory should behave like the noncompact thermal 1+1
dimensional theory with very massive scalars. This theory was analyzed
in \sz. There, it was argued that by integrating out all degrees of
freedom except the spatially dependent holonomy matrix (in the $R_2$
direction) $U(x)$, the
system reduces to the model
\eqn\szaction{
S = \int dx \left\{ { N \over 2 \lambda R_2} \tr(|\partial_x U|^2) - p
\sqrt{M \over 2 \pi R_2} e^{-M R_2} \tr(U(x)) \tr(U^\dagger(x)) \right\} \; .
}
Using collective field theory methods, they showed that this model
displays a first order deconfinement transition at a temperature
corresponding to
\eqn\rtwoptnn{
r_2 = {2 \over m} \ln(m)\left(1 + {\cal O}\left( {\ln(\ln(m)) \over \ln(m)}
\right)\right) \; .
}
Thus, our model will have a phase transition curve asymptoting to this
value for large $r_1$. Above the curve (corresponding to the
noncompact theory at low temperatures), the eigenvalues of the
temporal holonomy $U$ will be unclumped, while below it, they will be
clumped. By symmetry, there must be an additional phase boundary
asymptoting to $r_1 \sim {2 \over m} \ln(m)$ at large $r_2$, across which the
eigenvalues of the spatial holonomy $V$ clump.

\subsec{Large $M$, small volume}

We may now ask about the opposite limit, where $R_1$ is so small that
the theory is effectively one-dimensional. Assuming that the
eigenvalues of $V$ are clumped (we will see that this is the case for
small enough $R_1$), the theory will behave as a one dimensional
theory with $p$ massive scalars and one massless scalar (from the
spatial component of the gauge field) as long as the Kaluza-Klein
scale $1/R_1$ is much larger than the mass $M$, the temperature
$1/R_2$ and the scale $\sqrt{\lambda}$ of the gauge theory, yielding
conditions
\eqn\threeconds{
r_1 \ll r_2, \qquad \qquad r_1 \ll 1/m, \qquad \qquad r_1 \ll 1.
}
Here, the last relation is implied by the second for large $m$.

In the language of section 4, the corresponding one-dimensional theory
will have parameters $\tit = r_1^{1 \over 3}/r_2$ and $m_1 = m r_1^{1
\over 3}$. From the results of section 4.4, we expect that this theory
undergoes a phase transition at some $\tit_c(m_1)$ for all values of
$m_1$, where $\tit_c$ rises from some value $\tit_c^{(p+1)}$ of order
one for small $m_1$ and asymptotes to $2 m_1 / (3 \ln(m_1))$ for large
$m_1$. Expressing these results in terms of our two-dimensional
parameters, we predict a phase transition at
\eqn\smallphase{\eqalign{
r_2 &= {r_1^{1 \over 3} \over \tit_c(m r_1^{1 \over 3})}
}}
which gives
\eqn\smallphase{
r_1 = \left\{ \matrix{r_2^3 (\tit_c^{(p+1)})^3 &\qquad \qquad r_1
\ll {1 \over m^3} \cr
{r_2^2 \over p m} e^{2 m r_2} & \qquad \qquad  {1 \over m^3}
\ll r_1 \ll {1 \over m} } \right.
}
where to derive the last expression we used the equation in the second
line of \mmtrans.
For larger values of $r_1$, the conditions \threeconds\ are no
longer satisfied, so the theory is no longer well approximated by a
one-dimensional theory.

The transition line \smallphase\ separates a region with clumped $U$
eigenvalues (for smaller $r_2$) from a region with unclumped $U$
eigenvalues (for larger $r_2$). By symmetry, we will have an addition
phase boundary separating regions of clumped and unclumped $V$
eigenvalues, obtained by the replacement $r_1 \leftrightarrow r_2$ in
\smallphase.

\subsec{Large $M$, intermediate radius.}

At intermediate values of $R_1$, neither of the approximations so far
apply, but the theory is still simple enough to treat analytically as
long as the scalar masses are very large. In this case, we can
reliably integrate out the scalars at one-loop order, leading to an
effective action
\eqn\scveff{S=\frac{N}{2\lambda} \left( \int d^2x \, \tr(F_{12}^2) \right)+
\frac{p}{2}\ln\det(-D_{\mu}^2+M^2)}
for the gauge fields. The explicit expression for the determinant will
in general be quite complicated, with local terms built from the gauge
field strength and its covariant derivatives, together with non-local
terms involving the Wilson lines about the spatial and temporal
cycles. The former will be small relative to the tree-level $F^2$ term
as long as $m \gg 1$. The non-local terms involving the holonomy
around a cycle fall off exponentially with the radius of the cycle in
units of the inverse scalar mass, but we will see that these terms are
important even when this exponential is very small (equal to an
inverse power of $m$).

For sufficiently small $R_1$ and $R_2$ (we will be more explicit
below), terms involving the spatially varying modes of the fields will
be suppressed, and it is enough to consider the effective action
assuming that the components $A_1$ and $A_2$ of the gauge field
(and hence the holonomy matrices)
are spatially constant. In Appendix B, we show that up to commutator
terms, the result for the effective action in this case is
\eqn\logdet{\eqalign{S_{eff} &= {p
\over 2} \ln\det(-D_{\mu}^2+M^2)
\cr & =-p\sum_{(k,n) \ne (0,0)} \tr (U^n V^k) \tr(U^{-n} V^{-k} ) M R_1 R_2
{K_1\left( M \sqrt{R_1^2k^2+R_2^2 n^2}\right)
\over 2\pi \sqrt{R_1^2k^2+R_2^2n^2}}  \cr  
& \to - p\sum_{(k,n)\ne (0,0)}
\tr (U^n V^k) \tr(U^{-n} V^{-k} ) \sqrt{M} R_1 R_2 {\exp\left(-
M \sqrt{R_1^2k^2+R_2^2n^2}\right) \over
2 \sqrt{2 \pi} (R_1^2 k^2+R_2^2 n^2)^{3 \over 4}},}}
where in the last line we have taken the limit of large $MR_i$ and
used the asymptotic form of the
associated Bessel function $K_1$.\foot{When $U$ and $V$ commute, this
is precisely a pairwise potential between the $N$ points on the dual
torus formed by the simultaneous eigenvalues of $U$ and $V$.} Note
that if $M R_1$ and $M R_2$ are both sufficiently large, all terms
will be small relative to the $(k,n)=(0,\pm 1)$ and $(k,n)=(\pm 1,0)$
terms. In addition, the commutator terms that we have not written
necessarily involve at least two powers of $U$ and two powers of $V$,
so we expect them to be suppressed at least as strongly as the $U^2
V^2$ terms above.

Thus, when the spatially varying modes can be ignored and both $M R_1$
and $M R_2$ are large, the effective action from the scalars is well
approximated by
\eqn\seffuv{
S_{eff}(U,V) = f(U) + g(V),
}
where
\eqn\uvact{\eqalign{
f(U) &= -{p \over \sqrt{2 \pi}} \frac{R_1}{R_2}\sqrt{MR_2}e^{-MR_2}
\tr(U)\tr(U^{\dag}) \; ,\cr
g(V) &= -{p \over \sqrt{2 \pi}} \frac{R_2}{R_1}\sqrt{MR_1}e^{-MR_1}
\tr(V)\tr(V^{\dag}) \; .
}}
Note that $f$ comes from the second term in \szaction\ if we take the
integral to run over a finite range $R_1$ and ignore the non-zero
modes of $U(x)$. Both the kinetic term and the potential term in
\szaction\ will produce quadratic terms in the non-zero modes of
$U(x)$. By demanding that the relative coefficient for the potential
term is small, we obtain an explicit condition for when the non-zero modes
of $U(x)$ may be ignored,\foot{This condition is most appropriate in the
regime $r_1 \gg r_2$ where \szaction\ is valid, but this is the only
place we will need it in what follows. Similarly, the equation for the
non-zero modes of $V(x)$ is the
appropriate condition for the regime $r_2 \gg r_1$ . }
\eqn\umodecond{
r_1^2 \ll {1 \over \sqrt{m r_2}} e^{m r_2}.
}
Similarly, the non-zero modes of $V(x)$ can be ignored when
\eqn\vmodecond{
r_2^2 \ll {1 \over \sqrt{m r_1}} e^{m r_1}.
}
When all of our assumptions apply, the partition function becomes (using
\scveff\ and \pureym)
\eqn\notpure{
Z_{ym}=\int\,DU\,DV\,\sum_Rd_Re^{-\frac{\tilde{\lambda}}{2N}
C_2(R)}\chi_R(UVU^{-1}V^{-1})e^{-f(U)}e^{-g(V)}. }
Making the change of variables $U \to W U W^{-1}$ for unitary $W$, and
integrating over $W$ (more details in appendix D), we find
\eqn\notpureym{\eqalign{
Z_{ym} &=\int\,DU\,DV\,\sum_R e^{-\frac{\tilde{\lambda}}{2N}
C_2(R)} {d_R^2 \over d_R^2 - 1} e^{-f(U)}e^{-g(V)} \cr &
\qquad \left\{-1 + \chi_R(V) \chi_R(V^\dagger) + \chi_R(U) \chi_R(U^\dagger)
- {1 \over d_R^2} \chi_R(U) \chi_R(U^\dagger) \chi_R(V) \chi_R(V^\dagger)
\right\} .
}}
For large $N$, the coefficient $d_R^2/(d_R^2 - 1)$ can be set to one.

From this expression, we would like to understand the behavior of our
order parameters, $\ln(Z_{ym})/N^2$ and the expectation values for the
temporal and spatial Wilson loops. We note first that for the purposes
of computing the partition function or any observable depending on
either $U$ or $V$ alone, the last term in \notpureym\ can be ignored
relative to the other terms as long as $R_1 M$ and $R_2 M$ are large
enough. For, suppose we evaluate the partition function with an
operator ${\cal O}(U)$ inserted. Then in the last two terms of
\notpureym, the $V$ dependent terms may be collected inside the $U$
integral to obtain
\eqn\vsubint{
\int \,DV\ e^{-g(V)} \left\{1 - {1 \over d_R^2} \chi_R(V) \chi_R(V^\dagger)
\right\}
= Z_g \left\{ 1 - \langle {1 \over d_R^2} \chi_R(V) \chi_R(V^\dagger)
\rangle_g \right\}.
}
Here, $Z_g$ and $\langle \rangle_g$ are the partition function and
expectation values for the unitary matrix model with action
$g(V)$. But as long as the coefficient in $g(V)$ is small,
\eqn\rtwocond{
r_2 \ll {1 \over p} \sqrt{ r_1 \over m} e^{m r_1} \; , }
such a model will be dominated by the repulsive eigenvalue potential
from the measure, and display confining behavior with a uniform
eigenvalue distribution for the saddle point. In this case, the
expectation value in \vsubint\ will vanish for large $N$, since this
is just the norm squared of the Polyakov loop in the representation
$R$. In other words, term by term in the sum over $R$ in \notpureym,
the contribution from the last term will be negligible relative to the
contribution from the third term. Similarly, if we insert any operator
${\cal O}(V)$, the last term will be negligible compared to the second
term as long as
\eqn\ronecond{
r_1 \ll {1 \over p} \sqrt{ r_2 \over m} e^{m r_2} \; .
}

Let us now assume that both \rtwocond\ and \ronecond\ are satisfied
and consider first the partition function \notpureym\ with no operator
inserted. We then obtain
\eqn\simppart{
Z_{YM} = -Z_{M=\infty} Z_g Z_f + Z_f Z_{S_{\infty}(V) + g} + Z_g
Z_{S_{\infty}(U) + f} }
where $Z_{M=\infty}$ is the partition function
\ymres\ for the pure Yang-Mills theory, and $S_{\infty}$ is the action
in the expression \ymuact\ for the pure Yang-Mills partition function
reduced to a single matrix integral. Now, $Z_{M=\infty}$, $Z_g$, $Z_f$
are all finite for large $N$ (when our conditions are satisfied) since
they correspond to confining theories, so $Z_{YM}$ will show
deconfined behavior if and only if either $Z_{S_{\infty}(V) + g}$ or
$Z_{S_{\infty}(U) + f}$ does.

Using the large $N$ result \ymmarkp\ to get an explicit expression for
$S_{\infty}$ and rewriting in terms of the modes of the eigenvalue
distribution, we find
\eqn\ymseffu{
Z_{S_{\infty(U)+f}} = \int\,du_n d \bar{u}_n \,\exp\left(-N^2\sum_n
{1 \over n} (1 -e^{-\frac{r_1 r_2 n}{2}})^2 |u_n|^2 +
N^2 {p \over \sqrt{2 \pi}} \frac{r_1}{r_2}\sqrt{m r_2}e^{-m r_2}|u_1|^2
\right).
}
We see that the scalar effective potential gives a negative
contribution to the mass squared for the lowest Fourier mode of the
eigenvalue distribution. Thus, $u_1$ becomes tachyonic, and the free
energy in \ymseffu\ becomes of order $N^2$, when
\eqn\decu{
{p \over \sqrt{2 \pi}} \frac{r_1}{r_2}\sqrt{m r_2}e^{-m r_2} >
(1 -e^{-\frac{r_1 r_2}{2}})^2.
}
Since we assumed that the left-hand side is very small, this can only
happen when $r_1 r_2 \ll 1$.
Similarly, $Z_{S_{\infty}(V) + g}$ will show deconfining behavior when
\eqn\decv{
{p \over \sqrt{2 \pi}} \frac{r_2}{r_1}\sqrt{m r_1}e^{-m r_1} >
(1 -e^{-\frac{r_1 r_2}{2}})^2.
}
These two curves intersect at
\eqn\curveint{
r_1 = r_2 = {4 \over m} \ln(m) (1 +
{\cal O}\left( {\ln(\ln(m)) \over \ln(m)} \right)),
}
and together form part of the boundary of the region with free energy
of order one and both $U$ and $V$ unclumped. Note that the curve
\decu\ exits the region where \umodecond\ is satisfied when $r_1 \sim
m/\ln(m^2)$ and $r_2$ has decreased to be of order $2\ln(m)/m$. Beyond
this, we cannot trust the functional form \decu\ of the curve, but we
have already argued in section 5.4 that there should be a phase
boundary asymptoting to precisely this value, suggesting that our
result \decu\ matches on smoothly to the large $R_1$ behavior (and
similarly for \decv ). Noting that $r_1 r_2 \ll 1$ on the curve \decu\
everywhere that \umodecond\ is satisfied, we can write a more explicit
expression for the phase boundary in this region,
\eqn\decbound{\eqalign{
r_1 &= {4 p \over \sqrt{2
\pi}} {\sqrt{m} \over r_2^{5 \over 2}} e^{-m r_2} \qquad r_1 < r_2 \ll
{1 \over r_1} \cr
r_2 &= {4 p \over \sqrt{2 \pi}} {\sqrt{m} \over r_1^{5 \over 2}} e^{-m
r_1} \qquad r_2 < r_1 \ll {1 \over r_2}
}}
Note that the conditions \rtwocond\ and \ronecond\
are satisfied for all such values.

To see what happens on the deconfined side of this boundary, let us
now understand the behavior of the order parameters $\langle
|\tr(U)|^2/N^2 \rangle$ and $\langle |\tr(V)|^2/N^2 \rangle$, focusing
for now on the first one. As we have argued above, the last term in
\notpureym\ can be ignored in computing either of these, so by the
arguments leading to \simppart, we obtain
\eqn\vevusquared{
\langle {1 \over N^2} |\tr(U)|^2 \rangle = {-Z_{M=\infty} Z_g Z_f
\langle |u_1|^2 \rangle_f + Z_f Z_{S_{\infty}(V) + g}
\langle |u_1|^2 \rangle_f + Z_g Z_{S_{\infty}(U) + f}
\langle |u_1|^2 \rangle_{S_\infty(U) + f} \over -Z_{M=\infty} Z_g Z_f +
Z_f Z_{S_{\infty}(V) + g} + Z_g Z_{S_{\infty}(U) + f}}
}
Now, in the region of interest, at least one of $Z_{S_{\infty}(U) +
f}$ or $Z_{S_{\infty}(V) + g}$ behaves as $\exp(c N^2 )$ so the first
terms in both the numerator and denominator (which are at most of
order one in $N$) will be completely negligible here. In the remaining
expression,
\eqn\vevusquaredn{
\langle {1 \over N^2} |\tr(U)|^2 \rangle = { Z_f Z_{S_{\infty}(V) + g}
\langle |u_1|^2 \rangle_f + Z_g Z_{S_{\infty}(U) + f} \langle |u_1|^2
\rangle_{S_\infty(U) + f} \over Z_f Z_{S_{\infty}(V) + g} +
Z_g Z_{S_{\infty}(U) + f}}
}
even when both $Z_{S_{\infty}(U) + f}$ and $Z_{S_{\infty}(V) + g}$
have $\exp(c N^2)$ behavior, the larger will completely dominate
in both numerator and denominator, so
\eqn\choices{
\langle {1 \over N^2} |\tr(U)|^2 \rangle = \left\{ \matrix{ \langle |u_1|^2
\rangle_f = 0 & \qquad Z_{S_{\infty}(V) + g} > Z_{S_{\infty}(U) + f} \cr
\langle |u_1|^2 \rangle_{S_\infty(U) + f} \ne 0 & \qquad
Z_{S_{\infty}(U) + f} > Z_{S_{\infty}(V) + g} } \right.
}
The model with larger magnitude for the free energy will be the one
for which the $|u_1|$ mode is more tachyonic, and by inspection of
\uvact, this will be the $S_\infty(U) + f$ model for $R_1 > R_2$ and
the $S_\infty(V) + g$ model for $R_2 > R_1$.

Thus, from \choices\ and the analogous result for $\langle |v_1|^2
\rangle = \vev{|\tr(V)|^2/N^2}$,
we conclude that there is a phase boundary at $r_1=r_2$
starting from curve \decbound\ and continuing towards the origin, such
that the eigenvalues of $U$ are clumped and the eigenvalues of $V$ are
unclumped for $R_1 > R_2$ while the opposite is true for $R_2 >
R_1$. This phase boundary cannot continue all the way to the origin,
since we have argued in section 5.5 that both sets of eigenvalues are
clumped for small enough $r_1=r_2$. Indeed, in ignoring the higher
order terms in \logdet, we have assumed that $R_1 M$ and $R_2 M$ are
large, so we can only say with certainty that the phase boundary
exists in the region $1/m \ll r_1 = r_2 < 4\ln(m)/m$. For smaller
values of $r$, the phase boundary must bifurcate symmetrically, and we
expect that the two curves thus produced connect smoothly onto the two
phase boundaries which we argued in section 5.5 emanate from the
origin.

\fig{Phase diagram for Euclidean two-dimensional Yang-Mills theory with
massive adjoint scalars on $T^2$. Solid lines indicate phase
boundaries for which we have analytic
expressions.}{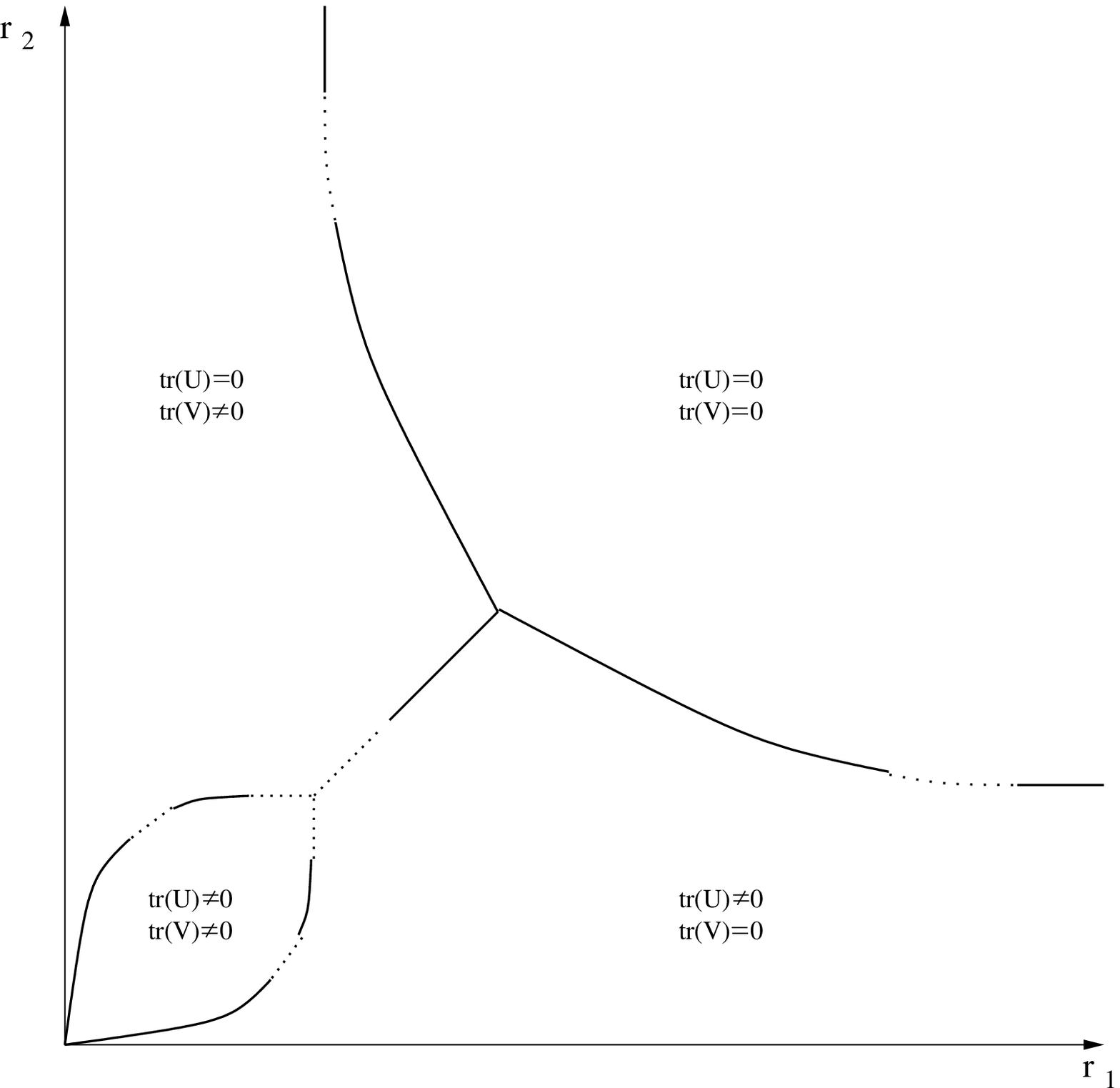}{3truein}
\figlabel{\twodmm}

\subsec{Summary of the large mass theory}

Combining all of our results, we conclude that the phase diagram for
large $N$ Yang-Mills theory with very massive adjoint scalars
compactified on a rectangular torus appears as in figure
\twodmm. Here, solid lines indicate regions where we have been able to
determine the analytic form of the phase boundary. Note that this
diagram displays the same qualitative behavior as one of the possible
completions of the massless phase diagram (figure \smMgen),
so it seems reasonable to
speculate that the phase structure is qualitatively identical for all
values of the scalar masses.

\fig{Phase diagram for 1+1 dimensional Yang-Mills theory with massive
adjoint scalars on $S^1$. The free energy is of order one in the shaded
region and of order $N^2$ elsewhere.}{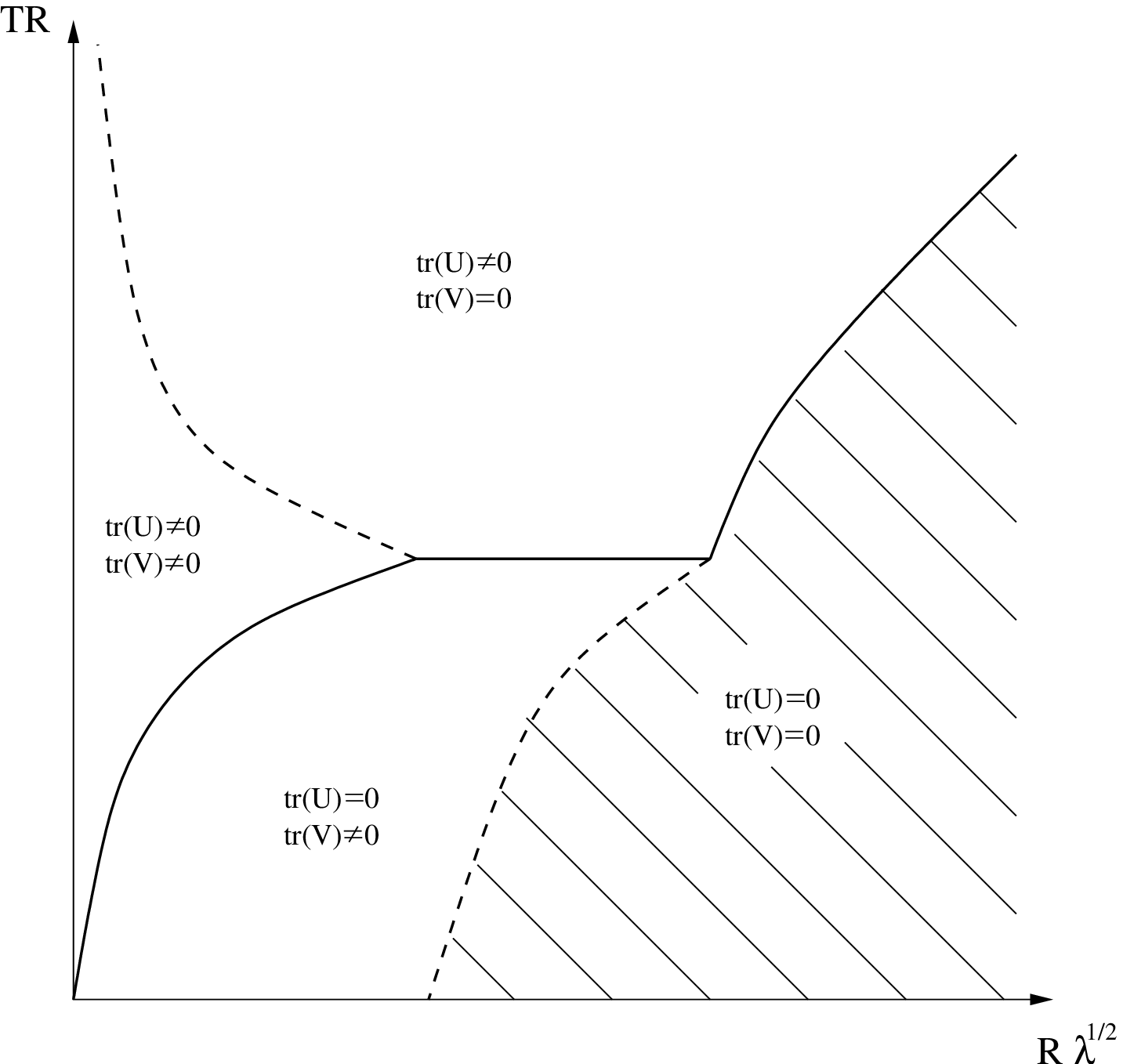}{3truein}
\figlabel{\like}

In figure \like, we present an alternate version of the diagram, with
axes labeled by the size of the spatial circle in units of the
coupling ($r_1 = R \sqrt{\lambda}$) and the temperature
in units of the spatial radius ($T R = r_1 / r_2$). Analogous
units were used in our analysis \refs{\sphere,\threelooppaper} of 3+1
dimensional Yang-Mills theory on a sphere, where our analysis
suggested a single first order transition extending from large volume
to a finite value of $TR$ in the zero volume limit. In the present
case, it is interesting to note that:

1) The solid line corresponding to deconfinement as measured by the
   temporal Wilson loop expectation value is not smooth.

2) While the deconfinement transition does extend all the way to zero
   volume as for the topologically trivial case, the transition
   temperature does not approach a finite value in units of the
   inverse spatial radius.

3) The large volume regions for both the confined and deconfined
   regions are separated by phase transitions from the small volume
   regions.

4) The two order parameters for confinement, $1/N^2$ times the free
   energy and the expectation value of the Polyakov loop, do not have
   the same behavior at small and large volumes in this example. More
   specifically, in the ``confined'' region below the solid line, the
   free energy is of order one at large volume, but it becomes of order
   $N^2$ to the left of the dotted line.

\newsec{Maximally Supersymmetric Quantum Mechanics}

In the rest of this paper we will study one and two dimensional
maximally supersymmetric gauge theories on circles and tori. As is
well known, Yang-Mills theories with $16$ supercharges display rather
different dynamics from their bosonic counterparts; in particular
the scalar effective potential in these theories has flat
directions, and the spectrum of these theories is not gapped. As a
consequence we expect (see section 2) that the phase diagrams of
these theories (which we will obtain)
will differ in significant aspects from the
equivalent diagrams for the purely bosonic theories of the previous
sections.

Partition functions of supersymmetric theories on tori are strongly
dependent on the boundary conditions of the fermions around cycles
of the tori. When all fermions are periodic, the path integral is
localized on supersymmetric configurations (and computes a Witten
Index \WittenDF). These path integrals depend weakly (if at all) on parameters,
and do not undergo phase transitions. In this paper we will focus on
the dynamically more interesting path integrals with anti-periodic
boundary conditions for fermions; these are relevant in particular
for studying finite temperature.

In this section we consider the maximally supersymmetric $SU(N)$ Yang-Mills
theory in one dimension, in the 't Hooft large $N$ limit\foot{It is
important to distinguish this from the M(atrix) theory large $N$
limit \BanksVH,
which is conjectured to describe the light-cone quantization of a
flat-space gravitational theory (M-theory)
and has a different thermodynamic behavior.
}. The Euclidean Lagrangian is given by
\eqn\onedact{
S = {N \over \lambda_1} \int dt \; \tr( {1 \over 2} D_0 \Phi_i D_0
\Phi_i + {i \over 2} \Psi^\dagger D_0 \Psi - {1 \over 4} [\Phi_i,
\Phi_j]^2 + {1 \over 2} \Psi^\dagger \gamma^i [\Phi_i, \Psi]),
}
where $i,j=1,\cdots,9$ and $\Psi$ is a real 16 component spinor. The time
direction is identified with a period $\beta=1/T$; the bosonic fields are
periodic in this direction and the fermions are anti-periodic. The
theory may be characterized by the single dimensionless parameter $\tit = T
/ \lambda_1^{1 /3}$, in terms of which the effective dimensionless
't Hooft coupling is $1/\tit^3$.

As the fermions are anti-periodic about the circle, all fermion modes
acquire large masses as the circle shrinks to zero size, so in the
small radius or large $\tit$ limit, our theory behaves identically to the
bosonic theory considered in section 3; in particular, the
eigenvalues of the holonomy matrix are clumped on the circle.

In the opposite, small $\tit$, limit the quantum mechanics is strongly
coupled; but in this case, we can use a dual description to
understand the dynamics.

We recall that the 't Hooft limit of massless maximally
supersymmetric 0+1 dimensional Yang-Mills theory is believed
\refs{\adscft,\ItzhakiDD}
to be dual to type IIA string theory in the near horizon
limit of the extremal black D0-brane solution of type IIA string
theory. This background has a reliable description in terms of type
IIA supergravity (small curvatures in string units and weak string
coupling) for a range of the radial coordinate corresponding to energy
scales $E \ll \lambda_1^{1/3} = (g_1^2 N)^{1 \over 3}$ in the quantum
mechanics.\foot{For large but finite $N$, the type IIA supergravity
description is also no longer valid at low energies,
when $E \ll g_1^{2/3} N^{1/7}$, but this
scale goes to zero in the 't Hooft large $N$ limit.}

Similarly, the finite temperature quantum mechanics in the 't Hooft
limit should be dual to a near-extremal version of this geometry.
This dual geometry contains a horizon, and type IIA supergravity is
a valid description near the horizon as long as $\tit \ll 1$. The
presence of a horizon in the Lorentzian geometry implies a
contractible thermal circle in the associated Euclidean geometry.
Thus, the expectation value of the (traced) Wilson line
around the circle in the gauge theory, which is mapped in the string
theory (after adding to it some dependence on scalar fields)
to the action of a string worldsheet ending on the thermal
circle \refs{\MaldacenaIM,\ReyIK,\wittent}
is non-zero. This implies that the eigenvalue distribution
is non-uniform.

Therefore, for the massless supersymmetric theory, the eigenvalues of
the Wilson line operator seem to be clumped (or at least non-uniformly
distributed) at both large and small values of $\tit$, so the phase
transition that we found in the bosonic theory appears to be absent
here.

This disparity of behavior between the bosonic and supersymmetric
theories at low temperatures should not come as a surprise; indeed,
their zero temperature dynamics is completely different. In the
purely bosonic theory, the classical flat directions in the
potential are completely removed by quantum fluctuations which
result in a linear potential between the spatial eigenvalues. The
resulting spectrum should be discrete. On the other hand,
for the supersymmetric theory, the quantum fluctuations generate
only the well-known $\dot{x}^4/|x|^7$ potential between eigenvalues,
and the spectrum is continuous.

As for the bosonic theory, we can consider more generally a
massive deformation. The simplest such deformation has equal masses
for all scalars and fermions (of course, this breaks supersymmetry). In
the limit of large mass, all fermion and scalar modes are weakly
coupled, so as for the bosonic theory, we may integrate out the
matter at one loop to reduce the partition function to a unitary
matrix model. The result is similar to \crittemp, but with $p = 17$
(assuming that all fields have the same mass). Thus,
for large enough masses, we again have a phase transition from
clumped eigenvalues at large $\tit$ to a uniform distribution for small
$\tit$, with $\tit_c = m/\ln(17) + {\cal O}(1/m^2)$. Since we have a
phase transition for $\tit \approx m/\ln(17)$ at large $m$ and no phase
transition at $m=0$, it must be that the phase transition line
intersects the $\tit=0$ axis either at $m=0$ or at some small non-zero
value of $m$. There is certainly a qualitative change in the theory
as soon as a mass is turned on, since the zero temperature potential
goes from being flat asymptotically (with a corresponding continuous
spectrum) to asymptotically harmonic (with a discrete spectrum).
While this motivates a possible phase transition, it is not clear
how to directly relate this information to the behavior of our order
parameter.

\newsec{Maximally Supersymmetric Yang-Mills Theory on $T^2$}

In this section, we would like to study the thermodynamics of
maximally supersymmetric 1+1 dimensional Yang-Mills theory on a
circle, corresponding to the Euclidean theory on $T^2$. In this case,
the fermions are anti-periodic about the thermal circle, but we have a
choice between anti-periodic and periodic boundary conditions for the
spatial circle. In fact, we will see that both possibilities are
included if we make the natural generalization to non-rectangular
tori.

\subsec{General tori: classification and fundamental regions}

In this subsection we will classify and describe the inequivalent
tori on which maximally supersymmetric Yang-Mills theory can be
compactified. We specify our torus by two identifications:
\eqn\torusid{z\sim z+L \sim z+L\tau} (where $\tau=\tau_1+i \tau_2$
is the modular parameter), and further specify that all fermions are
anti-periodic around both of these cycles. Below we will denote the
holonomy along the cycle $z\sim z+L$ as $V$ and the holonomy along
$z\sim z+L\tau$ as $U$. The phase diagram we wish to determine is a
function of three real dimensionless variables: the complex variable
$\tau$ and the dimensionless coupling constant $\tilde{\l}=g_{YM}^2
N A$, where $A=L^2 \tau_2$ is the area of the torus.

\fig{Fundamental region of the modular group for a theory with
fermions.}{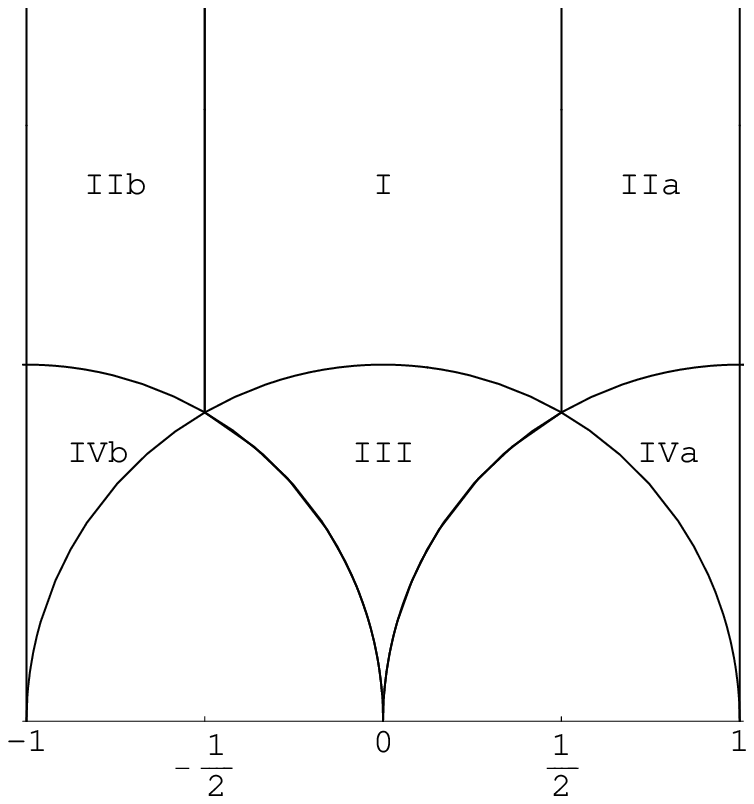}{4truein}
\figlabel{\moddomains}

Bosonic Yang-Mills theories on tori whose $\tau$ parameters are
related by an $SL(2, \IZ)$ transformation ($\tau \to (a\tau+b)/(c\tau+d)$
with integers $a,b,c,d$ satisfying $ad-bc=1$) are identical at equal
values of ${\tilde \lambda}$. It is thus sufficient to study such
theories in the familiar $SL(2,\IZ)$ fundamental domain, denoted by
the region I in figure \moddomains.

The supersymmetric Yang-Mills theory we will study in this section
includes fermions; the partition function for this theory will be
identical (at equal values of ${\tilde \lambda}$) only on those
tori that are related by the subgroup of $SL(2,\IZ)$ that preserves
our fermion boundary conditions. This group is generated by $\tau
\rightarrow -1/\tau$ and  $\tau \rightarrow \tau+2$. The
fundamental domain for this subgroup is the union of the regions I, II
and IV in figure \moddomains. In this section we will study the
Yang-Mills partition function on tori with modular parameters lying in this
fundamental domain.

To end this subsection we comment on the physical interpretation of
tori in regions II and IV. Tori on the two vertical lines that border
the diagram in figure \moddomains\ have modular parameters of the form
$\tau=\pm1+i\tau_2$ (with real $\tau_2$). These tori are best
thought of as rectangular tori with $\tau'= i\tau_2$, with periodic
boundary conditions on fermions along the $\tau'$ cycle (the fermion
boundary conditions remain anti-periodic along the `1' cycle). In
general, a torus in the region $IIb/a$  with modular parameter
$\tau$ may be thought of as a torus in the region I, with
$\tau'=\tau\pm 1$ and periodic boundary conditions along the $\tau'$
cycle. Similarly, a torus in the region $IVb/a$ may be reinterpreted
as a torus in the region I with $\tau'=-1 / (\tau\pm 1)$ and
periodic boundary conditions along the `1' cycle.

\subsec{Analysis at small ${\tilde \lambda}$}

As discussed in previous sections, at small ${\tilde \lambda}$ the
Kaluza Klein modes of the field theory on $T^2$ may reliably be
integrated out at one loop, provided that ${\tilde \lambda /
M^4}$ is small, where $M$ is the mass of the lightest KK mode in
units in which the area of the torus is one.
When this integrating out is legitimate, the resulting determinant
(see Appendix A) ensures that the eigenvalues for both holonomies $U$
and $V$ clump.

Expanding all fields (take a scalar field $\phi(z)$ as an example)
on the torus in Fourier modes subject to the periodicity
$\phi(z)=\phi(z+L) =\phi(z+L\tau)$ gives
\eqn\expansiontilted{\phi(z)=\sum_{m,n=-\infty}^{\infty}
\phi_{m,n}\exp\left[{2
\pi i \over {L \tau_2}}\left( \tau_2 m {\rm Re}(z) +(n - \tau_1
m){\rm Im}(z)\right)\right],}
leading to $M^2={4 \pi^2 \over \tau_2}
\left(m^2 |\tau|^2 +n^2-2m n \tau_1\right)$ (in units of $1/L$).
For bosonic fields $m$ and
$n$ are integers, while for fermions $m+1/2$ and $n+1/2$ are integers.

In regions I and II the lightest non-zero-mode for scalars
is the $(m=0,n=\pm 1)$ mode with
$M^2={4 \pi^2 / \tau_2}$.  So, we  can integrate out all the KK modes  at
one loop reliably in regions I and II if
\eqn\intout{\tilde{\l} \tau_2^2 \ll 1.}
For larger values of $\tau_2$ all modes with $m\neq 0$ (the boundary
conditions ensure that this includes all the fermions) are still very
heavy and decouple, so we can study the theory by including only the
$m=0$ modes. The resulting quantum mechanical theory is simply the
bosonic quantum mechanics of section 4, with $R= L \tau_2$ and coupling
$\l_1=\l/L$,
giving an effective dimensionless coupling $\tilde{\l}\tau_2^2$.
It follows from
the analysis of section 4 that the system undergoes a phase transition
at $\tau_2 =a/ \sqrt{\tilde{\l}}$, where $a$ is a number of
order unity, determined in the analysis of section 4. At larger
values of $\tau_2$ one of the $\IZ_N$ symmetries of the system is
restored, and $\vev{\tr(U)} = 0$.

In region  IVa the lightest mode is $(m=n=\pm 1)$, with mass
\eqn\masskkm{M^2={4 \pi ^2 \over \tau_2} \left(
(1- \tau_1)^2+\tau_2^2\right)}
(the IVb region is related to this by $(\tau_1\to -\tau_1, m\to -m)$).
The KK  modes are  weakly coupled and
can be integrated  out provided that \eqn\intoutag{{\tilde{\l}
\tau_2^2 \over ((1-\tau_1)^2 + \tau_2^2)^2}\ll 1.}
When \intoutag\
is not obeyed (but still ${\tilde \lambda} \ll 1$),
all modes with $n \neq m $ become very heavy and
decouple. The resulting effective quantum mechanical theory of the light
modes includes
fermions, and is, in fact, precisely the one dimensional
supersymmetric quantum mechanics of section 6. We
have argued that it is unlikely that this system undergoes a phase transition.
It seems that our system is in the $\vev{\tr(U)} \neq 0,\vev{\tr(
V)} \neq 0$ phase throughout region IV.

\fig{The phase transition line at $\tau_2 \sim {1/
\sqrt{\tilde{\l}}}$ and its images, for a specific value of $\tilde{\l}$.}
{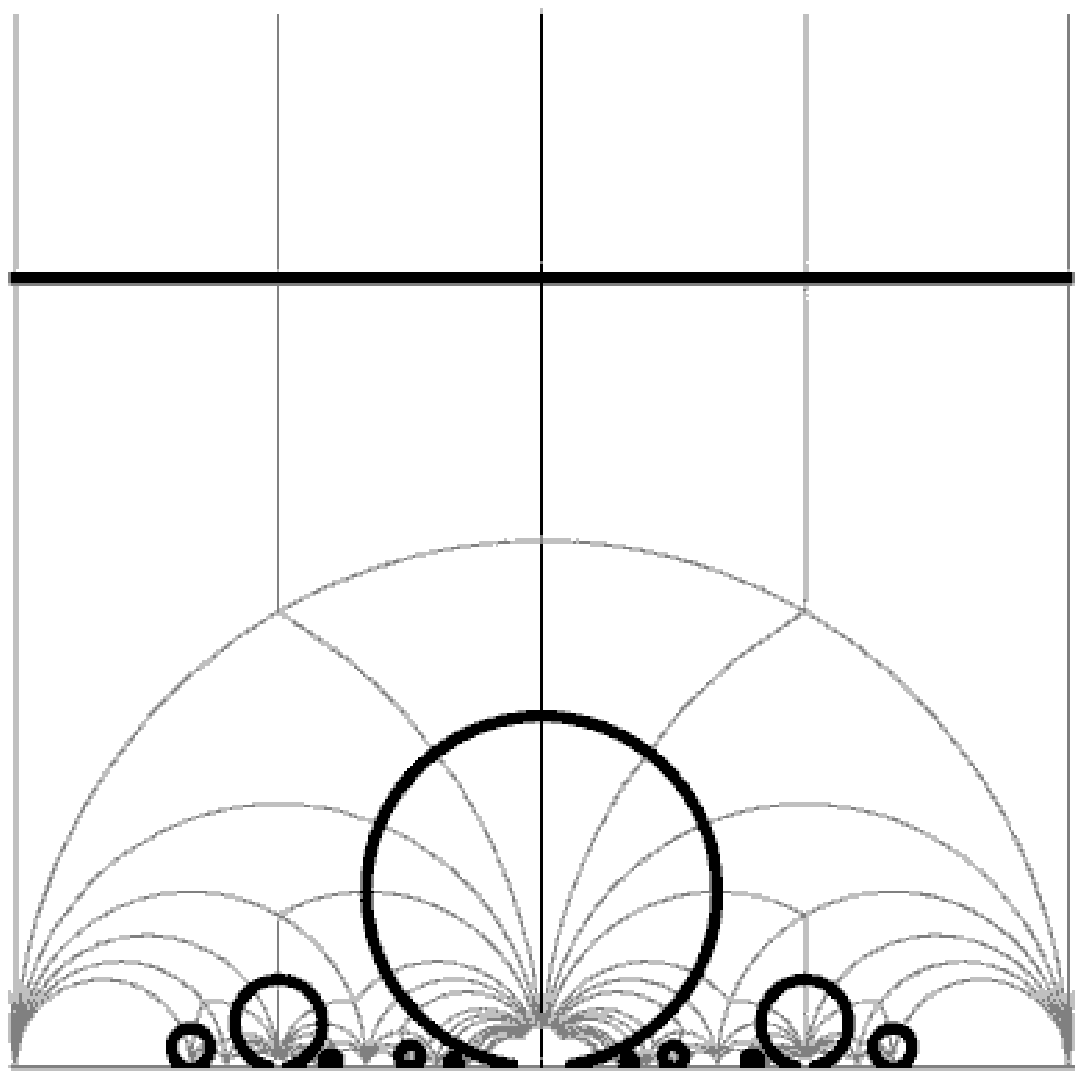}{4truein}
\figlabel{\modphase}

Mapping regions  I  and  II by  the (allowed)
modular transformations  we find  an infinite  number of  images of
the phase transition line  $\tau_2=a/\sqrt{\tilde{\l}}$, which are
drawn in figure \modphase. For
example, in region III this line is mapped by the $\tau  \r -1/\tau$
transformation to the line
$(\tau_2'-\sqrt{\tilde \lambda}/2a)^2 + \tau_1'^2 = {\tilde \lambda} / 4a^2$.
Above this line
both  holonomies are clumped, $\vev{\tr(U)}  \neq  0$ and
$\vev{\tr(V)}\neq  0$,
while below
the line (in region III) $\vev{\tr(V)}  =  0$.

To  summarize, a  torus  is  characterized  by  its  modular
parameter $\tau$. If  we  scale  the  't Hooft  coupling constant
appropriately  then  the partition function  of the theory  is a
modular invariant  function on the  $\tau$ plane. Thus, we  only need
to study  it in  the fundamental region  (which is  different  from
the standard  one  because of  the fermions). In this region we find
a single phase transition line where the eigenvalues of the Wilson
loop along the longer fundamental cycle clump. This  line has  an
infinite number  of  images on  the $\tau$ plane  through modular
transformations.
Above we gave the results for the case of two anti-periodic cycles for
the fermions, but the case of one periodic cycle (corresponding, in
particular, to the finite temperature theory) is simply related to
this by $\tau\to \tau+1$.

\subsec{Strong coupling from AdS/CFT}

When the effective coupling $\tilde \lambda$ is large,
the  theory cannot  be analyzed  using
perturbative techniques, but  we can use  the AdS/CFT correspondence
to  study the phase structure of the system.

The dual  of the $1+1$ dimensional
maximally supersymmetric Yang-Mills theory in the 't Hooft large
$N$ limit is the  near-horizon geometry of the D1-brane  metric of
type IIB string theory \ItzhakiDD. If we
put the gauge  theory on a  spatial circle  of circumference  $L$ then
the dual theory is  obtained by a periodic identification of the spatial
coordinate. If we
want  to study the  theory at temperature  $T$ we  have to  consider
the  near extremal D1-brane solution. The relevant Euclidean string
frame metric
and dilaton are:
\eqn\donemetric{ds^2=\apm \left[{u^3 \over \sqrt{d_1 \lp}}
\left(1-{u_0^6\over u^6}\right){dt'^2\over L^2} +{u^3 \over
\sqrt{d_1 \lp}} {d\theta'^2\over (2\pi)^2}+{\sqrt{d_1 \lp}\over
u^3\left(1-{u_0^6\over u^6} \right)}du^2 +u^{-1}\sqrt{d_1 \lp}
d\Omega_7^2\right],}
\eqn\dilatonsol{e^\phi = 2 \pi {\l ' \over N} \sqrt{d_1 \l ' \over
u^6},}
where $\l' = \l L^2,d_1=2^6\pi^3,u_0^2={16 \pi ^{5/2}
\over 3} TL \sqrt{\l '}$, we have to identify $t' \sim t'+ 1/
T$ to avoid the conical singularity at $u=u_0$, and the periodicity
of $\theta'$ is $2\pi$. There is also a 3-form field which we do not
explicitly write down. Defining $\rho = TL$ and
changing coordinates to $t =t' T$ and
$\theta={\rho \theta' \over 2 \pi}$, we can write the metric in the
form:
\eqn\metrictwo{ds^2=\apm \left[{u^3 \over \sqrt{d_1 \lp}\rho^2} \left[
\left(1-{u_0^6\over u^6}\right)dt^2 +d\theta^2\right]+{\sqrt{d_1
\lp}\over u^3\left(1-{u_0^6\over u^6} \right)}du^2 +u^{-1}\sqrt{d_1
\lp} d\Omega_7^2\right],}
with $u_0^2 = {16 \pi^{5/2}\over 3} \rho \sqrt{\l'}$.
Now, defining a complex coordinate $z=t +i \theta$, we have the
identifications $z\sim z+1\sim z+i \rho$, and the conformal boundary of
this geometry is a rectangular  $T^2$. The parameter $\rho$ appears both
in the identifications and explicitly in \metrictwo, but in fact we can
get rid of it in the metric by noting that the form of \metrictwo\ is
invariant under a rescaling $u \to \alpha u, u_0 \to \alpha u,
\lambda' \to \alpha^2 \lambda', \rho \to \alpha \rho$. If we use this
rescaling with $\alpha = \rho^{-1}$ we can rewrite \metrictwo\ as
\eqn\metricthree{ds^2=\apm \left[{u^3 \over \sqrt{d_1 \tl}} \left[
\left(1-{u_0^6\over u^6}\right)dt^2 +d\theta^2\right]+{\sqrt{d_1
\tl}\over u^3\left(1-{u_0^6\over u^6} \right)}du^2 +u^{-1}\sqrt{d_1
\tl} d\Omega_7^2\right],}
with $\tl = \lambda' / \rho^2$ and $u_0^2 = {16 \pi^{5/2}\over 3} \sqrt{\tl}$.
Since now $\rho$ appears only in the identifications, it is clear that
if  we  change  the
identifications  to $z\sim   z+1   \sim   z+\tau$ with
$\tau=\tau_1+i \tau_2$ an arbitrary complex number,  we   still
have   a   solution   to IIB supergravity (with the same $u_0$)
without any
singularities, whose asymptotic boundary is a torus of modular
parameter $\tau$.

If we want to compute the thermal partition function of the field theory
living on the boundary we have to consider all supergravity (or more
generally string theory) solutions $X_i$ with the appropriate
behavior at infinity. The partition function of the dual gauge
theory in the large $N$ limit is:
\eqn\zgauge{Z_{gauge}=\sum_i \exp(-I(X_i)),}
where $I$ is the Euclidean action of the supergravity (string theory)
solution.

Let's say that  we want to compute the partition function  of supersymmetric
Yang-Mills (SYM) theory on
a torus of modular parameter $\tau_f$. It is clear that solutions
of the form  \metricthree\ contribute  to  the partition  function
only if  the asymptotic  $T^2$  of the  supergravity  solution  can
be  conformally mapped to the  $T^2$ on which the gauge  theory
lives. This is possible if  and only  if the  modular parameter  of
the  torus of  the gravity $\tau_g$ is  related to $\tau_f$ by some
modular transformation. So we have  to start  from  the supergravity
solution with  $\tau_g=\tau_f$, consider  all modular
transformations  of $\tau_g$,  and then  sum the exponentials of the
Euclidean action of the corresponding supergravity solutions
\refs{\MaldacenaBW,\DijkgraafFQ}.

In general all solutions will contribute to the partition function.
But we notice that the Euclidean action scales as $N^2$, so in the
large $N$ limit  we get  the dominant  contribution from  the solution
with the smallest Euclidean action. This allows for
sharp phase transitions  in the  $N\r  \infty$  limit, if  the
solution with  the minimum action  changes discontinuously as  we
vary the  parameters of the theory.

If we try  to compute the Euclidean action of any  of the solutions
we find that  it diverges as we  integrate over all  space. However
we are not interested in the value of  the action but in comparing
the action of different solutions. One way to compare the  actions
is to match the solutions  (metric and other fields) at  some large
cutoff value of the radius, calculate the  action difference and then send
the value of the cutoff to infinity. The  divergent terms cancel
once the geometries are properly matched, and one obtains a finite
result. Alternatively one can add counter-terms to cancel the
divergences, leading to equivalent results.

Before we proceed we must clarify two subtle points:

\item{(a)}Not all solutions with different $\tau_g$ are different.
For example, it is easy  to see that $\tau_g$ and
$\tau_g\pm 1$ describe  the same supergravity  solution. The
asymptotic  torus of the  supergravity solutions  has a  cycle that
is  special, namely the $(1,0)$ cycle (labeled by $t$) which is
contractible in the interior. It is fairly  easy   to  see  that
all  inequivalent  geometries   can  be characterized by specifying
the cycle  $(p,q)$ of the gauge theory torus onto which the $(1,0)$
cycle of \metricthree\ is  mapped when  we conformally match the two
tori. The two integers $p$ and $q$ have to be relatively prime.

\item{(b)}We have to be careful about the boundary conditions of the fermions
and the spin structures of the supergravity solutions. If there is a
circle factor in the asymptotic geometry then we can choose either
periodic or anti-periodic boundary conditions for the fermions along
this circle if it is not contractible in the interior. However, if it
is contractible then only anti-periodic boundary conditions are
allowed. In our case the $(1,0)$ cycle of the gravity torus is
contractible, so it has to be mapped to an anti-periodic cycle of
the gauge torus. Since in our gauge theory analysis above we chose
anti-periodic boundary conditions for the fermions on the $(1,0)$ and
$(0,1)$ cycles of the gauge
theory, we conclude that the acceptable geometries of
(a) are those with $p+q$ odd.

Let us now analyze which solutions we have to include for a specific
gauge theory on a torus with parameter $\tau$ whose partition function
includes a contribution from \metricthree. Up to a change of coordinates
all these solutions should have the same asymptotics as \metricthree,
and they should involve identifications by some modular parameter $\tau'$
which is related by an $SL(2,\IZ)$ transformation to $\tau$.
It is easy to verify that for this to be the case the parameters of the
two metrics have to be related by $\tl' \tau_2' = \tl \tau_2$, and the
$u$ coordinates of the two metrics are related by $u' \sqrt{\tau_2'} =
u \sqrt{\tau_2}$. Note that the relation between the $\tl$'s is precisely
the one we expect from the field theory point of view, for the two theories
to have the coupling constant times area.

A straightforward computation of the (regularized) Euclidean action
of the IIB solution \metricthree\ gives:
\eqn\euclactiona{I={9 N^2 V_7 \over {64 \tl^2 \pi ^9} } \tau_2
\int_{u_0}^R du u^5,}
where $R$ is some very large radial position and
$V_7$ is the volume of the unit 7-sphere.
Using the results of the previous paragraph
we   see   that   if   we   compute $I(X_1)-I(X_2)$ for two
solutions  with $\tau$ and $\tau'$ related by a modular
transformation, the divergent parts cancel and we find:
\eqn\diffaction{I(X_1)-I(X_2)={32 N^2 V_7 \over 9 \sqrt{\tl \tau_2} \pi
^{3/2} } \left(-\tau_2^{3/2}+\tau_2'^{3/2}\right),}
where we used  $u_0^2={16 \pi ^{5/2} \over 3} \sqrt{\tl}$.
So, using  the  invariance  of  ${\tl \tau_2}$,
we conclude that  the solution with the  minimum
action is  the one which has the maximum value of $\tau_2$
(and is consistent  with $p+q=odd$).

Let us now see how  this works  in different  regions of  the
$\tau_f$ plane.   In  regions   I  and  II  the  geometry   that
dominates  is $\tau_g=\tau_f$.  It maps the $(1,0)$  cycle of the
gravity torus to the $(1,0)$ cycle of the gauge torus. Any other
solution with $\tau_g$ related by a modular transformation
has a smaller $\tau_2$  so it
has a bigger action. Since the $(1,0)$ cycle is the  only contractible
cycle in the gravity solution, the Wilson  loop around it will generically be
nonzero $\vev{\tr(U)}\neq 0$,
while  all Wilson loops  around any other cycle will be zero.
%
%
Though it takes a little more work to see it, region IV is also
dominated by the same geometry\foot{In region IVa there is the
geometry with the largest $\tau_2$ is the $\tau_g=-{1 \over
\tau_f-1}$.However this would map the contractible $(1,0)$ cycle to
$(1,\pm 1)$ which is a periodic cycle, so this is not an acceptable
solution.}.

So, in summary, a single saddle point \metrictwo\
dominates the thermodynamics of strongly coupled Yang-Mills theory on a
torus in the
fundamental region (the union of $I, II , IV$). The $(1,0)$ cycle is
contractible on this solution, so $\vev{\tr(V)}=0$ and $\vev{\tr(U)}
\neq 0$ in this phase. For other regions we can find the dominant solution
by mapping them into the fundamental region.

\subsec{Putting it together}

Let us summarize our understanding of the phase structure of this
system. We begin by discussing tori with $\tau=it$ where $t$ is
real. At weak coupling the system is in the $\vev{\tr(U)} \neq 0$,
$\vev{\tr(V)}=0$ phase at large $t$. At $t={a / \sqrt{{\tilde \lambda}}}$
the system undergoes a phase transition; at smaller values of $t$
the system is in the $\vev{\tr(U)}\neq 0$, $\vev{\tr(V)}\neq 0$ phase. At
$t=\sqrt{{\tilde \lambda}} / a$ the system undergoes another
phase transition, to the $\vev{\tr(U)}=0$, $\vev{\tr(V)}\neq 0$ phase.
On the other
hand, at strong coupling the analysis of the previous subsection shows that
the system undergoes exactly one phase
transition at $t=1$. When $t>1$, $\vev{\tr(U)}\neq0$ and $\vev{\tr(V)}=0$.
On the
other hand, when $t<1$ $\vev{\tr(V)} \neq 0$ and $\vev{\tr(U)}=0$.
\fig{Conjectured phase diagram for SYM on a
rectangular torus with anti-periodic boundary
conditions on both cycles.}{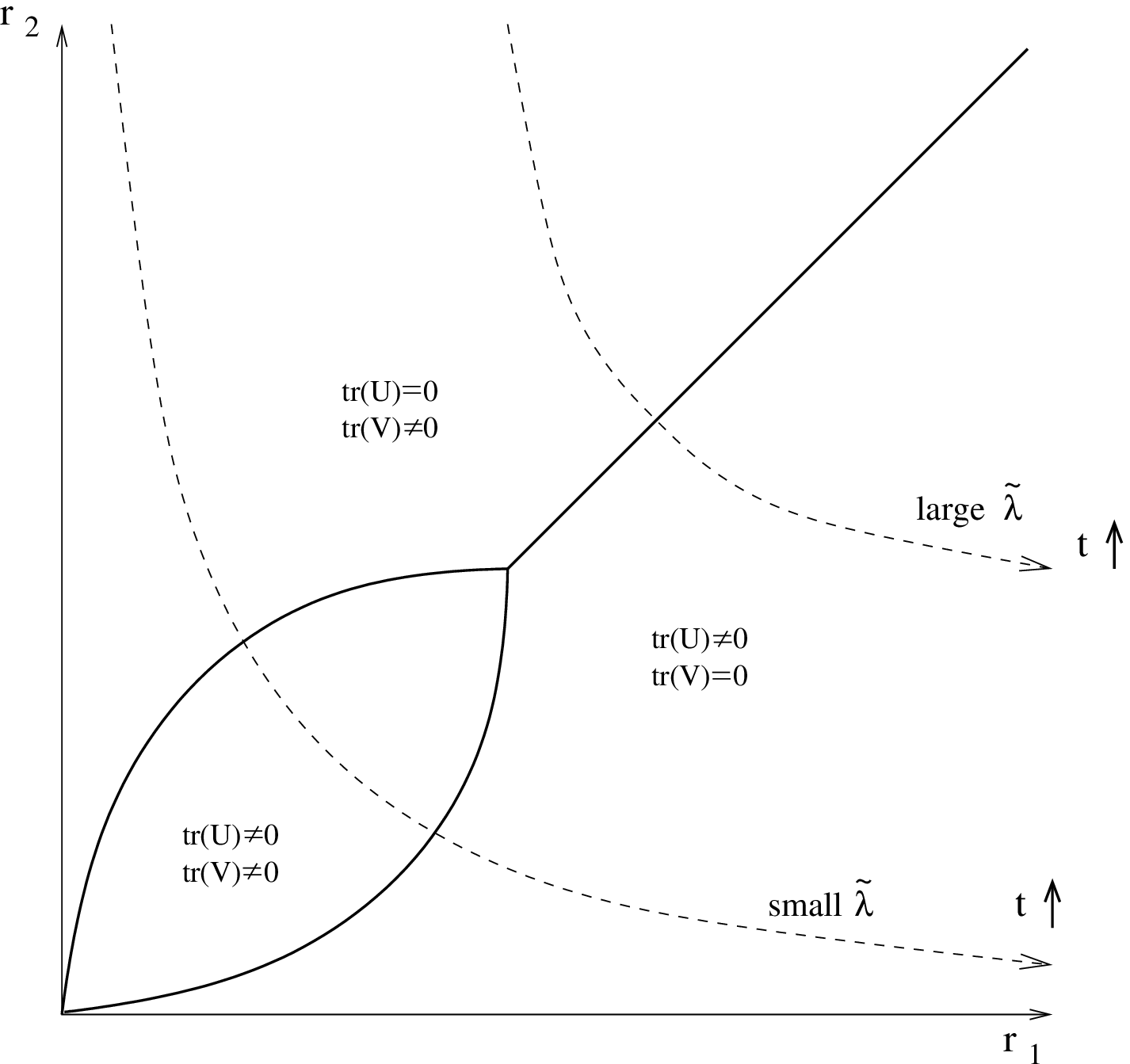}{4truein}
\figlabel{\rectorus}
The phase diagram in figure \rectorus\ summarizes this behavior,
and provides the simplest possible interpolation between these two limits.

\fig{Phase diagram for SYM on a torus with one periodic and one anti-periodic
boundary condition for the fermions. The large $\tilde\lambda$
transition here is seen in the dual IIA description as a
Gregory-Laflamme transition \glpaper.}{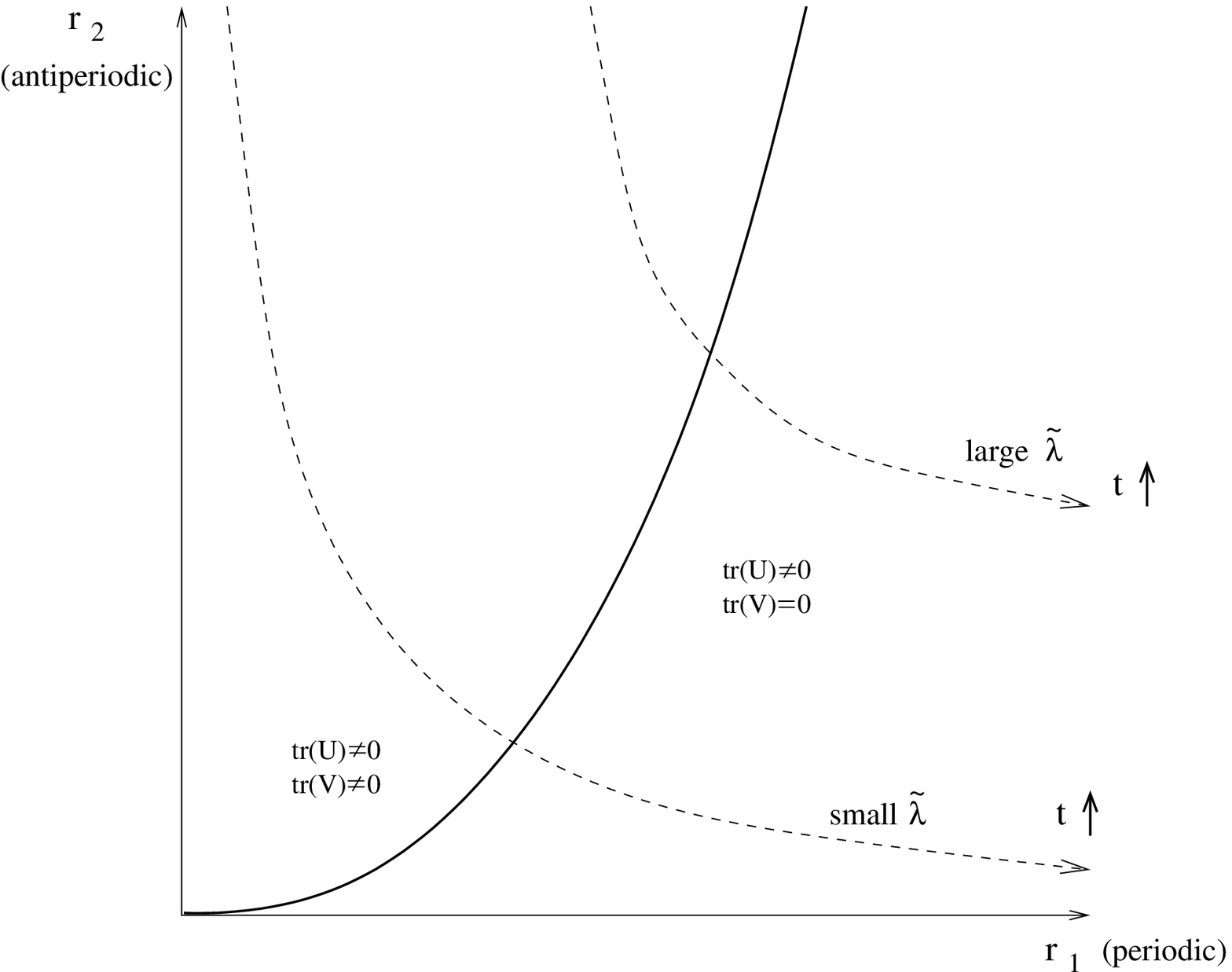}{4truein}
\figlabel{\perantiper}
Next, recall that a torus with ${\rm Re}(\tau)=1$ may equally well be
regarded as a rectangular torus with $\tau'=\tau-1$ but with periodic
boundary conditions along the $\tau$ axis. Following previous
discussion in \glpaper\ we expect a phase transition associated only
with the $V$ holonomy, seen as a Gregory-Laflamme transition in the
type IIA string description which is obtained by performing a T-duality on the
periodic torus direction. Hence, we expect the phase diagram
of figure \perantiper\ (or equivalently figure 1 in \glpaper). From the 
analysis in \glpaper\ we expect the phase transition to occur at $t
\sim 1/\tilde\lambda$ for large $\tilde\lambda$. Note that this phase
transition involves a new solution (localized in the T-dual direction)
which is not included in the discussion of the previous subsection.

\fig{Conjectured phase boundaries for the $V$ holonomy shown in the 
fundamental domain for general values of $\lambda$. The phase
transition associated with the blue surface involves only the $V$
holonomy, whilst the red surface corresponds also to a transition of
the $U$ holonomy.  }{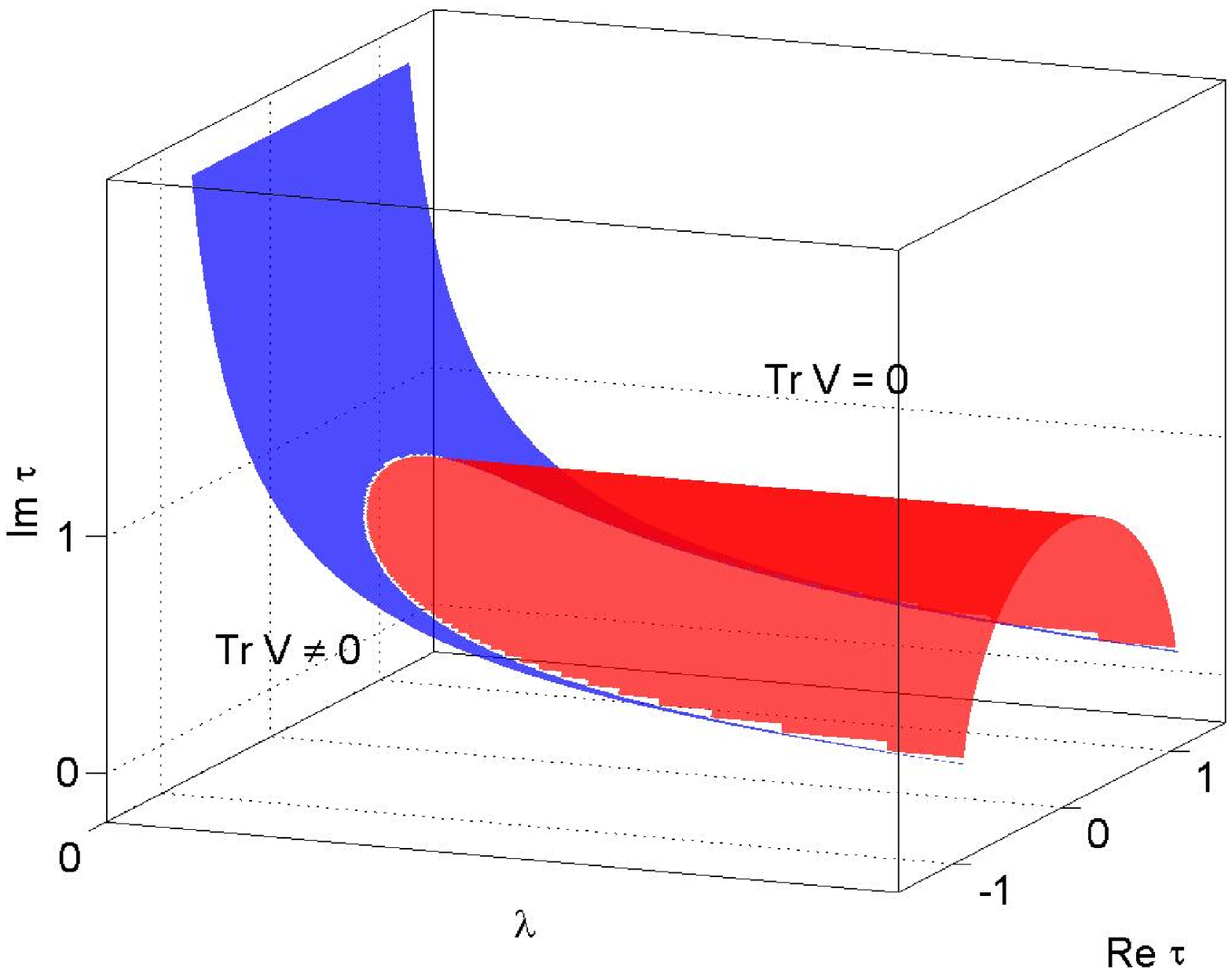}{6truein}
\figlabel{\gentau}
Using this information we conjecture the phase structure for a general
value of $\tau$. In figure \gentau\ we show the two phase boundaries
we expect for clumping of the $V$ holonomies in the fundamental domain
as a function of $\tau$ and $\tilde
\lambda$. For large $\tilde{\lambda}$,
the relevant boundary (shown in red) coincides with the boundary of
the fundamental domain and corresponds also to anti-clumping of the
$U$ holonomy. For small $\tilde{\lambda}$, the relevant boundary
(shown in blue) involves only the transition of the $V$ holonomy. As
discussed earlier, modular transformations of this latter surface give
other phase surfaces associated with transitions of different
holonomies; for example, the clumping of the $U$ holonomy is given by $\tau
\rightarrow -1/\tau$. The ${\rm Re}(\tau)=0$ slice reproduces the
previous figure
\rectorus\ when converting to $r_1, r_2$ coordinates and including this 
$U$ transition. Likewise, the  ${\rm Re}(\tau)=1$ slice
reproduces figure \perantiper. Note that we expect that the two phase
transition 
surfaces (red and blue) do not join smoothly for any ${\rm Re}(\tau)$,
and their intersection moves to larger $\tilde \lambda$ as we move
nearer ${\rm Re}(\tau) = 1$, eventually leaving only the two phases with
clumped $U$ holonomy that are seen in \glpaper.

By comparing figures \rectorus\ and \perantiper\ with the phase
diagrams \smMgen\ and \twodmm\ of section 5, it is clear that for
either choice of spatial boundary conditions, the supersymmetric
theory has qualitatively different behavior from the bosonic
theory. Curiously, the bosonic theory and the supersymmetric theory
with anti-periodic-anti-periodic, anti-periodic-periodic, and
periodic-periodic boundary conditions, have four, three, two, and one
distinct phases, respectively.

\centerline{\bf Acknowledgements}

We would like to thank Rajesh Gopakumar,  Dan Jafferis, Subhaneil
Lahiri, Andy Neitzke, Gordon Semenoff, Xi Yin for useful
discussions. JM would like to thank IAS and KITP for hospitality
during the course of this work. OA, SM, and KP would like to thank
KITP for hospitality while this work was in progress. The work of OA
was supported in part by the Israel-U.S. Binational Science
Foundation, by the Israel Science Foundation (grant number 1399/04),
by the Braun-Roger-Siegl foundation, by the European network
HPRN-CT-2000-00122, by a grant from the G.I.F., the German-Israeli
Foundation for Scientific Research and Development, and by Minerva.
The work of JM was supported in part by an NSF Graduate Research
Fellowship.  The work of SM was supported in part by DOE grant
DE-FG01-91ER40654, NSF career grant PHY-0239626, a Sloan fellowship,
and a Harvard Junior Fellowship.  The work of KP was supported in
part by DOE grant DE-FG01-91ER40654. The work of MVR has been
supported in part by the Natural Sciences and Engineering Research
Council of Canada and by the Canada Research Chairs Programme. TW is supported by NSF grant PHY-0244821.

\appendix{A}{One-Loop Effective Potentials for Zero Modes}

In this appendix, we shall derive the one-loop effective action for
the zero modes of $SU(N)$ gauge theories with adjoint scalar fields
in various dimensions, by integrating out all massive modes. We begin
in \S A.1 with an analysis of the effective action for the diagonal
modes in the $0$-dimensional matrix integral and its implications
for the matrix integral discussed in \S3. The same analysis applies
also to the KK zero modes of higher dimensional theories; in \S A.2
we add the contribution of the higher KK modes as well, and in \S A.3 we
consider the one-loop contribution for supersymmetric Yang-Mills
theories.

\subsec{The zero mode integral}

In this subsection we investigate the effective action for the matrix
integral with $p$ massless scalar fields discussed in section 3,
\eqn\intza{Z = \int{\cal{D}}\psi^{\alpha}\exp\left[-\frac{N}{4\lambda_0}
\sum_{\alpha,\beta}\tr(\left[\psi^{\alpha},\psi^{\beta}\right]^2)\right],}
where $\alpha,\beta=1,\cdots,p$. The same integral (except that some of
the $\psi^{\alpha}$'s become periodic) arises for the zero
modes of a $d$-dimensional gauge theory on $T^d$ with $p_d$ scalar fields and
coupling constant $\lambda_d$; in this context $p=p_d+d$ since additional
scalar fields arise from the gauge field, and $\lambda_0$ is $\lambda_d$
divided by the volume of the $T^d$.

A naive extrapolation of the formula \trphsq, giving the width of the
eigenvalue distribution for large masses, to small masses
would suggest that the eigenvalue
distribution is infinitely spread out at $m=0$. This estimate is too
crude since it  ignores the effect of the quartic restoring term in
\intza. Taking this term into account, a simple dimensional analysis suggests
that the characteristic width scale for the eigenvalue distribution
at the saddle point for $m=0$ is given by
$a = K
\lambda_0^{{1\over 4}}$, where $K$ is independent of $N$ and $\lambda_0$.
The $N$ dependence of this estimate follows from 't Hooft scaling,
while the $\lambda_0$-dependence is deduced from a simple change of
variables in \intza. However, this scaling argument cannot rule out the
possibility that $K=\infty$, a danger that seems real given the fact
that the quartic restoring term in \intza\ vanishes along a
noncompact `moduli space' along which all the matrices commute (and so
may simultaneously be diagonalized). Consequently, the question of
whether or not the eigenvalues in \intza\ clump at $m=0$ is a
dynamical issue \foot{See \Austing\ for a
rigorous proof of the convergence of this integral, and
\Austingthesis\ and references therein for a nice review and more
details.}. One way to analyze this is by an analysis of the quantum effective
action on the classical moduli space \refs{\Aoki, \Hotta,
\Staudacher} that is obtained by integrating out the off diagonal
modes. The generic effective mass of an off-diagonal mode is of order $a$,
the width of the eigenvalue distribution. Consequently, as in the
analysis of the massive theory in section 3,
the effective coupling for the off diagonal modes is $\sim {\lambda
/ a^4}$; these modes may accurately be integrated out at one-loop
provided that this coupling is small. This one-loop computation was
performed in \Hotta\ and is reproduced, for the convenience of the
reader, in the following paragraphs.

Let us expand around a configuration in which all matrices
$\psi^\alpha$ are diagonal with eigenvalues $\psi^\alpha_i$ ($i=1,\cdots,N$),
and integrate out the off-diagonal
components of the matrices in \intza\ at 1-loop.  Expanding the action
\intza\ to quadratic order in the off-diagonal modes, we find
\eqn\intzexp{S_{zeromode}^{quadratic}=
\frac{N}{4\lambda_0} \sum_{\alpha \neq \beta}
\sum_{i<
j}\left[\left(\Delta\psi^{\alpha}\right)_{ij}^2
\psi^{\beta}_{ij} \psi^{\beta}_{ji}
-\left(\Delta\psi^{\alpha}\right)_{ij}
\left(\Delta\psi^{\beta}\right)_{ij}\psi^{\alpha}_{ij}\psi^{\beta}_{ji}
\right],}
where we defined $(\Delta\psi^{\alpha})_{ij} = \psi^\alpha_i-\psi^\alpha_j$.
Naively, integrating out the $(ij)$'th off-diagonal components of all the
matrices simply yields
$\det(M)^{-1}$, where $M$ is the matrix
\eqn\matdet{M^{\alpha\beta}=\delta^{\alpha \beta} \sum_{\gamma}
(\Delta\psi^{\gamma})_{ij}^2-
(\Delta\psi^{\alpha})_{ij}(\Delta\psi^{\beta})_{ij}.}
However, the eigenvalues of the matrix $M$ are easily seen to be
$[\sum_{\gamma} (\Delta\psi^{\gamma})_{ij}^2]$ with degeneracy $p-1$ and
zero with degeneracy one.  The appearance of the zero eigenvalue is a
result of the fact that we have neglected the existence of flat
directions in the space of off-diagonal components, due to the remaining
$SU(N)$ gauge symmetry of \intza, which can be used to rotate the diagonal
components into off-diagonal components.

To correct for this, we eliminate these flat directions by using the
$SU(N)$ symmetry to diagonalize one of the $\psi^{\alpha}$, say
$\psi^1$, exactly.  Performing this
diagonalization by inserting a gauge-fixing $\delta$-function
introduces the Fadeev-Popov determinant factor
\eqn\fpzer{\det\left(\left[\psi^1,\ast\right]\right)\sim
\prod_{i\ne j}\left(\Delta\psi^1\right)_{ij}\sim\prod_{i<j}
\left(\Delta\psi^1\right)_{ij}^2,}
where we have evaluated the determinant at 1-loop.

Note that this factor is nothing more than the Vandermonde determinant
appearing in the change of variables from the matrix $\psi^1$ to its
eigenvalues.

Now there are no longer any off-diagonal components for $\psi^1$ in \intzexp.
This has the
effect of removing the first row and column from the matrix $M$ in
\matdet.  The eigenvalues of the resulting matrix are
$[\sum_{\gamma} (\Delta\psi^{\gamma})_{ij}^2]$ with degeneracy $p-2$ and
$\left(\Delta\psi^1\right)_{ij}^2$ with degeneracy one.  Having
computed the relevant determinant, we may now write the contribution
to the path integral from the gauge-fixing and from
integrating out the off-diagonal components of the zero modes as
\eqn\Zoffdiag{\frac{\prod_{i<
j}\left(\Delta\psi^1\right)^2_{ij}}{\left[\prod_{i<
j}\left(\Delta\psi^1\right)^2_{ij}\right]\left\{\prod_{i<
j}\left[\sum_{\gamma}
\left(\Delta\psi^{\gamma}\right)_{ij}^2\right]^{p-2}\right\}}.}
The Fadeev-Popov factor in the numerator cancels the contribution
from the eigenvalues depending only on $\psi^1$, leaving us with an
expression which (as expected)
is symmetric under permutations of the $\psi^{\alpha}$'s,
\eqn\Zoffdiagg{\prod_{i<
j}\left[\sum_{\gamma} \left(\Delta\psi^{\gamma}\right)_{ij}^2\right]^{2-p}.}

Thus, we find that the one-loop contribution to the matrix integral
may be expressed as an integral
over the eigenvalues of the matrices,
\eqn\intol{Z \propto \int
\prod_{\alpha, i} d \psi^\alpha_i { 1\over \prod_{i<j}
\left(\sum_\gamma (\psi_i^\gamma - \psi_j^\gamma )^2 \right)^{(p-2)}
} ~~~~{\rm valid\ when\ } a^4 \gg \lambda_0. }
In the large $N$ limit it is natural to rewrite this in terms of an
eigenvalue distribution function $\rho(\psi^\alpha)$,
and to evaluate the integral by saddle points (since the action is
proportional to $N^2$).

As all pairwise eigenvalue forces are attractive, the
minimum of the action is attained when all eigenvalues sit at a
point. Unfortunately this saddle point lies outside the domain of
validity of \intol, since it has $a=0$. Thus, the one-loop analysis does
not suffice to analyze the matrix integral, though it does reliably
show that the eigenvalues attract each other at large distances. A
more precise analysis shows that on the saddle point of
\intza\ the eigenvalues have a sharply localized distribution with
the scale $a \sim \lambda_0^{{1\over 4}}$, as suggested by the scaling
arguments presented earlier in this section. This is verified by the
Monte Carlo simulations shown in section 3, and was originally shown
by similar methods in \Staudacher.

In summary, the behavior of \intza\ is qualitatively unaffected as
the effective coupling constant ${\lambda_0 / m^4}$ is varied
from zero to infinity. The eigenvalue distribution of the matrix
${1\over N}(\sum_{\alpha}(\psi^{\alpha})^2)$
in \intza\ is governed by a saddle point that is always
strictly localized; the localization length is given by $a^2= {p
\lambda \over 4 m^2} f({\lambda \over m^4})$, where $f(0)=1$ and
$f(x) \propto {1 / \sqrt{x}}$ at large $x$.

\subsec{Integrating out KK modes at 1-loop}

We now consider a $d$-dimensional gauge theory compactified on a torus $T^d$
with circles of circumference $R_{\mu}$ ($\mu=1,\cdots,d$), and compute
the effect of integrating out the KK modes at 1-loop on the effective
action for the diagonal zero modes which we computed in the previous
subsection. Consider the $d$-dimensional bosonic action\foot{If
there are also fermionic fields, they decouple from the bosons
at 1-loop so we can consider them separately as we will do below.}
\eqn\bosact{S=\int d^dx\,\frac{N}{4\lambda}\tr\left[F_{\mu\nu}F_{\mu\nu}+
\sum_I 2D_{\mu}\phi^ID_{\mu}\phi^I- \sum_{I,J}
\left[\phi^I,\phi^J\right]^2\right],}
with $p-d$ scalar fields, such that the zero mode action is given
by \intza, where
$\psi^{\mu}$ ($\mu=1,
\cdots,d$) are the zero modes of $A_{\mu}$, and $\psi^{d+I}$ ($I=1,
\cdots,p-d$) are the zero modes of $\phi^I$.
We expand the fields $A_{\mu},\phi^I$ in KK modes
$A_{\mu,\{m_{\lambda}\},ij},\phi^I_{\{m_{\lambda}\},ij}$, with mode
numbers $m_{\lambda}\in \IZ$ around the $\lambda$'th circle.
Expanding the action to quadratic order, we obtain
\eqn\soexp{\eqalign{S=\frac{N}{4\lambda_0}\sum_{\{m^{\mu}\}}
\sum_{i<
j}&\left\{
\sum_{\nu}\left[\sum_{\mu \neq \nu}
\left(\left(\Delta\psi^{\mu}\right)_{ij}-{{2\pi m^{\mu}}\over R_{\mu}}\right)^2
+ \sum_I \left(\Delta\psi^{d+I}\right)_{ij}^2\right]
|A_{\nu,\{m_{\lambda}\},ij}|^2\right.\cr
&+\sum_I \left[ \sum_{\mu}
\left(\left(\Delta\psi^{\mu}\right)_{ij}-{{2\pi m^{\mu}}\over R_{\mu}}\right)^2
+\sum_{J \neq I} \left(\Delta\psi^{d+J}\right)_{ij}^2\right]
|\phi^I_{\{m_{\lambda}\},ij}|^2\cr
&-\sum_{\mu \neq \nu} \left(\left(\Delta\psi^{\mu}\right)_{ij}-
{{2\pi m^{\mu}}\over R_{\mu}}\right)
\left(\left(\Delta\psi^{\nu}\right)_{ij}-{{2\pi m^{\nu}}\over R_{\nu}}\right)
A_{\mu,\{m_{\lambda}\},ij}A_{\nu,\{m_{\lambda}\},ij}^{\ast}\cr
&-\sum_{I \neq J} \left(\Delta\psi^{d+I}\right)_{ij}
\left(\Delta\psi^{d+J}\right)_{ij}\phi^I_{\{m_{\lambda}\},ij}
\phi^{\ast,J}_{\{m_{\lambda}\},ij}\cr
&\left.-\sum_{\mu,I} \left(\left(\Delta\psi^{\mu}\right)_{ij}-
{{2\pi m^{\mu}}\over R_{\mu}}\right)
\left(\Delta\psi^{d+I}\right)_{ij}\left[A_{\mu,\{m_{\lambda}\},ij}
\phi^{\ast,I}_{\{m_{\lambda}\},ij}+c.c.\right]\right\}.}}

Let us now perform a naive first attempt at integrating out the KK modes at
one-loop.  Defining a vector $D^{\alpha}_{ik,\{m_{\lambda}\}}$ as
\eqn\Ddefn{D_{ik,\{m_{\lambda}\}}=\left(\left(\Delta\psi^{\mu}\right)_{ik}-
{{2\pi m_{\mu}}\over R_{\mu}},\left(\Delta\psi^{d+I}\right)_{ik}\right)^T,}
we see that performing the quadratic integral over KK modes in \soexp\
yields a factor of $\det(M')^{-1}$, with
\eqn\mprdefn{M'=\left(D^TD\right)I-DD^T,}
similar to what we found in the zero dimensional case.
This determinant is easily evaluated, as the eigenvalues of $M'$ are
again
$D^TD$ with degeneracy $p-1$ and zero with degeneracy one.  The appearance
of a zero eigenvalue is not surprising.  Indeed, it is expected as we
have yet to fix the gauge.  A convenient set of gauge-fixing
constraints to adopt is the following :
\eqn\gfix{\eqalign{\partial_1A_1=&0,\cr
\partial_2\int\,dx_1\,A_2=&0,\cr
\partial_3\int\,dx_1\,dx_2\,A_3=&0,\cr
\ldots&\cr
\partial_{d}\int\,dx_1\ldots dx_{d-1}
A_{d}=&0.}}
For each constraint, we must also insert an appropriate Fadeev-Popov
determinant factor.  For the generic constraint
\eqn\gcon{\partial_{n+1}\int\,dx_1\,dx_2\ldots dx_n\,A_{n+1}=0,}
the corresponding determinant takes the form
\eqn\gdet{\det^{(n)}\left(\partial_{n}\right)
\det^{(n)^\prime}\left(\partial_{n+1}-i\left[A_{n+1},\ast\right]\right),}
where
the superscript $(n)$ is used to indicate that the determinant is
taken over all modes of the gauge field that are constant in
$x_1,\ldots,x_n$, while the prime in the second determinant indicates
that it also includes only non-zero modes in $x_{n+1}$.  We neglect the
first, constant, determinant and evaluate the second one at one-loop, finding
\eqn\detev{\prod_{i\ne j}\prod_{m^{\mu}}
\left(\left(\Delta\psi^{n+1}\right)_{ij}-{{2\pi m^{n+1}}\over R_{n+1}}\right)
= \prod_{i<j}\prod_{m^{\mu}}
\left(\left(\Delta\psi^{n+1}\right)_{ij}-
{{2\pi m^{n+1}}\over R_{n+1}}\right)^2,}
where the product over $m^{\mu}$ is suitably constrained.
Summarizing, we find that our gauge-fixing procedure introduces a
factor of $\left(\left(\Delta\psi^{n}\right)_{ij}-2\pi m^{n}/R_{n}\right)^2$
for each mode of the gauge field which has
$m^{1}=\cdots =m^{n-1}=0$ and $m^{n}\ne 0$.

Returning now to the task of integrating out the KK modes in \soexp,
we write the gauge fixing constraints \gfix\ as
\eqn\gfixmom{\eqalign{A^1_{r,m_2,m_3,\ldots ,m_{d}}&=0\cr
A^2_{0,r,m_3,\ldots,m_{d}}&=0\cr
\ldots & \ldots\cr
A^{d}_{0,0,\ldots,0,r}&=0,}}
where $r\ne 0$ and the other $m_{\mu}$ are arbitrary.
It is now easy to see that, for every choice of mode
numbers $\{m_{\lambda}\}$, exactly one component of the gauge field,
say $A_{\nu,\{m_{\lambda}\}}$, is
eliminated by these constraints.  When the corresponding row and
column are removed from $M'$, the eigenvalues become $D^TD$ with
degeneracy $p-2$ and
$\left(\left(\Delta\psi^{\nu}\right)_{ij}-2\pi m^{\nu}/R_{\nu}\right)^2$ with
degeneracy one.  Each factor of
$\left(\left(\Delta\psi^{\nu}\right)_{ij}-2\pi m^{\nu}/R_{\nu}\right)^2$
that arises
for a gauge field component eliminated by the gauge-fixing conditions, though,
serves to cancel a corresponding
factor from the Fadeev-Popov determinants \detev,
so that the
final result obtained from gauge-fixing and
integrating out KK modes at 1-loop is given by $\exp(-S_{eff,bos})$,
where
\eqn\Znzmod{S_{eff,bos} = (p-2) \sum_{\vec{m}\in \IZ^{d}}\,\,\, \sum_{i< j}
\ln \left[\sum_{\mu} \left(\left(\Delta\psi^{\mu}\right)_{ij}-
{{2\pi m^{\mu}}\over R_{\mu}}\right)^2+
\sum_I \left(\Delta\psi^{d+I}\right)_{ij}^2\right].
}
We have included also the terms \Zoffdiagg\ obtained from
integrating out the off diagonal zero-modes. Like \intol, the result
corresponds to a pairwise logarithmic effective potential between the
eigenvalues, but now these live on the dual torus in the gauge field
directions, so we have additional interactions between the eigenvalues
and the infinite set of image eigenvalues. These image interactions
(corresponding to the sum over $\vec{m}$) ensure that the full result
is periodic. This sum may be evaluated explicitly using the results of
appendix B, and we find
\eqn\bossec{\eqalign{S_{eff,bos}&=-2(\prod_{\mu} R_{\mu})
(p-2)\sum_{i<j}\theta_{ij}^{d/2} (2\pi)^{-d/2}\cr
&\quad \sum_{\vec{k}\in
\IZ^{d}-\{\vec{0}\}}\frac{1}{\left(\sum_{\mu} k_{\mu}^2 R_{\mu}^2
\right)^{d/4}}e^{
i\sum_{\mu} k_{\mu}R_{\mu}\left(\Delta\psi^{\mu}\right)_{ij}}
K_{d/2}\left(\theta_{ij} \sqrt{\sum_{\mu} k_{\mu}^2 R_{\mu}^2}\right),}}
%
where $\theta^2_{ij}$ is defined as
\eqn\thdef{\theta^2_{ij}=\sum_I \left(\Delta\psi^{d+I}\right)^2_{ij}.}

\subsec{Generalization to supersymmetric field theories}

It is a simple matter to include also fermionic fields, if they exist.
Assuming that the Yukawa couplings to the scalars are
the same as the gauge couplings, as in supersymmetric gauge theories,
and that the fermions are periodic around $(d-1)$ of the circles but
anti-periodic around the $d$'th circle,
the appropriate determinant is easily computed and found to yield
\eqn\Zfd{\prod_{m^{\nu}}\prod_{i<j}\left[
\sum_{\mu=1}^{d-1} \left(\left(\Delta\psi^{\mu}\right)_{ij}-
{{2\pi m^{\mu}}\over R_{\mu}}\right)^2+
\left(\left(\Delta\psi^{d}\right)_{ij}-
{{2\pi (m^{d}+\frac{1}{2})}\over R_d}\right)^2
+\sum_I \left(\Delta\psi^{d+I}\right)^2_{ij}\right]^{p-2}.}
As with \Znzmod, this corresponds to a pairwise potential between each
eigenvalue and an infinite series of ``images'' of the remaining
eigenvalues, except that the potential has the opposite sign and
the coordinates of the images are shifted by
a half period in any direction for which the fermions are
anti-periodic.

Combining \Zoffdiagg, \Znzmod, and \Zfd, as appropriate for a supersymmetric
gauge theory on a torus, we arrive at the full result :
\eqn\Zeffa{\eqalign{Z=\int
d\psi^{\alpha}_{i} & \prod_{m^{\nu}}\prod_{i<j}\cr
&\left[\frac{\sum_{\mu=1}^{d-1}
\left(\left(\Delta\psi^{\mu}\right)_{ij}-
{{2\pi m^{\mu}}\over R_{\mu}}\right)^2+
\left(\left(\Delta\psi^{d}\right)_{ij}-
{{2\pi (m^{d}+\frac{1}{2})}\over R_d}\right)^2
+\sum_I\left(\Delta\psi^{d+I}\right)_{ij}^2}{\sum_{\mu=1}^{d-1}
\left(\left(\Delta\psi^{\mu}\right)_{ij}-
{{2\pi m^{\mu}}\over R_{\mu}}\right)^2+
\left(\left(\Delta\psi^{d}\right)_{ij}-{{2\pi m^{d}}\over R_d}\right)^2+
\sum_I\left(\Delta\psi^{d+I}\right)_{ij}^2}\right]^{p-2}.}}

Using the results of Appendix B we can again evaluate the product over
mode numbers and obtain an effective action for the eigenvalue
separations.  The contribution from the fermionic sector is identical
to the bosonic result \bossec, with the exception of an additional
factor of $(-1)^{k_{d}+1}$ in the sum.  As a result, the fermions
serve to eliminate half of the terms in the $k_{d}$ sum.  Our final
result for $S_{eff}$ thus becomes:
\eqn\olseff{\eqalign{S_{eff}&=-4(\prod_{\mu} R_{\mu})
(p-2)\sum_{i<j}\theta^{d/2}_{ij} (2\pi)^{-d/2}\cr
&\quad \sum_{k_{\mu}=-\infty}^{\infty}\sum_{k_{d}\,odd}
\frac{1}{\left(\sum_{\mu} k_{\mu}^2 R_{\mu}^2\right)^{d/4}}
e^{i \sum_{\mu} k_{\mu} R_{\mu} \left(\Delta\psi^{\mu}\right)_{ij}}
K_{d/2}\left(\theta_{ij}\sqrt{\sum_{\mu} k_{\mu}^2 R_{\mu}^2}\right).}}

\subsec{Analysis of the effective potentials for $d=2$}

We will now focus on the case $d=2$, and
argue that for both the bosonic and the supersymmetric cases,
the eigenvalues of the gauge field will be
clumped in all directions whenever the one-loop effective potentials
\Znzmod\ or \Zeffa\ are reliable.

The contribution of the zero modes (corresponding to the $\vec{m}=0$
terms in the bosonic contribution to the potential) was discussed in
\S A.1, and we concluded that this leads the eigenvalues to clump at a scale
\eqn\Meigclmp{a\sim \lambda_0^{1/4},}
where $\lambda_0 = \lambda / (R_1 R_2)$. The additional terms in the
effective potential correspond to the interactions with image
eigenvalues. These interactions will be negligible for the clumped
configuration as long as the scale \Meigclmp\ is much smaller than
either $1/R_1$ or $1/R_2$ (i.e. when the distance to the nearest image
charge is large compared with the distances between eigenvalues in the
clump).\foot{It is not immediately obvious that the infinite set of
image charges does not conspire to produce a larger effect, but
because of cancellations between $\vec{m}$ and $-\vec{m}$ terms in the
potential, it may be checked that the sum of contributions from all
charges is of the same order of magnitude as the contribution from the
nearest charge.} Since the KK-mode terms are important for
configurations that are not highly clumped, we may worry that the full
effective potentials \bossec\ or \olseff\ could have additional saddle point
configurations. However, a quick analysis shows that in either case,
the full effective potential (periodic in the directions corresponding
to the gauge field eigenvalues) is attractive both for the scalar and
for the gauge field zero modes,\foot{There is one qualitative difference
between the bosonic and supersymmetric cases. In the supersymmetric
case, the potential between two eigenvalues actually goes to infinity
as the eigenvalues approach antipodal points in the anti-periodic
directions, since in this configuration there is a (repulsive) image
charge from one eigenvalue sitting on top of the other
eigenvalue.} so the eigenvalues are driven towards
the clumped saddle point for which the KK-mode contributions are
negligible. We conclude that the eigenvalues are clumped on a scale
\Meigclmp\ whenever
\eqn\scales{
\left( {\lambda \over R_1 R_2} \right)^{1 \over 4} \ll {1 \over R_1}
\qquad {\rm and} \qquad  \left(
{ \lambda  \over R_1 R_2 }\right)^{1 \over 4} \ll {1 \over R_2}.
}
We see that the eigenvalues spread out relative to the sizes of the
dual circles as either $R_1$ or $R_2$ increases, suggesting a possible
phase transition when either of the inequalities in \scales\ is
violated. However, the effective coupling of the lightest KK-modes is
$\lambda_0 /m_{KK}^4$, so when the relations \scales\ are not satisfied,
either the $R_1$ or the $R_2$ KK-modes become strongly coupled and we
can no longer trust our perturbative results for the effective
potential. Therefore, other methods (discussed in the main text) are
required to deduce the presence (or not) of a phase transition as the
eigenvalues spread.

\subsec{Integrating out massive scalars}

Before closing this section, we describe the result of integrating out
a set of $p$ very massive adjoint scalar fields in a two dimensional gauge
theory on $T^2$. In general, the one-loop result is given by \scveff,
where $D_\mu$ is the covariant derivative for the adjoint
representation. Here, we consider the special case where the gauge
fields are constant commuting matrices, so that the holonomies are
$U = e^{i R_2 A_2}$ and $V = e^{i R_1 A_1}$. Then
\eqn\scalarea{\eqalign{
S_{eff} & = {p \over 2} \ln(\det(-D_\mu^2 + M^2)) \cr
&= {p \over 2} \tr(\ln(-D_1^2 - D_2^2 + M^2)) \cr
&={p \over 2} \tr \left(\sum_{m,n} \ln \left(\left({2 \pi n \over R_1} + A_1
\right)^2 + \left({2 \pi m \over R_2} + A_2\right)^2 + M^2\right)\right) \cr
&=- {p \over 2 \pi} M R_1 R_2 \sum_{k,l} \tr(U^l V^k) \tr(U^{-l} V^{-k})
{K_1( M \sqrt{(kR_1)^2 + (l R_2)^2}) \over \sqrt{(k R_1)^2 + (l R_2)^2} }
}}
where in the last line, we have used a result from appendix B.

\appendix{B}{Infinite products}

In evaluating the determinants involved in integrating out
Kaluza-Klein modes and massive scalars in the main text and in appendix A,
we encounter infinite
products whose logarithm gives an infinite sum of the form
\eqn\pveca{
P(\vec{a}) = \sum_{\vec{m}} \ln(\theta^2 + (\vec{m} + \vec{a})^2).
}
Here, the sum runs over all vectors $\vec{m}$ in
$d$ dimensions with integer components. Now, $P$ is clearly periodic in
each component of $\vec{a}$, with period 1. Thus, we can write
\eqn\pform{
P(\vec{a}) = \sum_{\vec{k}} e^{2 \pi i \vec{k} \cdot \vec{a}} P_{\vec{k}},
}
and we can compute the Fourier transform of $P$ :
\eqn\eval{\eqalign{
P_{\vec{k}} &= \int_0^1 d{a_1} \cdots \int_0^1 d{a_d}
\sum_{\vec{m}} \ln(\theta^2 + (\vec{m} + \vec{a})^2) \; e^{-2 \pi i \vec{k}
\cdot \vec{a}} \cr
&= \sum_{\vec{m}} \int_{m_1}^{m_1+1} d{a_1} \cdots
\int_{m_d}^{m_d + 1} d{a_d} \; \ln(\theta^2 + \vec{a}^2)\;  e^{-2 \pi i
\vec{k} \cdot \vec{a}} \cr
&= \int d \vec{a}\; \ln(\theta^2 + \vec{a}^2) \;
e^{-2 \pi i \vec{k} \cdot \vec{a}} \cr
&= lim_{\epsilon \to 0}
(-\ln(\epsilon) - \int_\epsilon^\infty {d \alpha \over \alpha} e^{-\alpha
\theta^2} \int d \vec{a}\; e^{-\alpha \vec{a}^2 - 2 \pi i \vec{k} \cdot
\vec{a}}) \cr
&= C_\infty - \pi^{d \over 2} \int_0^\infty {d \alpha \over
\alpha^{{d \over 2} + 1}} e^{-\alpha \theta^2 - {\pi^2 k^2 \over \alpha}}
\cr
&= C_\infty - 2 {\theta^{d \over 2} \over |\vec{k}|^{d \over 2}} K_{d
\over 2} (2 \pi |\vec{k}| \theta). }}
Here, $C_\infty$ is an infinite constant independent of $k$ and $\theta$.

\appendix{C}{Effective Action for the Wilson Line in $d=1$ Gauge
Theories}

In this appendix, we consider the one dimensional Euclidean gauge theory
on a circle of circumference $R$ with $p$ scalars and action
\eqn\appaction{
S = \int dt \tr( {1\over 2} D_t \Phi_i D_t \Phi_i + {M\over 2}
\Phi_i \Phi_i - {g^2 \over
4} [\Phi_i,\Phi_j][\Phi_i,\Phi_j] ).  }
The analysis here follows that of \jh, which considered the special
case of two scalar fields.
Note that here we have rescaled the gauge field and the scalar fields
by a factor of $g$ compared
to our previous analysis, so as to have canonical kinetic terms.
We would like to integrate out the scalar
fields for small values of the dimensionless 't Hooft coupling $g^2 N
/ M^3$, to obtain an effective action in terms of the Wilson line of
the gauge field around the circle. We choose the gauge $\partial_t A_0
= 0$, so that $A_0$ is a $t$-independent Hermitian matrix $A_0 =
\alpha$. With this choice, the Wilson line is simply given by
\eqn\wilzerodef{
U = e^{i \alpha R} \; .
}
We define
\eqn\seffdef{
\exp(-S_{eff}(U)) = \int [d \Phi_i] e^{-S(\Phi, \alpha)},
}
in terms of which the partition function is given by
\eqn\partfundef{
Z = \int DU \exp(-S_{eff}(U)) \; .
}
As explained in section 4 of \sphere, the Haar measure $DU$ arises
from the initial measure $[d A_0]$ upon introducing the Fadeev-Popov
determinant associated with the gauge fixing condition $\partial_t A_0 = 0$.

By gauge invariance, the effective action must be some function of the
variables
\eqn\defun{
u_n \equiv {1 \over N} \tr(U^n) \; .  }
At one-loop order, the
calculation of $S_{eff}$ was described in \sphere, with the result
\eqn\resforqm{
S_{eff}^{1-loop}(U)  =  N^2 \sum_{n=1}^\infty {1 \over n} (1-p x^n) |u_n|^2,
}
where $x = e^{- R M}$. Here, the $x$-independent term is a rewriting
of the Vandermonde determinant obtained in writing the Haar measure in
terms of eigenvalues. For $x < 1/p$, the one-loop effective action is
positive definite, minimized by the saddle-point configuration $u_n =
0$. As $x$ passes $x_c=1/p$, the mode $u_1$ becomes unstable and condenses
to its maximum allowed value (as long as all other $u_n=0$)
$u_1 = 1/2$, giving rise to a first order large $N$
phase transition in the strict $g^2N = 0$ limit.

As described in section 6 of \sphere, in order to determine the nature
of the phase transition for weak but non-zero coupling, it is
necessary to take into account higher order terms in the effective
action. The relevant physics may be deduced easily from the effective
action for $u_1$, obtained by integrating out both the scalar fields
and all the modes $u_{n>1}$ near the transition. This takes the general form
\eqn\genseff{
S_{eff} (u_1) = N^2(m_1^2(x, \lambda) |u_1|^2 +
b(x, \lambda) |u_1|^4 + {\cal O}(\lambda^4)) \; ,
}
where in perturbation theory $b$ starts at order $\lambda^2$.
If the coefficient $b$ is positive at the value $x_c$ of $x$ where $m_1^2$
drops to zero, we will have a second order phase transition with the
eigenvalue distribution for $U$ changing continuously. On the other
hand if (as we will find below) the coefficient $b$ is negative at
this value $x=x_c$, the potential develops a second minimum which
is lower than the first already at some slightly lower value of $x$, the
eigenvalue distribution changes discontinuously, and we have a
first order phase transition.

The leading order contribution to $b$ is given by
\eqn\bdef{
b = D_3 - {C_2^2 \over B_1},
}
where $B_1$, $C_2$, and $D_3$ are the leading coefficients of the
terms $|u_2|^2$, $(u_2 u_{-1}^2 + u_{-2} u_1^2)$, and $|u_1|^4$ in the
effective action obtained by integrating out the scalars; these terms
first arise at
one, two, and three-loop order, respectively. We now proceed to
compute these coefficients, together with the corrections to the
coefficient $A=N^2 m_1^2$ of $|u_1|^2$, needed to determine how the phase
transition temperature varies with the coupling constant.

The higher loop corrections to the effective action are given by
\eqn\genpert{
S_{eff}^{pert} = \langle - e^{-{g^2 \over 4} \tr([\Phi^i, \Phi^j]^2)}
\rangle_{connected}.
}
This may be evaluated in perturbation theory using the propagator
\eqn\propdef{
\langle (\Phi_i)_{kl}(t_1) (\Phi_j)_{mn}(t_2) \rangle = \Delta_{kn}(t_1 - t_2,
\alpha) \delta_{lm} - \delta_{kn} \Delta_{lm}(t_1 - t_2, \alpha),
}
where the matrix $\Delta$ is defined by
\eqn\propdefn{
\Delta(t, \alpha) = {e^{i \alpha t} \over 2M} \left( {e^{-t M} \over 1 -
e^{-M \beta} e^{i \alpha R}} - {e^{t M} \over 1 - e^{M \beta} e^{i \alpha R}}
\right).  }
More details about the perturbative evaluation and our conventions
may be found in \threelooppaper.
Setting $M=1$ for now, the contribution to the effective action from
the two-loop figure eight diagram is given by
\eqn\figeight{
S_{eff}^{2\; loop} =
{g^2 \over 2} \beta (p^2 - p) \tr( \Delta(0, \alpha_{ab})
\Delta(0, \alpha_{ac}) ) \; ,
}
where we have introduced the notation $\alpha_{ab}= \alpha_a - \alpha_b$, with
\eqn\alphadef{
\alpha_a = \alpha \otimes 1 \otimes 1, \qquad
\alpha_b = 1 \otimes \alpha \otimes 1, \qquad
\alpha_c = 1 \otimes 1 \otimes \alpha.
}
This leads to quadratic terms
\eqn\quadeightterms{
S_{quad}^{2 \; loop} = -{1 \over 4} N^2 \lambda (p^2 - p) \ln(x) \sum_n
(x^{2n} + 2 x^n) |u_n|^2,
}
and leading order cubic terms
\eqn\cubiceight{
S_{cubic}^{2 \; loop} = -{1 \over 8} N^2 \lambda (p^2-p) \ln(x) (x^2 + 2 x^3)
(u_2 u_{-1}^2 + u_{-2} u_1^2)+ \dots \; .
}

\fig{Three-loop diagrams contributing to the effective action.}
{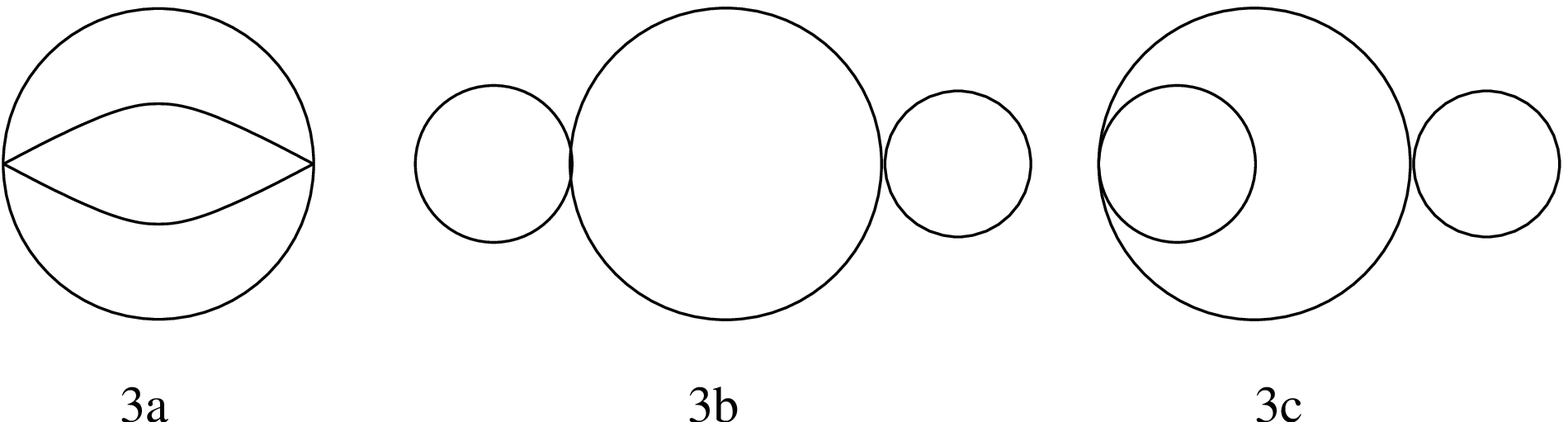}{4truein}
\figlabel{\diags}

At three loops, the diagrams 3a, 3b, and 3c shown in figure \diags\ give
the following contributions
\eqn\threeloop{
\eqalign{
S_{eff}^{3a} &= -{1 \over 2} g^4 (p^2 - p) \beta \int dt
\tr( \Delta(t,\alpha_{ab})
\Delta(t, \alpha_{bc}) \Delta(t, \alpha_{cd}) \Delta(t, \alpha_{da})), \cr
S_{eff}^{3b} &= -{1 \over 2} g^4 p(p-1)^2 \beta \int dt
\tr(\Delta(0, \alpha_{ab})
\Delta(t, \alpha_{ac}) \Delta(t, \alpha_{ca}) \Delta(0, \alpha_{ad})), \cr
S_{eff}^{3c} &= -{1 \over 2} g^4 p(p-1)^2 \beta \int dt
\tr(\Delta(0, \alpha_{ab})
\Delta(t, \alpha_{ac}) \Delta(t, \alpha_{ca}) \Delta(0, \alpha_{cd})).
}}
These contribute to the quadratic term for the lowest mode
\eqn\massthree{
\eqalign{
S_{quad}^{3a} &= -{1 \over 16} N^2 \lambda^2 |u_1|^2 (p^2 - p)
\ln(x) x (2x \ln(x) + x^2 - x - 3), \cr
S_{quad}^{3b} &= -{1 \over 16}  N^2 \lambda^2 |u_1|^2 p(p-1)^2
\ln(x) x (\ln(x)(2x+1)-3(x+1)),\cr
S_{quad}^{3c} &= -{1 \over 16} N^2 \lambda^2 |u_1|^2 p(p-1)^2
\ln(x) x(\ln(x)(x+1)^2 - (x^2 + 2x + 3)).
}}
The same diagrams also give the leading $|\tr(U)|^4$ terms,
\eqn\massthree{
\eqalign{
S_{quart}^{3a} &= -{1 \over 32} N^2 \lambda^2 |u_1|^4 (p^2 - p)
\ln(x) x^2 (2 \ln(x) + 2x^2 - 5), \cr
S_{quart}^{3b} &= -{3 \over 16}  N^2 \lambda^2 |u_1|^4 p(p-1)^2
\ln(x) x^3 (\ln(x)-1),\cr
S_{quart}^{3c} &= -{1 \over 16} N^2 \lambda^2 |u_1|^4 p(p-1)^2
\ln(x) x^2 (2 x^2 \ln(x) - x^2 - 2).
}}

Collecting our results, we have
\eqn\newseff{
S_{eff} = A |u_1|^2 + B_1 |u_2|^2 + C_2 (u_2 u_{-1}^2 + u_{-2} u_1^2) +
D_3 |u_1|^4 + \dots
}
where
\eqn\hag{\eqalign{
A/N^2 =& (1-px) -{1 \over 4} \lambda (p^2 - p) \ln(x) (x^2+2x) \cr
&- {1 \over 16} \lambda^2(p^2-p) x \ln(x) (2x \ln(x) +x^2-x-3) \cr
& -{1 \over 16} \lambda^2p(p-1)^2 x \ln(x)(\ln(x)(x^2+4x +2) -x^2-5x-6)  +
{\cal O}(\lambda^3), \cr
B_1/N^2 =& {1 \over 2}(1-px^2) + {\cal O} (\lambda), \cr
C_2/N^2 =& -{1 \over 8}  \lambda (p^2-p)
\ln(x) (x^2 + 2 x^3) + {\cal O} (\lambda^2),\cr
D_3/N^2 =&  -{1 \over 32}  \lambda^2 (p^2 - p) \ln(x) x^2 (2 \ln(x) + 2x^2 - 5) \cr
&-{1 \over 16} \lambda^2 p(p-1)^2 \ln(x) x^2 (\ln(x)(2x^2+3x)-x^2-3x-2) +
{\cal O}(\lambda^3).
}}
From these above expressions and from \bdef\ we find that
\eqn\finalb{
b(x = x_c= 1/p)/N^2 = -{1 \over 32} {(p-1) \ln(p) \over p^3}
( \ln(p) (9 p^2 + 2 p) + 4 p^3 + 7 p^2 - 4 p - 4) \lambda^2,
}
which is negative for all $p>1$. Thus, the phase transition is of first order
at weak coupling.

To determine the precise transition point as a function of $\lambda$,
we note that to order $\lambda^2$, the second minimum $|u_1| = 1/2$
will become dominant at the point $x_c$ where
\eqn\newtranspoint{
m_1^2(x) {1 \over 2^2} + b(x)  {1 \over 2^4}  = 0.
}
Solving perturbatively for $x_c$ as a function of $\lambda$, we find
\eqn\xsol{\eqalign{
 x_c &= {1 \over p} + \lambda {(p-1)(2p+1)\ln(p) \over 4 p^2} \cr
&+ \lambda^2 {(p-1) \ln(p) \over 128 p^4} ((16p^4-33p^2-10p)\ln(p)
-80p^4-20p^3+41p^2+28p+4)+{\cal O}(\lambda^3).
}}
Expressing this as an equation for $\tit = 1/(R \lambda^{1/3})$ as a
function of $m = M/ \lambda^{1/3}$ gives the relation \crittemp\ .

\appendix{D}{Pure and Deformed Yang-Mills Partition Functions on $T^2$}

In this appendix, we review the exact solution of pure Yang-Mills
theory on $T^2$ \MigdalAB, and use techniques developed by Gross and Taylor
\refs{\GrossTU,\GrossHU,\GrossYT} in order to write the result at large $N$
in terms of an effective action for one of the two holonomies, $U$ and
$V$.  We then proceed to study the partition function for a class of
deformations which are encountered in section 5.

\subsec{Pure Yang-Mills theory on $T^2$}

An exact expression for the partition function of pure Yang-Mills
theory on $T^2$ is easily obtained, following Migdal \MigdalAB, if we use a
particular lattice regularization of the theory.
We put unitary matrices $U_L$ on the links on the lattice, and
consider the partition function
\eqn\pureymz{Z=\int\,\prod_L\,dU_L\prod_{P}Z_P(U_P),}
where $P$ denotes the
plaquettes of the lattice, $U_P$ is a product of the $U_L$'s around the
plaquette $P$, and $Z_P$ is a plaquette action chosen so
that the continuum theory coincides with Yang-Mills theory (for instance,
the Wilson action $Z_P = \exp({1\over g_{YM}^2} \tr(U_P+U_P^{-1}))$).

Migdal noted that, upon integrating out various links, one is left
with an action of the form \pureymz\ with a plaquette action for the
remaining (larger) plaquettes that approaches
a weighted sum of characters $\chi_R$, in the
representation $R$ of the gauge group :
\eqn\heatkern{Z_P(U_P)=
\sum_Rd_R\chi_R(U_P)e^{-\frac{\lambda A}{2N}C_2(R)},}
where the representation $R$ has dimension $d_R$ and quadratic Casimir
$C_2(R)$, $\lambda$ is the dimensionless 't Hooft coupling,
and $A$ is the area of the plaquette{\foot{In \MigdalAB,
Migdal considered a more general situation in which \pureymz\ was not
required to yield pure Yang-Mills theory in the continuum limit.  As
such, he originally wrote the factor in the exponent in terms of a
representation-dependent coupling, $g_R$, noting that
$g_R^2\rightarrow g_{YM}^2C_2(R)$ gives the correct Coulomb law.  One can also see this directly by applying Migdal's analysis to the Wilson action.}}.  Using standard
orthogonality relations, it is easy to verify the additivity property
of \heatkern,
\eqn\hkadd{\int\,DU
Z_{P}(V_1U, \lambda A_1 )Z_{P'}(U^{\dag}V_2, \lambda A_2)=
Z_{P+P'}(V_1 V_2, \lambda(A_1 + A_2)),
}
at which point
\pureymz\ can be reduced to an integral over two matrices, $U$ and
$V$, corresponding to the products of the $U_L$'s over the
links along the two non-trivial cycles of the torus :
\eqn\pureym{Z_{ym}=\int\,DU\,DV\,\sum_Rd_R
e^{-\frac{\tilde{\lambda}}{2N}C_2(R)}\chi_R(UVU^{-1}V^{-1}) \;,}
where $\tilde{\lambda}$ is the 't Hooft coupling times the area of the $T^2$.
We can then use the result
\eqn\orthres{\int\,DU\,\chi_R(UAU^{-1}B)=\frac{1}{d_R}\chi_R(A)\chi_R(B)}
to integrate out one matrix, say $U$, from \pureym, obtaining
\eqn\ymuonly{Z_{ym}=\int\,DV\,\sum_R
e^{-\frac{\tilde\lambda}{2N}C_2(R)}\chi_{R\otimes\bar{R}}(V).}
This integral is also trivial to perform, since $R\otimes \bar{R}$
contains the identity precisely once, yielding the well-known
expression for $Z_{ym}$ purely as a function of $\tilde{\lambda}$,
\eqn\ymnouv{Z_{ym}=\sum_Re^{-\frac{\tilde{\lambda}}{2N}C_2(R)}.}

In the rest of this subsection, we shall be interested in determining,
in the limit of large $N$, a more explicit form for the effective
action, $S_{eff}(V)$, corresponding to the partition function
\ymuonly.

Thus, we seek a large $N$ expansion of the quadratic Casimir and
character appearing there.
As pointed out by Gross and Taylor in \refs{\GrossTU-\GrossYT},
one may obtain the
correct large $N$ expansion of \ymuonly\ by summing independently over
representations formed from tensor products of finitely many
fundamentals and anti-fundamentals. The Young tableaux for any such
representation may be obtained uniquely by adjoining the Young tableau
for some representation obtained from the product of finitely many
anti-fundamentals with the Young tableau for some representation
obtained from the product of finitely many fundamentals. Following the
notation in \refs{\GrossTU,\GrossYT} we denote these representations
by $\bar{S}$ and $R$ respectively, and denote the representation
corresponding to the original tableau, termed the ``composite''
representation of $R$ and $S$, by $T=\bar{S}R$.

Letting $n_R$ ($n_S$) denote the number of boxes associated to the
representation $R$ ($S$), we may express the quadratic Casimir of the
composite representation $T$ in terms of those of its components as:
\eqn\quadcas{C_2(T) = C_2(R) + C_2(S)+\frac{2n_Rn_S}{N^2},}
where
\eqn\ctwor{
C_2(R) = N n_R + {\cal O}(N^0)\; , \qquad C_2(S) = N n_S + {\cal O}(N^0) \; .
}
Denoting the set of Young tableaux with $n$ boxes by $T_n$, and
denoting both an arbitrary element of $T_n$ and the representation
corresponding to it by $Y_n\in T_n$, we may now write \ymuonly\ at
large $N$ as
\eqn\vonemat{Z_{ym}=\int\,DV\,\sum_{n,n'}\sum_{Y_n\in T_n}
\sum_{Y'_{n'}\in T_{n'}}
e^{-\left(n+n'\right){\tilde{\lambda}\over 2}}
\chi_{(\bar{Y}'_{n'}Y_n)\otimes(\overline{\bar{Y}_nY'_{n'}})}(V).}

We now seek to expand the characters appearing in the integrand as a
polynomial in traces.  To proceed, we first define some further
notation.  To any given tableau, $Y_n\in T_n$, we may associate not
only an irreducible representation, also denoted by $Y_n$, of $SU(N)$,
but also an irreducible representation, $\hat{Y}_n$, of the
permutation group $S_n$.  We use $\chi_{Y_n}(V)$ to denote the
character of the $SU(N)$ representation $Y_n$ evaluated on $V\in
SU(N)$, and $\chi_{\hat{Y}_n}(\sigma)$ to denote the character of the
$S_n$ representation $\hat{Y}_n$ evaluated on $\sigma\in S_n$.  To
each tableau, $Y_n\in T_n$, we also associate a conjugacy class,
$\rho_{Y_n}$, of $S_n$, which can also be labeled by the numbers
$\{\sigma_i\}$, which specify the number of cycles of length $i$ in an
element of $\rho_{Y_n}$.  Finally, we let $\Upsilon_{\sigma}$ denote
the Schur function associated to $\sigma\in S_n$,
\eqn\schdef{\Upsilon_{\sigma}(V)=\sum_{Y_n\in T_n}
\chi_{\hat{Y}_n}(\sigma)\chi_{Y_n}(V)=\prod_i \tr(V^i)^{\sigma_i}.}
The character of any representation $Y_n\in T_n$ is easily expanded in terms of Schur functions
\eqn\frob{\chi_{Y_n}(V)=\frac{1}{n!}\sum_{\sigma\in S_n}\chi_{\hat{Y}_n}(\sigma)\Upsilon_{\sigma}(V).}

To study \ymuonly, we shall need the following generalized version of
\frob\ for composite representations that was worked out by Gross and
Taylor \GrossYT:
\eqn\genfrob{\chi_{\bar{Y}'_{n'}Y_n}(V)=\frac{1}{n!n'!}
\sum_{\sigma\in S_n}\sum_{\tau\in S_{n'}}\chi_{\hat{Y}_n}(\sigma)
\chi_{\hat{Y}'_{n'}}(\tau)\Upsilon_{\bar{\tau}\sigma}(V,V^{\dag}),}
where
\eqn\genschur{\eqalign{\Upsilon_{\bar{\tau}\sigma}(V,V^{\dag})&=
\sum_{Y_n\in T_n}\sum_{Y'_{n'}\in T_{n'}}\chi_{\hat{Y}_n}(\sigma)
\chi_{\hat{Y}'_{n'}}(\tau)\chi_{\bar{Y}'_{n'}Y_n}(V)\cr
&=\prod_l\Upsilon_{\bar{\tau}^{(l)}\sigma^{(l)}}(V,V^{\dag}),\cr
\Upsilon_{\bar{\tau}^{(l)}\sigma^{(l)}}(V,V^{\dag})&=
\sum_{k=0}^{\min(\sigma_l,\tau_l)}{\sigma_l\choose k}{\tau_l\choose k}
(-1)^kl^kk!\tr(V^l)^{\sigma_l-k}\tr(V^{-l})^{\tau_l-k}.}}
Using this result, we may now write \ymuonly\ as
\eqn\veff{\eqalign{Z_{ym}=\int\,DV\,\sum_{n,n'}\sum_{\sigma,\sigma'\in S_n} &
\sum_{\tau,\tau'\in S_{n'}}\sum_{Y_n\in T_n}\sum_{Y'_{n'}\in T_{n'}}
e^{-\left(n+n'\right){\tilde{\lambda}\over 2}} \times \cr
&\frac{\chi_{\hat{Y}_n}(\sigma)\chi_{\hat{Y}_n}(\sigma')
\chi_{\hat{Y}'_{n'}}(\tau)\chi_{\hat{Y}'_{n'}}(\tau')}{(n!)^2(n'!)^2}
\times \Upsilon_{\bar{\tau}\sigma}(V,V^{\dag})
\Upsilon_{\bar{\sigma}'\tau'}(V,V^{\dag}).}}
The sum over Young tableaux can be performed using the completeness relation
\eqn\compl{\frac{1}{n!}\sum_{Y_n\in T_n}\chi_{\hat{Y}_n}(\sigma)
\chi_{\hat{Y}_n}(\sigma')=\frac{1}{|\rho_{\sigma}|}
\delta_{\rho_{\sigma},\rho_{\sigma'}},}
where $\rho_{\sigma}$ denotes the conjugacy class to which $\sigma$
belongs and $|\rho_{\sigma}|$ its dimension.  Using this result
we obtain
\eqn\vefff{\eqalign{Z_{ym}=\int\,DV\,\sum_{n,n'}\sum_{\sigma\in S_n} &
\sum_{\tau\in S_{n'}} \frac{e^{-\frac{\tilde{\lambda}}{2}(n+n')}}{n!n'!}
\Upsilon_{\bar{\tau}\sigma}(V,V^{\dag})
\Upsilon_{\bar{\sigma}\tau}(V,V^{\dag})\cr
=\int\,DV\,\sum_{n,n'}\sum_{\sigma\in S_n} & \sum_{\tau\in S_{n'}}
\frac{e^{-\frac{\tilde{\lambda}}{2}(n+n')}}{n!n'!}
\prod_l\{\left[\tr(V^l)\tr(V^{-l})\right]^{\sigma_l+\tau_l} \times \cr
& \sum_{k,k'=0}^{\min(\sigma_l,\tau_l)}
{\sigma_l\choose k}{\sigma_l\choose k'}{\tau_l\choose k}{\tau_l\choose k'}
(-l)^{k+k'}k!k'!\left[\tr(V^l)\tr(V^{-l})\right]^{-(k+k')}\}.}}
We now replace the sums over $\sigma,\tau$ with (appropriately
weighted) sums over the $\sigma_i,\tau_j$ which label conjugacy
classes.  Typically, such sums must be subject to the restrictions
$\sum_ii\sigma_i=n$, $\sum_jj\tau_j=n'$ and are difficult to evaluate.
Fortunately, the sums over $n,n'$ serve to lift these restrictions,
leaving us with the following expression
\eqn\veffsimp{\eqalign{Z_{ym}&=\int\,DV\,
\prod_l\left\{\sum_{\sigma_l,\tau_l=0}^{\infty}
\sum_{k,k'=0}^{\min(\sigma_l,\tau_l)}\left(
\frac{e^{-\frac{\tilde{\lambda}l}{2}}}{l}\right)^{\sigma_l+\tau_l}
\left[\tr(V^l)\tr(V^{-l})\right]^{\sigma_l+\tau_l}\right.\cr
&\left.\qquad\times\frac{\sigma_l!\tau_l!}
{k!k'!(\sigma_l-k)!(\sigma_l-k')!(\tau_l-k)!(\tau_l-k')!}
(-l)^{k+k'}\left[\tr(V^l)\tr(V^{-l})\right]^{-(k+k')}\right\},}}
which can be rewritten as
\eqn\veffmrsp{\eqalign{Z_{ym}&=\int\,DV\,
\prod_l\left\{\sum_{\sigma,\tau=0}^{\infty}
\sum_{k,k'=0}^{\infty}\left(\frac{e^{-\frac{\tilde{\lambda}l}{2}}}{l}
\right)^{\sigma+\tau+2\max(k,k')}\frac{(-l)^{k+k'}}{\sigma!\tau!k!k'!}
\right.\cr
&\left.\qquad\qquad\times
\left[\frac{\left[\sigma+\max(k,k')\right]!\left[\tau+\max(k,k')\right]!}
{\left[\sigma+|k-k'|\right]!\left[\tau+|k-k'|\right]!}\right]
\left[\tr(V^l)\tr(V^{-l})
\right]^{\sigma+\tau+|k-k'|}\right\}.}}

We now consider the sum at fixed $l$, which may be rewritten, after a
little algebra, as
\eqn\veffrew{\eqalign{&\sum_{\sigma,\tau,k'=0}^{\infty}
\sum_{k=-\infty}^{\infty}\frac{1}{\sigma!\tau!k'!}
\left(\frac{e^{-\frac{\tilde{\lambda}l}{2}}}{l}\right)^{\sigma+\tau}
\left(-\frac{e^{-\tilde{\lambda}l}}{l}\right)^{|k|}
\left(e^{-\tilde{\lambda}l}\right)^{k'}
\left[\frac{(\sigma+|k|+k')!(\tau+|k|+k')!}
{(\sigma+|k|)!(\tau+|k|)!(|k|+k')!}\right]\cr
&\qquad\qquad\qquad\qquad\times\left[\tr(V^l)\tr(V^{-l})\right]^{\sigma+\tau+|k|}.}}

In our large $N$ limit, \veffrew\ becomes
\eqn\sumkpln{\exp\left[\left(\frac
{2e^{-\frac{\tilde{\lambda}l}{2}}-e^{-\tilde{\lambda}l}}{l}\right)
\tr(V^l)\tr(V^{-l})\right],}
with corrections that are subleading in $N$ in the exponent.  As a
result we find that \vefff\ can be written as
\eqn\zymeffac{Z_{ym}=\int\,DV\,\exp\left\{-S_{eff}(V,\tilde{\lambda})\right\},}
where
\eqn\seffvdef{S_{eff}(V,\tilde{\lambda})=\sum_{l=1}^{\infty}\left[
\left(e^{-\frac{\tilde{\lambda}l}{2}}-1\right)^2-1\right]
\left(\frac{\tr(V^l)\tr(V^{-l})}{l}\right).}

To study this result, we write it in terms of the moments $u_n \equiv
\tr(V^n)/N$ of the
eigenvalue distribution,
%
\eqn\zymwron{Z_{ym}\sim\int\,(\prod_n d^2 u_n)\,\exp\left\{-N^2
\sum_{n=1}^{\infty}
\frac{\left(e^{-\frac{n\tilde{\lambda}}{2}}-1\right)^2}{n}|u_n|^2\right\}.}
A model similar to this was studied in \sphere.  Here, the masses of
all moments, $u_n$, are positive for all positive values of the
coupling, $\tilde{\lambda}$, and thus the dominant eigenvalue
distribution is the uniform one with $u_n=0$ for all nonzero $n$.  At
large $N$, the $u_n$ are essentially independent variables and can be
integrated out to yield
\eqn\zymcheck{Z_{ym}\sim\prod_n\left(1-
e^{-\frac{\tilde{\lambda}n}{2}}\right)^{-2}.}
This is in agreement with the known partition function for pure
Yang-Mills theory on $T^2$ at large $N$.

\subsec{Deformed Yang-Mills theory on $T^2$}

In this appendix, we derive a formula relevant to our study in
section 5 of pure
Yang-Mills theory deformed by very massive scalars. In section 5.5, we
argue that in a certain regime of parameter space, integrating out
very massive adjoint scalars gives, to a good approximation, an
expression \notpure\ for the partition function, where $f$ and $g$
depend only on the eigenvalues of $U$ and $V$ respectively. Now,
changing variables $U \to WUW^{-1}$ and integrating over $W$ (which
doesn't change the result since the integrand cannot depend on $W$) we
obtain
\eqn\varchange{
Z_{ym}=\int\,DW\,DU\,DV\,\sum_Rd_Re^{-\frac{\tilde{\lambda}}{2N}
C_2(R)}\chi_R(WUW^{-1}VWU^{-1}W^{-1}V^{-1})e^{-f(U)}e^{-g(V)}. }
Here, we have used the invariance of both the measure $DU$ and of the
function $f(U)$ under the transformation used in the change of
variables. To proceed further, we seek to evaluate the integral
\eqn\desint{I(D_1,D_2)=\int\,DC\,\chi_R(CD_1C^{-1}D_2CD_1^{-1}C^{-1}D_2^{-1}).}
Using $SU(N)$ symmetry, the form of $I(D_1,D_2)$ is restricted to
\eqn\restint{I(D_1,D_2)=\alpha+\beta\left[\chi_R(D_1)\chi_R(D_1^{-1})+
\chi_R(D_2)\chi_R(D_2^{-1})\right]+\gamma\chi_R(D_1)\chi_R(D_1^{-1})
\chi_R(D_2)\chi_R(D_2^{-1}).}
To compute $I(D_1,D_2)$, we thus need only to determine the three numbers
$\alpha,\beta,\gamma$.  To constrain their values, let us first look
at $I(1,D_2)$.  Equating \desint\ and \restint\ we obtain
\eqn\fircons{\alpha+\beta\left[d_R^2+\chi_R(D_2)\chi_R(D_2^{-1})\right]+
\gamma d_R^2\chi_R(D_2)\chi_R(D_2^{-1})=d_R.}
We next consider integrating $\int\,DD_1 I(D_1,D_2)$.  Comparing the result obtained by interchanging the integrations over $D_1$ and $C$ yields
\eqn\seccons{\alpha+\beta\left[1+\chi_R(D_2)\chi_R(D_2^{-1})\right]+
\gamma\chi_R(D_2)\chi_R(D_2^{-1})=\frac{1}{d_R}\chi_R(D_2)\chi_R(D_2^{-1}).}
Equations \fircons\ and \seccons\ give 4 equations relating
the coefficients $\alpha,\beta,\gamma$ :
\eqn\conssys{\eqalign{\alpha+\beta d_R^2&=d_R,\cr
\beta+\gamma d_R^2&=0,\cr
\alpha+\beta &= 0,\cr
\beta+\gamma&=\frac{1}{d_R}.}}
This system has a unique solution
\eqn\syssol{\eqalign{\alpha&=-\frac{d_R}{d_R^2-1},\cr
\beta&=\frac{d_R}{d_R^2-1},\cr
\gamma&=-\frac{1}{d_R(d_R^2-1)}.}}
Applying \restint\ with these values to our expression \varchange, we
may rewrite $Z_{YM}$ as \notpureym.

\appendix{E}{More about the Monte-Carlo simulations}

The Monte-Carlo simulations presented here were written using an
elementary implementation of the Metropolis algorithm. The ensembles
were fully thermalized between samplings with various time-time
correlators being checked.

In the case of the 0+0 matrix integrals the implementation is
extremely simple. For the 0+1 matrix quantum mechanics we require a
lattice of $L$ spatial sites, with circle topology, and each site is
equipped with the scalar adjoint matter matrices, with the gauge field
living as usual on the links. Since the gauge dynamics is trivial in
this low dimension we may perform a gauge transformation to make the
unitary link variables equal on all sites. Furthermore we may use up
the remaining gauge freedom by diagonalizing this unitary link
matrix. The remaining diagonal components are pure phases, and are
physical, giving the eigenvalues of the Polyakov loop when raised to
the power $L$, the number of lattice sites. We carefully ensure that
the Jacobian introduced by this gauge fixing is properly
implemented. This unitary matrix measure is the non-perturbative
version of the Vandermonde determinant and is given by $\Pi_{i<j}
\sin^2 \frac{L}{2} (\theta_i -
\theta_j)$ for $i=1,\cdots,N$, where $e^{i \theta_i}$ are the
eigenvalues of the unitary link. This measure factor is implemented by
taking its log and introducing it as a potential term in the
action. We automatically adjust the Metropolis step size to ensure
decent acceptance rates, and use independent step sizes for the
unitary link eigenvalues, and for both the scalar adjoint matter
diagonal, and off-diagonal components.

For the large $N$ behaviour we study we require relatively few lattice
points to accurately capture the continuum behaviour. For $p$ scalars
with $p = 2, 4$ the data presented here uses 10 lattice points. For
the $p = 9$ data 5 lattice sites were used. For individual values of
$\lambda$ and $M$ we checked that this was sufficient for the
quantities we measured, finding that doubling or quadrupling the
number of lattice sites did not change the results at the level of
one percent.

\listrefs

\end